\documentclass[%
    aps,
    prd,
    letterpaper,
    superscriptaddress,
    preprintnumbers,
    floatfix,
    twocolumn,
    nofootinbib,
    showpacs,
    hyphens,
    10pt
]{revtex4-2}  

\usepackage[normalem]{ulem}
\usepackage{subfigure}
\usepackage{amsfonts}
\usepackage{graphicx}
\usepackage{dcolumn}
\usepackage{siunitx}
\usepackage{bm}
\usepackage{bbm} 
\usepackage{hyperref}
\usepackage{amsmath,amsfonts,amssymb,amsthm}
\usepackage{enumitem}

\usepackage[pdftex]{xcolor}
\usepackage{hyperref}
\hypersetup{
    colorlinks=true,     
    linkcolor=blue,      
    citecolor=blue,      
    filecolor=blue,      
    urlcolor=blue        
}

\usepackage{cleveref}
\crefname{equation}{Eq.}{Eqs.}
\crefname{figure}{Fig.}{Figs.}
\crefname{table}{Tab.}{Tabs.}
\crefname{section}{Sec.}{Secs.}
\crefname{appendix}{App.}{Apps.}
\Crefname{table}{Table}{Tables}
\Crefname{figure}{Figure}{Figures}

\newcommand{\ket}[1]{\left| #1 \right>} 
\newcommand{\bra}[1]{\left< #1 \right|} 
\newcommand{\braket}[2]{\left< #1 \vphantom{#2} \right|
 \hspace{-2pt} \left. #2 \vphantom{#1} \right>} 
\newcommand{\mbraket}[3]{\left< #1 \vphantom{#2#3} \right|
 #2 \left| #3 \vphantom{#1#2} \right>} 
\newcommand{\avg}[1]{\left< #1 \right>}
\usepackage{slashed}

\newcommand{\iket}[1]{\big| #1 \big>} 
\newcommand{\iavg}[1]{\big< #1 \big>} 

\newcommand{\Tm}[0]{T^{(m)}}

\newcommand{\CWm}[0]{\widetilde{T}^{(m)}}

\newcommand{\Tr}{{\mathrm{Tr}}}

\newcommand{\DCW}[0]{\Delta^{\mathrm{CW}(m)}_k}
\newcommand{\DZCW}[0]{\Delta^{\mathrm{ZCW}(m)}_k}
\newcommand{\eCW}[0]{\varepsilon^{\mathrm{CW}}}
\newcommand{\eZCW}[0]{\varepsilon^{\mathrm{ZCW}}}
\newcommand{\eRe}[0]{\varepsilon^{\mathrm{float}}}
\newcommand{\betaT}[0]{N_t}

\begin{document}
\title{Block Lanczos algorithm for lattice QCD spectroscopy and matrix elements}

\author{Daniel C. Hackett}
 \affiliation{Fermi National Accelerator Laboratory, Batavia, IL 60510, USA}
\author{Michael L. Wagman}
 \affiliation{Fermi National Accelerator Laboratory, Batavia, IL 60510, USA}  

\preprint{FERMILAB-PUB-24-0875-T}

\DeclareRobustCommand{\Eq}[1]{Eq.~\eqref{eq:#1}}
\DeclareRobustCommand{\Eqs}[2]{Eqs.~\eqref{eq:#1} and \eqref{eq:#2}}
\DeclareRobustCommand{\fig}[1]{Fig.~\ref{fig:#1}}
\DeclareRobustCommand{\figs}[2]{Figs.~\ref{fig:#1} and \ref{fig:#2}}
\DeclareRobustCommand{\app}[1]{App.~\ref{app:#1}}
\DeclareRobustCommand{\sec}[1]{Sec.~\ref{sec:#1}}
\DeclareRobustCommand{\secs}[2]{Secs.~\ref{sec:#1} and \ref{sec:#2}}
\DeclareRobustCommand{\tbl}[1]{Table~\ref{tbl:#1}}
\DeclareRobustCommand{\refcite}[1]{Ref.~\cite{#1}}
\DeclareRobustCommand{\refcites}[1]{Refs.~\cite{#1}}

\date{\today}

\begin{abstract}
Recent work~\cite{Wagman:2024rid,Hackett:2024xnx} introduced a new framework for analyzing correlation functions with improved convergence and signal-to-noise properties, as well as rigorous quantification of excited-state effects, based on the Lanczos algorithm and spurious eigenvalue filtering with the Cullum-Willoughby test.
Here, we extend this framework to the analysis of correlation-function matrices built from multiple interpolating operators in lattice quantum chromodynamics (QCD) by constructing an oblique generalization of the block Lanczos algorithm, as well as a new physically motivated reformulation of the Cullum-Willoughby test that generalizes to block Lanczos straightforwardly.
The resulting block Lanczos method directly extends generalized eigenvalue problem (GEVP) methods, which can be viewed as applying a single iteration of block Lanczos.
Block Lanczos provides qualitative and quantitative advantages over GEVP methods analogous to the benefits of Lanczos over the standard effective mass, including faster convergence to ground- and excited-state energies, explicitly computable two-sided error bounds, straightforward extraction of matrix elements of external currents, and asymptotically constant signal-to-noise.
No fits or statistical inference are required.
Proof-of-principle calculations are performed for noiseless mock-data examples as well as two-by-two proton correlation-function matrices in lattice QCD.
\end{abstract}

\maketitle

\section{Introduction}
\label{sec:intro}

Numerical lattice quantum chromodynamics (QCD) is an integral part of the particle and nuclear physics toolkit.
In lattice QCD, Euclidean-time path integrals are discretized and evaluated stochastically with Monte Carlo methods to estimate QCD correlation functions, allowing first-principles calculation of hadronic energies and matrix elements.
This approach has been employed with great success to study the dynamics of QCD, including its spectrum of states, scattering amplitudes, and various aspects of hadron structure, as well as to constrain the Standard Model.
However, in practice, the mutually compounding issues of exponentially decaying signal-to-noise and excited-state contamination limit what is possible.
This has lead to ongoing research into improved methods to alleviate these issues.

Challenges in reliably extracting the spectrum of QCD from single correlation functions (``correlators'') have lead to the development of methods which analyze correlation-function matrices (``correlator matrices'') defined from sets of multiple interpolating operators (``interpolators''), each providing complementary information.
This includes the variational method based on solving generalized eigenvalue problems (henceforth, GEVP methods or GEVP)~\cite{Fox:1981xz,Michael:1982gb,Luscher:1990ck,Blossier:2009kd,Fleming:2023zml,Fischer:2020bgv}, the block generalization~\cite{Fleming:2023zml} of Prony's method~\cite{Prony,Fleming:2004hs,Lin:2007iq,Fleming:2009wb}, and variations on generalized pencil of functions (GPOF)~\cite{Aubin:2010jc,Aubin:2011zz,Green:2014xba,Ottnad:2017mzd,Ottnad:2020qbw,Fischer:2020bgv}.
GEVP methods are the present standard and state of the art for spectroscopy.
In practice these methods provide better control over excited-state contamination than that available with single-correlator analyses, in both application to spectroscopy as well as in their generalizations to the extraction of matrix elements.
Importantly, GEVP energy levels admit rigorous interpretation as variational upper bounds, allowing determination of excited-state energies with controlled uncertainties.
However, these bounds are one-sided, leading to various well-known difficulties in interpretation.

Recent work~\cite{Wagman:2024rid,Hackett:2024xnx} has explored a new analysis framework based on the Lanczos algorithm.
In this framework, bounds on numerical Lanczos convergence properties can be reframed to provide rigorous two-sided bounds on energy levels in correlator analyses.
Separately, tools developed to treat numerical noise from finite-precision arithmetic can be repurposed to discard spurious states that arise in application to noisy correlator data.
The resulting Lanczos estimators have improved signal-to-noise properties, without the exponential degradation in Euclidean time separation present in ``effective mass'' estimators.
However, the method has thus far been limited to analysis of single (scalar) correlation functions.

In this work, we extend the Lanczos analysis formalism to treat correlator matrices.
The framework is built around the block Lanczos algorithm~\cite{Golub:1973,Cullum:1974,Golub:1977,Saad:1980,Parlett}, which generalizes the standard (scalar) Lanczos algorithm to use a basis of multiple vectors.
We construct its oblique generalization~\cite{Kim:1988,Kim:1990,Bai:1999,Abdel-Rehim:2014}, which uses distinct left and right bases, in the form of recusion relations that can be applied to treat noisy correlator data~\cite{Wagman:2024rid,Hackett:2024xnx}.
This method yields simple block generalizations of the recursion relations and change-of-basis matrices that arise for oblique Lanczos in the scalar case, allowing similar constructions of estimators for energies, overlap factors, and operator matrix elements.

The key element to the improved signal-to-noise properties of the Lanczos framework is filtering of spurious states that arise as noise artifacts.
We reformulate the methods introduced in Refs.~\cite{Wagman:2024rid,Hackett:2024xnx}---the Cullum-Willoughby and Hermitian subspace tests---to filter spurious noise-artifact states and generalize them to the block case.
In the process, we find that the Cullum-Willoughby test used in previous examples admits a simple physical interpretation as identifying states that have zero overlap with the entire set of interpolating operators and therefore have ``wrong quantum numbers'' that should not arise from application of the physical transfer matrix.
With this interpretation, the entire state filtering scheme can be understood in physical terms as restricting to the subspace of Hilbert space of physical interest.

We test the oblique block Lanczos method on noiseless mock data and a noisy LQCD $2\times 2$ nucleon correlator matrix.
We demonstrate that the improved convergence and signal-to-noise properties observed in the scalar case extend to the block case as well.
We find that the block analysis provides nontrivial advantages over separate applications of the scalar analysis, including cleaner state resolution and access to relative phases of transition matrix elements.

We find that GEVP is the one-step limit of block Lanczos, and thus that block Lanczos is a direct extension and superset of GEVP methods.
In comparisons on noiseless and noisy data, we find that block Lanczos offers similar advantages over GEVP as does scalar Lanczos over effective energies and other effective estimators.

The remainder of this paper proceeds as follows.
In \cref{sec:formalism}, we construct an oblique block Lanczos algorithm and corresponding recursion relations for analyzing correlator matrices.
In \cref{sec:noiseless}, we apply the recursive block Lanczos analysis formalism to extraction of the spectrum, overlap factors, and matrix elements in noiseless mock-data examples, comparing its performance with both GEVP and scalar Lanczos.
In \cref{sec:bounds}, we discuss the convergence properties of this algorithm and construct two-sided bounds allowing rigorous determination of energies.
In \cref{sec:lattice}, we discuss the physics of the Cullum-Willoughby test and spurious eigenvalue filtering and demonstrate how block Lanczos can be applied to noisy data, extracting the spectrum and overlap factors.
In \cref{sec:gevp}, we review the GEVP formalism and discuss how it coincides with single-iteration block Lanczos.
We conclude in \cref{sec:conclusions}.

Several appendices complement the main text.
\cref{app:ortho} presents further details of the derivation of the oblique block Lanczos algorithm.
\cref{app:KPS} translates notation from Ref.~\cite{Saad:1980} and derives the form of the KPS bound presented in \cref{sec:bounds}.
\cref{app:scalar-res-bounds} presents additional analysis of noiseless mock data and addresses the potential for ``pseudo-plateau'' behavior.
\cref{app:real} proves that correlator matrices constructed with most commonly used interpolating operators are real in expectation.
\cref{app:leftGEVP} discusses the left GEVP and differences with the standard right GEVP that can arise for noisy data applications.
\cref{app:ci} discusses Gaussianity tests and uncertainty assumptions for bootstrap-median and sample-mean estimators. 
\cref{app:PGEVM} proves that block Lanczos is not equivalent to the correlator matrix analysis method introduced in Ref.~\cite{Fischer:2020bgv} based on applying Prony's method to GEVP results.

\section{Block Lanczos formalism}
\label{sec:formalism}

This section presents the oblique block Lanczos algorithm for rank-$r$ correlator matrices of the form
\begin{equation}\label{eq:Cab}
  C_{ab}(t) = \left< \chi_a(t) \psi^\dagger_b(0) \right> = \mbraket{\chi_a}{T^t}{\psi_b},
\end{equation}
where $t \in \{0,\ldots,\betaT -1\}$, $T$ is the transfer matrix, and $\chi_a(t)$ and $\psi_a(t)$ are interpolating operators chosen to excite states
\begin{equation}
     \ket{\psi_b} = \psi_b^\dagger \ket{\Omega}
     \quad \text{and} \quad 
     \ket{\chi_a} = \chi_a^\dagger \ket{\Omega}
\end{equation}
with the quantum numbers of physical interest from the vacuum $\ket{\Omega}$. 
For a zero-temperature lattice quantum field theory, the transfer matrix is typically Hermitian and positive-definite.\footnote{This implies real, positive eigenvalues. 
Complex transfer-matrix eigenvalues can arise in theories with improved actions,  but they are typically exponentially suppressed in the continuum limit compared to a sector of (strictly positive) physical eigenvalues~\cite{Luscher:1984is}.
Negative eigenvalues can arise for theories with staggered fermions, where they can be treated analogously to thermal eigenvalues~\cite{Wagman:2024rid}.}
For lattice gauge theories, $T$ is infinite dimensional~\cite{Kogut:1974ag}, in which case objects like $\ket{\psi_a}$ and $\bra{\chi_a}$ are vectors in the infinite-dimensional Hilbert space of states.
The (non-thermal)\footnote{For a finite-temperature theory, such as Euclidean LQCD with a finite temporal extent, $T$ should be augmented to include ``backwards-propagating'' thermal states as described in Appendix A of Ref.~\cite{Wagman:2024rid}. 
The resulting thermal eigenvalues are eigenvalues of the augmented transfer matrix with magnitudes larger than unity~\cite{Wagman:2024rid}. Their determination is discussed in \cref{sec:thermal}.} eigenvalues of $T$ are denoted $\lambda_n$ and ordered such that $1\geq \lambda_0 \geq \lambda_1 \geq \ldots \geq 0$.
The spectrum of LQCD energies is related to the transfer matrix elements by $E_n = -\ln \lambda_n$ where units in which the lattice spacing is set to unity are used here and below.

In this section, Hermiticity is \emph{not} assumed for either $T$ or $C_{ab}(t)$.
Restrictions to Hermitian $T$ and to real, symmetric $C_{ab}(t)$ will be made in Sec.~\ref{sec:bounds} and Sec.~\ref{sec:lattice}, respectively.
To simplify the discussion of applications to noisy LQCD data it is convenient to begin by constructing the oblique block Lanczos algorithm for a general infinite-dimensional operator $T$ and time series of $r \times r$ invertible matrices $C_{ab}(t)$.
For $r = 1$, oblique block Lanczos reduces to the (scalar) oblique Lanczos algorithm~\cite{Saad:1982}.

To avoid proliferation of indices below, we use boldface matrix notation for block indices $a,b,c,d \in \{1,\dots,r\}$ while keeping all other indices explicit. 
When a symbol like 
$C_{ab}(t)$ is bolded, $\bm{C}(t)$, it indicates the same quantity as a matrix in the missing indices $a,b$. 
Brackets with subscripts around a matrix expression indicate taking the indices of the result, e.g.
\begin{equation}
    [\bm{a}_j]_{ab} \equiv \alpha_{jab}
    \quad \text{or} \quad 
    \left[ \bm{A} \bm{B} \right]_{ab} \equiv \sum_c A_{ac} B_{cb} ~ .
\end{equation}
Bolded symbols are always matrices and never vectors. 
All sums are taken explicitly, with repeated indices never indicating summation.

The practical steps required to carry out a Lanczos correlator matrix analysis are summarized here for convenience:
\begin{enumerate}
    \item Apply oblique block Lanczos recursion relations to $\bm{C}(t)$ to compute the elements of the block-tridiagonal matrix $\bm{T}^{(m)}_{ij}$ (\cref{sec:formalism:recursion}).
    \item Diagonalize $\bm{T}^{(m)}_{ij}$ to obtain Ritz values $\lambda^{(m)}_k$ and the eigenvectors $\omega^{(m)}_{iak}$ (\cref{sec:formalism:ritzes}). These provide all information necessary to estimate energies, overlaps (\cref{sec:formalism:overlaps}), and for spurious state filtering in the noisy case (\cref{sec:lattice}).
    \item If residual bounds (\cref{sec:bounds}) or operator matrix elements are desired, evaluate an auxiliary recursion to compute the Krylov coefficients $\bm{K}^{R/L}_{tj}$ (\cref{sec:formalism:krylov}).
    \item If operator matrix elements are desired, compute the Ritz coefficients $P^{R/L(m)}_{kta}$ (\cref{sec:formalism:ritz-rot}), normalize the Ritz vectors by computing overlap factors (\cref{sec:formalism:overlaps}), and evaluate the matrix element estimators (\cref{sec:formalism:MEs}).
    \item For noisy data, identify and filter spurious states using the Hermitian-subspace test (\cref{sec:Hermitian}) and the ``block Z-factor Cullum-Willoughby (ZCW) test'' (\cref{sec:CW}).
    \item Estimate uncertainties idiomatically using (nested) bootstrapping (\cref{sec:median}).
    \item Extract thermal modes separately using a thermal ZCW test if desired (\cref{sec:thermal}).
\end{enumerate}

\subsection{Oblique block Lanczos}
\label{sec:formalism:oblock}

An oblique block Lanczos algorithm iteratively constructs a sequence of left- and right-Lanczos vectors $\bra{v_{ia}^L}$ and $\ket{v_{jb}^R}$ indexed by $i,j = 1 \ldots m$ where $m$ is the iteration count.
The Lanczos vectors are defined to satisfy the generalized bi-orthogonality condition $\braket{v_{ia}^L}{v_{jb}^R} = \delta_{ij}\delta_{ab}$ along with the three-term recurrence relations
\begin{equation}\label{eq:recurrence}
  \begin{split}
  T \ket{v^R_{ja}} &=  \sum_b \left(  \ket{v^R_{jb}}\alpha_{jba} + \ket{v^R_{(j-1)b}}\beta_{jba}  \right. \\
  &\hspace{40pt} \left. + \ket{v^R_{(j+1)b}}\gamma_{(j+1)ba} \right), \\
   \bra{v^L_{ja}} T  &=  \sum_b \left( \alpha_{jab} \bra{v^L_{jb}} + \gamma_{jab}  \bra{v^L_{(j-1)b}} \right. \\
   &\hspace{40pt} \left. + \beta_{(j+1)ab} \bra{v^L_{(j+1)b}} \right),
  \end{split}
\end{equation}
where the block-tridiagonal matrix elements $\alpha_{jab}$, $\beta_{jab}$, and $\gamma_{jab}$ are defined as
\begin{equation}\label{eq:alpha_beta_gamma}
  \begin{split}
    \alpha_{jab} &\equiv \bigl< v^L_{ja} \big| T \big| v^R_{jb} \bigr>, \\
    \beta_{jab} &\equiv \bigl< v^L_{(j-1)a} \big| T \big| v^R_{jb} \bigr>, \\
    \gamma_{jab} &\equiv \bigl< v^L_{ja} \big| T \big| v^R_{(j-1)b} \bigr>.
  \end{split}
\end{equation}
The consistency of these definitions with each other and with the bi-orthonormality condition $\bigl< v^L_{ia} \big| v^R_{jb} \bigr> = \delta_{ij} \delta_{ab}$ is derived in Appendix~\ref{app:ortho}.

To construct each new set of Lanczos vectors, the algorithm 1) applies $T$ and orthogonalizes with previous iterations to make a new set of left and right residual vectors, then 2) biorthonormalizes the new left and right residual vectors to obtain the Lanczos vectors.
How to initialize the iteration is discussed below.
Step 1) defines the residual vectors for $j > 1$ as
\begin{equation}\label{eq:lr_vecs}
\begin{split}
    \bigl| r^R_{(j+1)a} \bigr> \equiv  T \ket{v^R_{ja}} - \sum_b \left( \ket{v^R_{jb}} \alpha_{jba}  +  \bigl| v^R_{(j-1)b} \bigr> \beta_{jba} \right), \\
    \bigl< r^L_{(j+1)a} \bigr| \equiv \bra{v^L_{ja}} T -  \sum_b \left( \alpha_{jab} \bra{v^L_{jb}}  + \gamma_{jab} \bigl< v^L_{(j-1)b} \bigr| \right),
\end{split}
\end{equation}
where terms involving $j-1$ are omitted for the case $j=1$.
In each equation, the first term applies $T$ to expand the Krylov space, and the remaining terms implement orthogonality with all previous Lanczos vectors.
However, the new sets of residual vectors are not biorthonormal, as quantified by the residual norm,
\begin{equation}\label{eq:Delta-res-norm-def}
    \Delta_{jab} \equiv \braket{r^L_{ja}}{r^R_{jb}} ~ .
\end{equation}
Step 2) thus decomposes the residual norm as
\begin{equation}\label{eq:Delta-beta-gamma-decomp}
    \Delta_{jab} = \sum_c \beta_{jac} \gamma_{jcb}
    \quad \left(\text{i.e., }
    \bm{\Delta}_j = \bm{\beta}_j \bm{\gamma}_j
    \right)
\end{equation}
to define the new set of Lanczos vectors as
\begin{equation}\label{eq:lr_norms}
\begin{aligned}
    \bigl| v^R_{ja} \bigr> &\equiv \sum_b  \bigl| r^R_{jb}\bigr> \gamma^{-1}_{jba}, \\
    \bigl< v^L_{ja} \bigr| &\equiv \sum_b \beta^{-1}_{jab} \bigl< r^L_{jb} \bigr|,
\end{aligned}
\end{equation}
for which $\braket{v^L_{ja}}{v^R_{jb}} = \delta_{ab}$.
As in the case of scalar oblique Lanczos, there is no unique correct choice for the decomposition $\bm{\Delta}_j = \bm{\beta}_j \bm{\gamma}_j$. 
The same feature\footnote{The left and right Lanczos vectors $\bigl< v^L_{(j+1)a} \bigr|$ and $\bigl| v^R_{(j+1)a} \bigr>$ are defined so that their spans are identical to those of the residuals $\bigl< r^L_{(j+1)a} \bigr|$ and $\bigl| r^R_{(j+1)a} \bigr>$.
This ensures that the action of $T$ on the $j$-th Lanczos vectors results only in terms proportional to the $(j+1)$-th Lanczos vectors in addition to the terms proportional to the $(j-1)$-th Lanczos vectors explicitly appearing in Eq.~\eqref{eq:lr_vecs}.
This feature is what leads to the three-term recurrence in Eq.~\eqref{eq:recurrence}.} leading to a three-term recurrence is achieved for any definition of $\bigl| v^R_{(j+1)a} \bigr>$ with $\text{span}\{ \bigl| v^R_{(j+1)a} \bigr> \} = \text{span}\{ \bigl| r^R_{(j+1)a} \bigr> \}$.
Thus, any choice for which $\bm{\beta}_j$ and $\bm{\gamma}_j$ are both invertible is a valid convention.

We present several options for block oblique conventions for concreteness, but first emphasize that all physical results are necessarily independent of convention; other than insight, different choices can only offer advantages in numerical precision and efficiency.
One simple choice is to take $\bm{\beta}_j = \bm{\Delta}_j$ such that $\bm{\gamma}_{j} = \bm{1}$; this is invoked as a simple way to connect block Lanczos to GEVP methods in \cref{sec:gevp}.
Roots of matrices offer another set of options, as well as any factorization applicable for complex, non-Hermitian matrices; this includes LU and QR, but not Cholesky.
Another class of conventions follows from the eigendecomposition
\begin{equation}
    \Delta_{jab} = \sum_c \Omega_{jac}^{-1} \Lambda_{jc} \Omega_{jcb},
\end{equation}
which for any choice of $\rho_{ja} \tau_{ja} = \Lambda_{ja}$ can be used to define
\begin{equation}
    \bm{\gamma}_j = \bm{\rho}_j \bm{\Omega}_j^{-1}, \hspace{20pt}
    \bm{\beta}_{j} = \bm{\Omega}_j \bm{\tau}_j ,
\end{equation}
where $\bm{\rho}_j$ and $\bm{\tau}_j$ are diagonal matrices with entries $\rho_a \delta_{ab}$ and $\tau_a \delta_{ab}$, respectively.
This convention intuitively separates orthogonalization, implemented by $\Omega$ and $\Omega^{-1}$, and normalization, implemented by $\rho$ and $\tau$; the remaining ambiguity in the choice of $\rho, \tau$ is the same ambiguity as arises in the scalar case, which is recovered identically for $r=1$ where $\bm{\Omega}_j = 1$.
This convention is used to produce the numerical results below.

In practice, the initial blocks of right and left states, $\ket{\psi_a}$ and $\bra{\chi_a}$, will typically not satisfy biorthonormality and thus cannot be used immediately as the initial Lanczos vectors $\iket{v^{R/L}_{1a}}$.
They instead must be treated as with the residual vectors $\iket{r^{R/L}_{ja}}$ above and orthonormalized before proceeding.
It is cleanest to define this similarly to subsequent biorthonormalizations, in terms of the decomposition
\begin{equation}
    \braket{\chi_a}{\psi_b} \equiv [\bm{\Delta}_1]_{ab} = [ \bm{\beta}_1 \bm{\gamma}_1 ]_{ab} ~ ,
    \label{eq:bg1_def}
\end{equation}
with a similar choice of convention required.
This then defines the initial Lanczos vectors as
\begin{equation}
  \begin{split}
    \ket{v_{1a}^R} 
    &= \sum_b \ket{ \psi_b } \gamma_{1ba}^{-1} \,, \\
    \bra{v_{1a}^L}
    &= \sum_b \beta_{1ab}^{-1} \bra{ \chi_b } ~ .
  \end{split}
  \label{eq:initial-lanczos-vectors}
\end{equation}
These relations are used ubiquitously below to relate expressions in terms of Lanczos vectors to correlator data via
\begin{equation}\begin{aligned}
    \mbraket{v^L_{1a}}{T^t}{v^R_{1b}}
    &= \sum_{cd} \beta^{-1}_{1ac} \mbraket{\chi_c}{T^t}{\psi_d} \gamma^{-1}_{1db}
    \\&= [\bm{\beta}^{-1}_1 \bm{C}(t) \bm{\gamma}^{-1}_1]_{ab},
\end{aligned}\end{equation}
and similar expressions for three-point functions and quantities defined with e.g.~right Lanczos vectors on both sides, etc.
However, we emphasize that $\bm{\beta}_1$ and $\bm{\gamma}_1$ do not appear explicitly in the tridiagonal matrix $T^{(m)}$ defined below, and instead generalize the normalization factors $|\psi|$ and $|\chi|$ from the scalar case.

\subsection{Recursion relations}
\label{sec:formalism:recursion}

To apply oblique block Lanczos to analyze LQCD correlator matrices, the iterative procedure on infinite-dimensional Hilbert-space vectors defined above must be used to derive recursion relations on finite-dimensional matrix elements---i.e., correlation function data of the form \cref{eq:Cab}.

To begin, we first compute $\bm{\beta}_1$ and $\bm{\gamma}_1$ by decomposing (with whatever choice of convention)
\begin{equation}\label{eq:C0_bg}
    C_{ab}(0) = \braket{\chi_a}{\psi_b} = [\bm{\beta}_1 \bm{\gamma}_1]_{ab}\,,
\end{equation}
per \cref{eq:bg1_def} above, recognizing the definition \cref{eq:Cab}.
When $r=1$, these reduce to the ubiquitous factors of $C(0)$ in the scalar formalism.
The first block of the Lanczos approximation to the transfer matrix is then given by
\begin{equation}
    \begin{split}
      \alpha_{1ab} \equiv \mbraket{v_{1a}^L}{T}{v_{1b}^R} 
      = \sum_{cd} \beta^{-1}_{1ac} C_{cd}(1) \gamma^{-1}_{1db} ,
    \end{split}
\end{equation}
or in matrix form
\begin{equation}\label{eq:alpha1}
  \bm{\alpha}_1 = \bm{\beta}_1^{-1} \bm{C}(1) \bm{\gamma}_1^{-1}.
\end{equation}
As discussed further in \cref{sec:gevp:vs-block}, the eigenvalues of $\bm{\alpha}_1$ are the same as the generalized eigenvalues provided by the GEVP method, given appropriately aligned definitions.

Further iterations allow improving this initial approximation of the transfer matrix.
The recursions are most naturally defined in terms of the residuals
\begin{equation}
    \Delta_{jab} = \braket{r^L_{ja}}{r^R_{jb}},
\end{equation}
and the generalized correlators
\begin{equation}
  \begin{split}
    A_{jab}(t) &\equiv \mbraket{v^L_{ja}}{T^t}{v^R_{jb}}, \\
    G_{jab}(t) &\equiv \mbraket{v^L_{ja}}{T^t}{v^R_{(j-1)b}}, \\
    B_{jab}(t) &\equiv \mbraket{v^L_{(j-1)a}}{T^t}{v^R_{jb}}.
  \end{split}
\end{equation}
The $j=1$ and $j=2$ terms form the recursion base case.
For $j=1$, the matrix elements defining $\bm{G}_{1}$ and $\bm{B}_{1}$ are not well-defined and only 
\begin{equation}\label{eq:Ak1}
  \bm{A}_1(t) = \bm{\beta}_1^{-1} \bm{C}(t) \bm{\gamma}_1^{-1}.
\end{equation}
is required.
The residual norm after the first iteration can then be computed and decomposed as
\begin{equation}\label{eq:sr2}
        \bm{\Delta}_{2}  
        = \bm{A}_{1}(2)  - \bm{\alpha}_{1} \bm{\alpha}_{1}
        \equiv \bm{\beta}_2 \bm{\gamma}_2,
\end{equation}
to define $\bm{\beta}_2$ and $\bm{\gamma}_2$ using some convention for decomposing $\bm{\Delta}_{2}$ into a product of matrices  as discussed above.
The next step proceeds by computing
\begin{equation}
    \begin{split}
      \bm{G}_{2}(t)&= \bm{\beta}^{-1}_{2}\left[ 
            \bm{A}_{1}(t+1) -  \bm{\alpha}_{1} \bm{A}_{1}(t) \right],
            \end{split}
\end{equation}
\begin{equation}
    \begin{split}
       \bm{B}_{2}(t) &= \left[ \bm{A}_{1}(t+1) -  \bm{A}_{1}(t) \bm{\alpha}_{1}  \right] 
            \bm{\gamma}_{2}^{-1}  ,
            \end{split}
\end{equation}
\begin{equation}\label{eq:diag_regular_2}
    \begin{split}
      \bm{A}_{2}(t)  &=    \bm{\beta}_{2}^{-1} \left[ 
         \bm{A}_{1}(t+1)  +  \bm{\alpha}_{1} \bm{A}_{1}(t) \bm{\alpha}_{1} \right. \\
        &\hspace{20pt} \left. -  \left( \bm{\alpha}_{1}  \bm{A}_{1}(t+1) +  \bm{A}_{1}(t+1) \bm{\alpha}_{1} \right)    
        \right] \bm{\gamma}_{2}^{-1} .
    \end{split}
\end{equation}

Once these initial computations are completed to seed the recursion, the calculation of quantities for subsequent $j\geq 3$ proceeds regularly.
In each step, first, $\bm{\beta}_{j+1}$ and $\bm{\gamma}_{j+1}$ are obtained by decomposing the residual norm
\begin{equation}\label{eq:sr}
        \bm{\Delta}_{j+1}  
        = \bm{A}_{j}(2)  - \bm{\alpha}_{j} \bm{\alpha}_{j} - \bm{\gamma}_{j} \bm{\beta}_{j} 
        \equiv \bm{\beta}_{j+1} \bm{\gamma}_{j+1} ~ .
\end{equation}
Then, the next set of generalized correlators can be computed as
\begin{equation}
    \begin{split}
      \bm{G}_{j+1}(t) &= \bm{\beta}_{j+1}^{-1} \left[ 
            \bm{A}_{j}(t+1) -  \bm{\alpha}_{j} \bm{A}_{j}(t) - \bm{\gamma}_{j} \bm{B}_{j} (t) \right],
            \end{split}
\end{equation}
\begin{equation}
    \begin{split}
       \bm{B}_{j+1}(t) &= \left[ \bm{A}_{j}(t+1) -  \bm{A}_{j}(t) \bm{\alpha}_{j}  - \bm{G}_{j}(t) \bm{\beta}_{j}  \right] 
            \bm{\gamma}_{j+1}^{-1} ,
            \end{split}
\end{equation}
\begin{equation}\label{eq:diag_regular}
    \begin{split}
      \bm{A}_{j+1}(t) &=    \bm{\beta}_{j+1}^{-1} \left[ 
         \bm{A}_{j}(t+2)  -  \left( \bm{\alpha}_{j}  \bm{A}_{j}(t+1) +  \bm{A}_{j}(t+1) \bm{\alpha}_{j} \right) \right. \\
         &\hspace{20pt} \left. +  \bm{\alpha}_{j} \bm{A}_{j}(t) \bm{\alpha}_{j} + \bm{\gamma}_{j} \bm{A}_{(j-1)}(t) \bm{\beta}_{j}  \right. \\
        &\hspace{20pt}  - \left( \bm{\gamma}_{j} \bm{B}_{j}(t+1) + \bm{G}_{j}(t+1) \bm{\beta}_{j} \right) \\
        &\hspace{20pt} \left. +  \bm{\gamma}_{j} \bm{B}_{j}(t) \bm{\alpha}_{j} + \bm{\alpha}_{j} \bm{G}_{j}(t) \bm{\beta}_{j}
        \vphantom{H_{jed}}
        \right] \bm{\gamma}_{j+1}^{-1}.
    \end{split}
\end{equation}
The next diagonal block of the Lanczos transfer matrix approximation is then given by
\begin{equation}
  \begin{split}
    \bm{\alpha}_{j+1} &= \bm{A}_{j+1}(1).
  \end{split}
\end{equation}
Although $\bm{\beta_j}$ and $\bm{\gamma_j}$ are computed otherwise, the off-diagonal blocks satisfy
\begin{equation}
  \begin{split}
    \bm{\gamma}_{j+1} = \bm{G}_{j+1}(1)
    \text{  and  }
    \bm{\beta}_{j+1} = \bm{B}_{j+1}(1),
  \end{split}
\end{equation}
which can be useful for consistency checks.

\subsection{Ritz values and vectors}
\label{sec:formalism:ritzes}

After $m$ iterations of oblique block Lanczos, the block-tridiagonal matrix
\begin{equation}\label{eq:Tmdef}
  \Tm_{iajb} \equiv \mbraket{v^L_{ia}}{T}{v^R_{jb}}, 
\end{equation}
expressing matrix elements of $T$ in the Lanczos-vector basis is given by
\begin{equation} \label{eq:untri}
  \Tm_{iajb} = \begin{pmatrix} \alpha_{1ab} & \beta_{2ab} &  & & & 0 \\ 
    \gamma_{2ab} & \alpha_{2ab} & \beta_{3ab} & & & \\
    & \gamma_{3ab} & \alpha_{3ab} & \ddots & &  \\
    & & \ddots & \ddots & \beta_{(m-1)ab} & \\
    & & & \gamma_{(m-1)ab} & \alpha_{(m-1)ab} & \beta_{mab} \\
  0 & & & & \gamma_{mab} & \alpha_{mab} \end{pmatrix}_{ij}.
\end{equation}
This matrix may be understood as the matrix elements of the true transfer matrix $T$ in the Krylov subspace of Hilbert space, i.e.,
\begin{equation}\label{eq:TmME}
    \mbraket{v^L_{ia}}{T^{(m)}}{v^R_{jb}} \equiv  \begin{cases} \Tm_{iajb}, & i, j \leq m \\
    0, & \text{otherwise} \end{cases}.
\end{equation}
It can be used to define a Hilbert-space operator
\begin{equation}\label{eq:Tm_braket}
\begin{aligned}
    \Tm &= \sum_{i,j = 1}^m \sum_{ab} \ket{v^R_{ia}} \mbraket{v^L_{ia}}{T}{v^R_{jb}} \bra{v^L_{jb}}\\
    &= \sum_{ijab} \ket{v^R_{ia}} T^{(m)}_{iajb} \bra{v^L_{jb}},
\end{aligned}
\end{equation}
which can be identified as $T$ multiplied by projection operators $\sum_{ia} \ket{v^R_{ia}} \bra{v^L_{ia}}$ that restrict its left- and right-action to left- and right-Krylov spaces defined by 
\begin{equation}
    \begin{split}
        \mathcal{K}^{L(m)} &= \text{span}\{ \ket{v^L_{ia}} \ | \ i=1,\ldots,m,\ a=1,\ldots,r \}, \\
        \mathcal{K}^{R(m)} &= \text{span}\{ \ket{v^R_{ia}} \ | \ i=1,\ldots,m,\ a=1,\ldots,r \}.
    \end{split}
\end{equation}
Thus, $T^{(m)}$ provides a Krylov-space approximation to the Hilbert-space operator $T$.

In this sense, the eigenvalues and eigenvectors of the Hilbert-space operator $\Tm$ provide the Lanczos algorithm's optimal approximations of the true transfer matrix eigenvalues and eigenvectors in the Krylov subspace.
Viewing $(ia)$ and $(jb)$ as composite indices, the matrix may be diagonalized as
\begin{equation}\label{eq:Tm_diag}
  \Tm_{iajb} =  \sum_k \omega^{(m)}_{iak} \, \lambda^{(m)}_k \, (\omega^{-1})^{(m)}_{kjb},
\end{equation}
where $\omega^{-1}$ is the matrix inverse of $\omega_{k(jb)}$.
The eigenvalues $\lambda^{(m)}_k$ are the Ritz values, from which Lanczos energy estimators are obtained via
\begin{equation}\label{eq:Em}
    E_k^{(m)} \equiv -\ln \lambda_k^{(m)}.
\end{equation}
These are the primary quantities required for LQCD spectroscopy analyses.
Calculations of matrix elements and residual bounds require the Ritz vectors, $\ket{y^{R/L(m)}}$, which decompose the Hilbert-space $\Tm$ as
\begin{equation}\label{eq:Tm_Ritz_decomp}
    \Tm = \sum_k \ket{y^{R(m)}_k} \lambda^{(m)}_k \bra{y^{L(m)}_k} ~ .
\end{equation}
Inserting \cref{eq:Tm_diag} into \cref{eq:Tm_braket} and comparing with \cref{eq:Tm_Ritz_decomp}, we may identify the right and left eigenvectors $\omega^{(m)}_k$ and $(\omega^{-1})^{(m)}_k$ of $\Tm_{iajb}$ as change-of-basis matrices relating Lanczos and Ritz vectors,
\begin{equation}\label{eq:Ritz_vecs}
  \begin{split}
    \bigl| y^{R(m)}_k \bigr> &\equiv \mathcal{N}^{(m)}_{k} \sum_{i=1}^m \sum_a \ket{ v^R_{ia} } \omega^{(m)}_{iak}, \\
    \bigl< y^{L(m)}_k \bigr| &\equiv  \frac{1}{\mathcal{N}^{(m)}_{k}} \sum_{i=1}^m \sum_a (\omega^{-1})^{(m)}_{kia} \bra{ v^L_{ia} } ,
  \end{split}
\end{equation}
where the $\mathcal{N}^{(m)}_{k}$ are constants used to normalize the Ritz vectors as discussed further below.
We may verify this construction by applying the Hilbert-space $\Tm$ to both sides of Eq.~\eqref{eq:Ritz_vecs}.
We first use Eq.~\eqref{eq:Tm_braket} and $\iavg{v_{ia}^L|v_{jb}^R} = \delta_{ij}\delta_{ab}$ to find the action of $\Tm$ on the Lanczos vectors,
\begin{equation}\label{eq:Tmv}
\begin{split}
    T^{(m)}\ket{v_{jb}^R} &= \sum_{i=1}^m \sum_b \ket{v_{ia}^R} T^{(m)}_{iajb}, \\
    \bra{v_{ia}^L} T^{(m)} &= \sum_{j=1}^m  \sum_b T^{(m)}_{iajb} \bra{v_{jb}^L}.
    \end{split}
\end{equation}
It is then straightforward to show that
\begin{equation}\label{eq:Ritz_vec_val}
\begin{split}
    \Tm \ket{y_k^{R(m)}} &=  \mathcal{N}^{(m)}_{k} \sum_{j=1}^m \sum_b \Tm \ket{v^R_{jb}} \omega^{(m)}_{jbk} \\
    &=  \mathcal{N}^{(m)}_{k} \sum_{j=1}^m \sum_{ab} \ket{v^R_{ia}} \Tm_{iajb} \omega^{(m)}_{jbk} \\
    &=  \mathcal{N}^{(m)}_{k} \sum_{i=1}^m  \sum_a \ket{v^R_{ia}} \omega^{(m)}_{iak} \lambda_k^{(m)} \\
    &= \ket{y_k^{R(m)}} \lambda_k^{(m)},
    \end{split}
\end{equation}
i.e., that $\ket{y^{R(m)}_k}$ is a right eigenvector of $\Tm$.
An identical proof applies to $\bra{y_k^{L(m)}}$.
The properties and uses of Ritz vectors are discussed further in the remainder of this section and in \cref{sec:bounds}.

\subsection{Krylov coefficients}
\label{sec:formalism:krylov}

Matrix element calculations essentially require performing changes of basis from the ``Krylov basis'' in which correlation functions are naturally defined to the Lanczos- and Ritz-vector bases~\cite{Hackett:2024xnx}.
The block Lanczos change-of-basis matrices are straightforward generalizations of their scalar Lanczos counterparts and explicitly constructed in this and the next subsection.

Krylov-basis vectors are defined by
\begin{equation}
    \ket{k^R_{ta}} \equiv T^t \ket{v^R_{1a}}, 
        \qquad  \bra{k^L_{ta}} \equiv \bra{v^L_{1a}} T^t.
\end{equation}
The Krylov coefficients $\bm{K}^R_{tj}$ and $\bm{K}^L_{jt}$ are defined to relate the Lanczos and Krylov bases as
\begin{equation}
\begin{split}
    \ket{v^R_{ja}} &= \sum_{t=0}^{j-1}  \sum_b \ket{k^R_{tb}} K^R_{tjba},
    \\
    \bra{v^L_{ja}} &= \sum_{t=0}^{j-1}  \sum_b K^L_{jtab} \bra{k^L_{tb}}  .
    \label{eq:krylov-coeff-def}
    \end{split}
\end{equation}
As in the scalar case~\cite{Hackett:2024xnx}, these may be computed from the elements of $\Tm_{iajb}$ by deriving an auxiliary recursion.
The $j=1$ Krylov coefficients follow directly from these definitions,
\begin{equation}\label{eq:K_0}
    K^{R}_{t1 ab} = K^L_{1t ab} = \delta_{t0} \delta_{ab}.
\end{equation}
The $j=2$ case can be computed by inserting these definitions into the recurrence Eq.~\eqref{eq:recurrence}, obtaining
\begin{equation}
\begin{split}
    \bm{K}^R_{t2} &= \delta_{t1} \bm{\gamma}^{-1}_{2} - \delta_{t0} \bm{\alpha}_{1} \bm{\gamma}^{-1}_{2}, \\
    \bm{K}^L_{2t} &= \delta_{t1} \bm{\beta}^{-1}_{2} - \delta_{t0}  \bm{\beta}^{-1}_{2} \bm{\alpha}_{1}
    \end{split}
\end{equation}
Krylov coefficients with ${j > 2}$ can be computed from these $j \in \{1,2\}$ results using the  recursion relations
\begin{equation}\label{eq:K_j}
\begin{split}
    \bm{K}^R_{t(j+1)} &= \left( \bm{K}^R_{(t-1)j} - \bm{K}^R_{tj} \bm{\alpha}_{j}  - \bm{K}^R_{t(j-1)} \bm{\beta}_{j} \right) \bm{\gamma}^{-1}_{j+1}, \\
    \bm{K}^L_{(j+1)t} &= \bm{\beta}^{-1}_{j+1} \left( \bm{K}^L_{j(t-1)}  - \bm{\alpha}_{j} \bm{K}^L_{jt} - \bm{\gamma}_{j} \bm{K}^L_{(j-1)t} \right).
    \end{split}
\end{equation}

Krylov coefficients can be used to compute Lanczos vector norms as
\begin{equation}
\begin{gathered}
\begin{aligned}
    &\braket{v^R_{ia} }{ v^R_{jb} } 
    = \sum_{s=0}^{i-1} \sum_{t=0}^{j-1} \sum_{cd} K^{R*}_{sica} \mbraket{v^R_{1c} }{ (T^\dagger)^s   T^t }{ v^R_{1d} } K^R_{tjdb} 
    \\ &\hspace{10pt}=  \sum_{st} \left[  [\bm{K}^R_{si}]^\dagger [\bm{\gamma}_1^{-1}]^\dagger   \bm{C}(s+t) \bm{\gamma}_1^{-1}  \bm{K}^{R}_{tj} \right]_{ab},
\end{aligned}
\\
\begin{aligned}
    &\braket{v^L_{ia} }{ v^L_{jb}}
    = \sum_{s=0}^{i-1} \sum_{t=0}^{j-1} \sum_{cd} K^L_{isac} \mbraket{v^L_{1c} }{ T^s (T^\dagger)^t }{ v^L_{1d} }  K^{L*}_{jtbd} 
    \\ &\hspace{10pt}= \sum_{st} \left[  \bm{K}^L_{is} \bm{\beta}_1^{-1}   \bm{C}(s+t) [\bm{\beta}_1^{-1}]^\dagger [\bm{K}^{L}_{jt}]^\dagger \right]_{ab}.
\end{aligned}
\end{gathered}
\label{eq:lanczos:wv}
\end{equation}
Both of these expressions reduce to $\delta_{ij}\delta_{ab}$ in noiseless applications where $\Tm$ is Hermitian.
They are useful in particular for computing the residual bounds described below.

\subsection{Ritz coefficients}
\label{sec:formalism:ritz-rot}

The block-tridiagonal matrix eigenvectors and Krylov coefficients may be combined to compute the Ritz coefficients
\begin{equation}
\begin{aligned}
    P^{R(m)}_{kta} &\equiv \mathcal{N}^{(m)}_{k} \sum_{i=1}^m \sum_{b} K^R_{taib} \omega^{(m)}_{ibk} , \\
    P^{L(m)}_{kta} &\equiv \frac{1}{\mathcal{N}^{(m)}_{k}} \sum_{i=1}^m  \sum_{b} (\omega^{-1})^{(m)}_{kib} K^L_{ibta} ,
\end{aligned}
\label{eq:ritz-coeffs}
\end{equation}
which directly relate the Ritz and Krylov vectors as
\begin{equation}
\begin{aligned}
    \ket{y^{R(m)}_k} &= \sum_{t=0}^{m-1} \sum_a P^{R(m)}_{kta} \ket{k^R_{ta}}, \\
    \bra{y^{L(m)}_k} &= \sum_{t=0}^{m-1} \sum_a \bra{k^L_{ta}} P^{L(m)}_{kta}~ .
\end{aligned}
\end{equation}
The factors $\mathcal{N}^{(m)}_k$ are arbitrary but can be set to enforce unit normalization using overlap factors as described in the next subsection.

It can be useful to define Hilbert-space operators called Ritz rotators, polynomials in the transfer matrix $T$ built from the Ritz coefficients as
\begin{equation}
    P^{R/L(m)}_{ka} \equiv \sum_{t=0}^{m-1} P^{R/L(m)}_{kta} T^t,
\end{equation}
which excite the Ritz vectors from the initial block of Lanczos vectors as
\begin{equation}
\begin{aligned}
    \ket{y^{R(m)}_k} 
        &= \sum_a P^{R(m)}_{ka} \ket{v^R_{1a}}, \\
    \bra{y^{L(m)}_k} 
        &= \sum_a \bra{v^L_{1a}} P^{L(m)}_{ka} ~ .
\end{aligned}
\label{eq:ritz-rotators}
\end{equation}
Inserting \cref{eq:initial-lanczos-vectors} obtains versions which can be applied to the starting states prior to biorthonormalization,
\begin{equation}
\begin{aligned}
    \ket{y^{R(m)}_k} 
        &= \sum_{ab} [ \gamma^{-1}_{1ba} P^{R(m)}_{ka} ] \ket{\psi_b}, \\
    \bra{y^{L(m)}_k} 
        &= \sum_{ab} \bra{\chi_b} [P^{L(m)}_{ka} \beta^{-1}_{1ab}] ~ .
\end{aligned}
\end{equation}
Because the Ritz vectors as well as $\ket{\psi_a}$ and $\ket{\chi_a}$ are convention-independent, it follows that the combinations
\begin{equation}
    \sum_a \gamma^{-1}_{1ba} P^{R(m)}_{kta}
    \quad \text{and} \quad
    \sum_a P^{L(m)}_{kta} \beta^{-1}_{1ab},
\end{equation}
must be convention-independent as well, which can be useful for consistency checks.

\subsection{Overlap factors}
\label{sec:formalism:overlaps}

Overlap factors between Ritz vectors and the starting interpolating operators can be computed as
\begin{equation}
\begin{aligned}[]
    [Z^{R(m)}_{ka}]^* &\equiv \braket{\chi_a }{ y^{R(m)}_k} \\
    &= \mathcal{N}^{(m)}_{k} \sum_{i=1}^m \sum_{bc}  \beta_{1ab} \braket{v^L_{1b} }{ v^R_{ic} } \omega^{(m)}_{ick}
    \\ 
    &= \mathcal{N}^{(m)}_{k} \sum_b  \beta_{1ab} \omega^{(m)}_{1bk},
    \\
    Z^{L(m)}_{ka} &\equiv \braket{y^{L(m)}_k }{ \psi_a} \\
    &= \frac{1}{\mathcal{N}^{(m)}_{k}} \sum_{j=1}^m \sum_{bc} (\omega^{-1})^{(m)}_{kjc} \braket{v^L_{jc} }{ v^R_{1b} }  \gamma_{1ba} 
    \\ 
    &= \frac{1}{\mathcal{N}^{(m)}_{k}} \sum_b (\omega^{-1})^{(m)}_{k1b}   \gamma_{1ba}.
\end{aligned}
\label{eq:lanczos:Z}
\end{equation}

As in the scalar case~\cite{Hackett:2024xnx}, the overlap factors provide a convenient means of computing the Ritz vector normalization constants $\mathcal{N}^{(m)}_k$ in the diagonal case where $\ket{\chi_a} = \ket{\psi_a}$.
The true transfer matrix is Hermitian, with degenerate left and right eigenvectors. Physical interpretability of Lanczos outputs requires that this holds also for the Lanczos approximation to these eigenvectors, i.e., that right and left Ritz vectors coincide as $\iket{y^{R(m)}_k} = \iket{y^{L(m)}_k}$.
When $\ket{\chi_a} = \ket{\psi_a}$, this requires $Z^{R(m)}_{ka} = Z^{L(m)}_{ka}$, which in turn implies the $\mathcal{N}^{(m)}_k$ can be computed as
\begin{equation}\label{eq:norm_calc}
    |\mathcal{N}^{(m)}_{k}|^2 = \frac{ \sum_b (\omega^{-1})^{(m)}_{k1b}  \gamma_{1ba}  }{ 
    \left[ \sum_c \beta_{1ac} \omega^{(m)}_{1ck} \right]^* } \hspace{20pt} \forall a,
\end{equation}
where there is no summation over $a$ implied, and the right-hand-side must be independent of $a$ so that the equality is true for all $a$ as indicated.
For noiseless data and a symmetric oblique convention $\bm{\beta}_j = \bm{\gamma}_j$, these normalization factors will equal unity automatically.
For asymmetric conventions and in applications to noisy LQCD data, the $|\mathcal{N}^{(m)}_{k}|^2$ can be computed directly using Eq.~\eqref{eq:norm_calc}.
In the off-diagonal case $\ket{\chi_a} \neq \ket{\psi_a}$, $Z^{R(m)}_{ka} \neq Z^{L(m)}_{ka}$ and moreover right and left Ritz vectors can only be approximately equal, so (at least this version of) the norm trick does not apply for spurious state filtering and may distort results if applied naively. 

The overlap factors together with the Ritz values provide all information necessary to diagnose and discard spurious states which arise in applications to noisy LQCD data.
As in the scalar case, \cref{eq:norm_calc} provides a means to diagnose noise-artifact states which arise due to non-Hermiticity of $\Tm$: when the computation yields a non-real or non-positive $|\mathcal{N}^{(m)}_{k}|^2$, it signals an inconsistency, specifically that $\iket{y^{R(m)}_k} \neq \iket{y^{L(m)}_k}$.
The $a$-independence of Eq.~\eqref{eq:norm_calc} provides a separate, non-trivial constraint not present in the scalar case that only holds for physical states.
Finally, we find the Cullum-Willougby test as employed in Refs.~\cite{Wagman:2024rid,Hackett:2024xnx} can be reformulated as a cut on anomalously small overlap factors.
These points are discussed in further detail in \cref{sec:lattice} where they may be explored with example data.

Once $|\mathcal{N}^{(m)}_{k}|^2$ is determined, normalized Ritz coefficients provide an alternate means to compute the overlap factors as
\begin{equation}
\begin{aligned}[]
    [Z^{R(m)}_{ka}]^* &= \sum_{t=0}^{m-1} \sum_b \mbraket{\chi_a }{T^t}{ v^R_{1b} } P^{R(m)}_{ktb} \\
    &= \sum_{t=0}^{m-1} \sum_{bc} C_{ac}(t) \gamma^{-1}_{1cb} P^{R(m)}_{ktb} ,
    \\
    Z^{L(m)}_{ka} &= \sum_{t=0}^{m-1}  \sum_b P^{L(m)}_{ktb} \mbraket{v^L_{1b} }{T^t}{ \psi_a} \\
    &= \sum_{t=0}^{m-1} \sum_{bc} P^{L(m)}_{ktb} \beta^{-1}_{1bc} C_{ca}(t) .
\end{aligned}
\label{eq:lanczos:Z-Ritz-rotator}
\end{equation}
This can be useful for consistency checks.

\subsection{Operator matrix elements}
\label{sec:formalism:MEs}

Matrix elements of generic (local or nonlocal) operators $J$ can be computed from three-point correlation function matrices,
\begin{equation}
    C^{\text{3pt}}_{ab}(\sigma,\tau) \equiv \mbraket{\psi'_a}{T^\sigma J T^\tau }{\psi_b},
\end{equation}
where primes denote quantities for the sector of final states, which could be distinct from those of the initial states (e.g.~in the case of off-forward matrix elements with different initial and final momenta).
As in Ref.~\cite{Hackett:2024xnx}, the strategy is simply to compute $\mbraket{y'^{(m)}_f }{J}{ y^{(m)}_i }$, i.e., the matrix elements in the basis of Ritz vectors.
The block generalization adds no formal complications over the scalar case.
Inserting the definition of the Ritz rotators, \cref{eq:ritz-rotators}, yields the prescription
\begin{equation}\begin{aligned}
    &\mbraket{y'^{L(m)}_f }{J}{ y^{R(m)}_i } 
    \\
    &\; =  \sum_{\sigma \tau ab} P'^{L(m)}_{f\sigma a} \mbraket{v^{L'}_{1a} }{ T^\sigma J T^\tau }{ v^R_{1b} } P^{R(m)}_{i\tau b}
    \\
    &\; =  \sum_{\sigma \tau ab} P'^{L(m)}_{f\sigma a}   [\bm{\beta}^{'-1}_{1} \bm{C}^{\text{3pt}}(\sigma,\tau) \bm{\gamma}^{-1}_{1}]_{ab} P^{R(m)}_{i\tau b}
    ~,
\end{aligned}
\label{eq:ritz-rotator-ME}
\end{equation}
where unprimed and primed quantities are computed from separate diagonal two-point correlator matrices with initial- and final-state quantum numbers,
\begin{equation}
    C_{ab}(t) = \mbraket{\psi_a}{T^t}{\psi_b}
    \;\, \text{and} \;\,
    C'_{ab}(t) = \mbraket{\psi'_a}{T^t}{\psi'_b} \,.
\end{equation}
Thus, as in the scalar case, once two-point data has been analyzed, oblique block Lanczos allows explicit computation of matrix elements for (approximate) energy eigenstates by simple matrix multiplication of three-point correlator matrices with no additional analysis choices.
Assuming the initial and final states of interest are in the physical subspace of Krylov space, the choice of $L$ vs $R$ labels in Eq.~\eqref{eq:ritz-rotator-ME} is irrelevant.

Note that, in certain cases, block Lanczos matrix-element estimators are numerically identical to GEVP estimators (see \cref{sec:gevp}) as well as estimators based on generalized pencil of function (GPOF) methods~\cite{Aubin:2010jc,Aubin:2011zz,Green:2014xba,Ottnad:2020qbw,Ottnad:2017mzd}.
These and other coincidences between methods will be explored more deeply in future work~\cite{future}.

\subsection{Correlator decomposition}
\label{sec:formalism:corr-decomp}

Lanczos energies and overlap factors provide an exact representation of correlation functions, even for noisy Monte Carlo data. 
Specifically, 
\begin{equation}\label{eq:Lanczos_reco}
\begin{aligned}
    C_{ab}(t) = \mbraket{\chi_a}{T^t}{\psi_b}
    &= \mbraket{\chi_a}{[\Tm]^t}{\psi_b}
    \\ &= \sum_k Z^{R(m)*}_{ka} Z^{L(m)}_{kb} [\lambda^{(m)}_k]^t,
\end{aligned}
\end{equation}
for all $t \leq 2m - 1$.
The first equality is the definition of $\bm{C}(t)$.
The third can be shown by inserting
\begin{equation}
    [\Tm]^t = \sum_k \ket{y^{R(m)}_k} [\lambda^{(m)}_k]^t \bra{y^{L(m)}_k} ~ ,
\end{equation}
which follows from the eigendecomposition \cref{eq:Tm_Ritz_decomp} and $\iavg{y^{L(m)}_k | y^{R(m)}_l} = \delta_{kl}$, 
and recognizing the definitions of the overlap estimators $Z^{L/R(m)}_{ka}$, \cref{eq:lanczos:Z}.
We prove the remaining second equality below.
The result and proof thereof generalize the scalar result found in Ref.~\cite{Hackett:2024xnx}.

Some preliminary results are necessary before demonstrating the second equality of \cref{eq:Lanczos_reco}.
Comparing the action of $T$, following from the three-term recurrence \cref{eq:recurrence}, with the action of its Krylov-space approximation $\Tm$, following from Eq.~\eqref{eq:Tmv} and Eq.~\eqref{eq:untri}, shows that
\begin{equation}
  \begin{split}
   \Tm  \ket{v^R_{ja}} &= \begin{cases} T  \ket{v^R_{ja}}, & j < m, \\ 
   \sum_b \left( \ket{v^R_{jb}}\alpha_{jba} + \ket{v^R_{(j-1)b}}\beta_{jba} \right), & j=m  \end{cases}.
  \end{split}
\end{equation}
The difference between these operators therefore acts on Lanczos vectors as
\begin{equation}\label{eq:Tdiff-lanczos}
\begin{aligned}[]
    [T - \Tm] \ket{v^R_{ja}} &= \delta_{jm} \sum_b \ket{ v^R_{(m+1)b }} \gamma_{(m+1)ba},
    \\
    \bra{v^L_{ja}} [T - \Tm] &= \delta_{jm} \sum_b \beta_{(m+1)ab} \bra{v^L_{jb}} ~ .
\end{aligned}
\end{equation}
These results are also used in the explicit construction of residual bounds below.
\Cref{eq:Tdiff-lanczos} implies that
\begin{equation}\label{eq:Tv_is_Tmv_t}
    T^t \ket{v^R_{1a}} = [\Tm]^t \ket{v^R_{1a}}
    \; \text{and} \;
    \bra{v^L_{1a}} T^t = \bra{v^L_{1a}} [\Tm]^t,
\end{equation}
for all $t < m$, and
\begin{equation}\label{eq:Tv_is_Tmv_m}
\begin{aligned}
    T^m \ket{v^R_{1a}} &= [\Tm]^m \ket{v^R_{1a}} + \sum_b \ket{v^R_{(m+1)b}} \gamma_{(m+1)ba},
    \\
    \bra{v^L_{1a}} T^m &= \bra{v^L_{1a}} [\Tm]^m + \sum_b \beta_{(m+1)ab} \bra{v^L_{1b}},
\end{aligned}
\end{equation}
for $t=m$, with the second term on each RHS representing a contribution outside the $rm$-dimensional Krylov space.
Finally, note that the three-term recurrence \cref{eq:recurrence} implies 
\begin{equation}\label{eq:Tmv_stays_in_kspace}
    [\Tm]^{m-1} \ket{v^R_{1a}} = \sum_{j=1}^{m} \sum_b \ket{v^R_{jb}} c^{R(m)}_{jba},
\end{equation}
where $c^{R(m)}_{jab}$ are some (in-principle computable) coefficients.
Their values are not important, but rather that the vector only has support inside the $rm$-dimensional Krylov space (i.e.~the sum runs over ${j\in [1,m]}$).
An analogous left expression holds as well.

We are now equipped to prove the second equality of \cref{eq:Lanczos_reco}.
Note first that 
\begin{equation}
    \mbraket{\chi_a}{T^t}{\psi_b} 
    = \sum_{cd} \beta_{1ac} \mbraket{v^L_{1c}}{T^t}{v^R_{1d}} \gamma_{1db},
\end{equation}
by \cref{eq:initial-lanczos-vectors}, so it is sufficient to show $\mbraket{v^L_{1c}}{T^t}{v^R_{1d}} = \mbraket{v^L_{1c}}{[\Tm]^t}{v^R_{1d}}$ for all $t \leq 2m -1$.
To do so, we must consider two cases, in each decomposing $T^t = T^{t_L} T^{t_R}$ with some choice of $t = t_L+t_R$.
In the case $t \leq 2m-2$, we may choose both of $t_L, t_R \leq m-1$.
It follows immediately from \cref{eq:Tv_is_Tmv_t} that
\begin{equation}
\begin{aligned}
    \mbraket{v^L_{1c}}{T^{t_L} T^{t_R}}{v^R_{1d}}
    &= \mbraket{v^L_{1c}}{[\Tm]^{t_L} [\Tm]^{t_R}}{v^R_{1d}}
    \\&= \mbraket{v^L_{1c}}{[\Tm]^{t}}{v^R_{1d}} ~ .
\end{aligned}
\end{equation}
For the remaining case $t = 2m-1$, consider $t_L=m$ and $t_R=m-1$, such that by \cref{eq:Tv_is_Tmv_m}
\begin{equation}
\begin{split}
    \mbraket{v^L_{1c}}{T^{m} T^{m-1}}{v^R_{1d}}
    = \mbraket{v^L_{1c}}{ [\Tm]^{2m-1} }{v^R_{1d}}
    \\+ \sum_b \beta_{(m+1)cb} \mbraket{v^L_{(m+1)b}}{[\Tm]^{m-1}}{v^R_{1d}} ~ .
\end{split}
\end{equation}
The second term is zero by \cref{eq:Tmv_stays_in_kspace}, i.e.,
\begin{equation}\begin{gathered}
    \mbraket{v^L_{(m+1)b}}{[\Tm]^{m-1}}{v^R_{1d}} 
    = \sum_{j=1}^{m} \sum_c \braket{v^L_{(m+1)b}}{v^R_{jc}} c^{R(m)}_{jcd}
    \\
    = \sum_{j=1}^{m} \sum_b \delta_{j,(m+1)} \delta_{bc} \, c^{R(m)}_{jcd}
    = 0,
\end{gathered}\end{equation}
using the biorthonormality of the Lanczos vectors, so the desired equality holds for $t=2m-1$ as well.
This completes the proof of \cref{eq:Lanczos_reco}.

\section{Noiseless demonstration}
\label{sec:noiseless}

This section works through an application of the block Lanczos algorithm to noiseless mock-data examples to assess its performance in the infinite-statistics limit.
We verify the validity of the formalism presented in \cref{sec:formalism} and compare block Lanczos extractions of energies, overlaps, and matrix elements with those computed using standard effective estimators and scalar Lanczos.
Comparison with GEVP methods is deferred to \cref{sec:gevp}.
The same mock-data examples defined in this section are also used in \cref{sec:bounds} and \cref{sec:gevp}.

\begin{figure}
    \includegraphics[width=\linewidth]{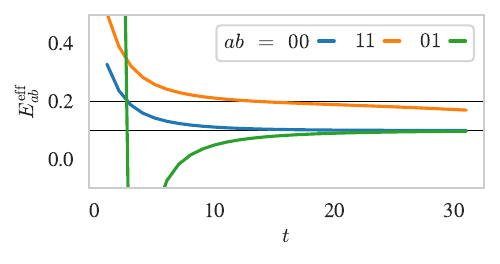}
    \caption{
        Effective energies $E^\mathrm{eff}_{ab} = -\ln C_{ab}(t)/C_{ab}(t-1)$ for the elements of the correlator matrix for noiseless example as defined in \cref{eq:noiseless:ex-def}.
        Not shown is $E^\mathrm{eff}_{10}$, which is identical to $E^\mathrm{eff}_{01}$.
        Black lines show the exact $E_0$ and $E_1$.
    }
    \label{fig:noiseless-meff}
\end{figure}

\begin{figure*}
    \includegraphics[width=\linewidth]{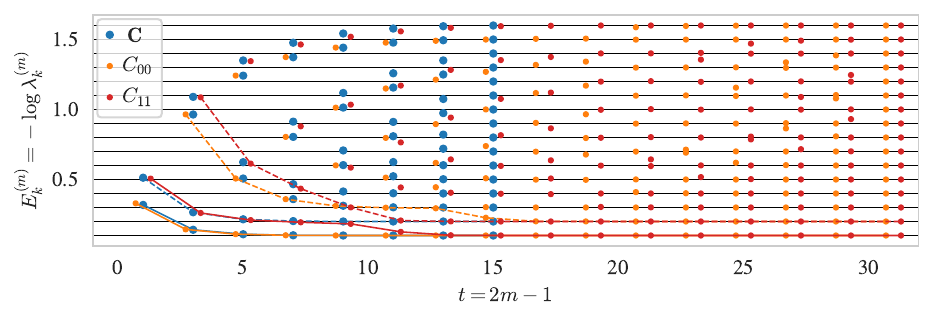}
    \caption{
        Spectrum extracted by block Lanczos (blue) applied to the noiseless correlator matrix \cref{eq:noiseless:ex-def} compared with the spectra extracted by scalar Lanczos (orange, red) applied to each diagonal correlator, as a function of the number of Lanczos iterations $m$.
        Solid lines connect the lowest energies in each extraction, and dashed lines the second-lowest.
        Block Lanczos solves for the true eigenvalues exactly after $8$ steps, and scalar Lanczos after $16$ steps.
        The black horizontal lines indicate the true energies $E_n = 0.1 (n+1)$.
    }
    \label{fig:noiseless-spectrum-block-vs-scalar}
\end{figure*}

For the demonstration of spectroscopy and overlap factor extractions, we construct a correlator of length $\betaT=32$ with 16 states,
\begin{equation}
    C_{ab}(t) = \sum_{k=0}^{15} Z^*_{ka} Z_{kb} e^{-E_k t},
\end{equation}
where $a,b \in \{0,1\}$ and
\begin{equation}
    \begin{gathered}
        E_k = 0.1 (k+1), \\
        \widetilde{Z}_{k0} = \begin{cases}
            1, & k~\mathrm{even} \\
            0.1, & k~\mathrm{odd}
        \end{cases}, \\
        \widetilde{Z}_{k1} = \begin{cases}
            -0.1, & k=0 \\
            1, & k~\mathrm{odd} \\
            0.1, & k>0~\mathrm{even}
        \end{cases}, \\
        Z_{ka} = \frac{\widetilde{Z}_{ka}}{\sqrt{2 E_k}} ~ .
    \end{gathered}
    \label{eq:noiseless:ex-def}
\end{equation}
The overall factor of $1 / \sqrt{2 E_k}$ on the overlaps mimics the single-particle relativistic normalization of states.
These definitions are engineered to produce common pathologies that arise in analysis of correlator matrix, visible in the effective energies
\begin{equation}
    E_{ab}^\mathrm{eff}(t) \equiv - \ln \frac{C_{ab}(t)}{C_{ab}(t-1)},
\end{equation}
shown in \cref{fig:noiseless-meff}.
The overlaps $Z_{k0}$ for $\ket{\psi_0}$ are defined to have large overlap with even-$k$ states including the ground state, resulting in the clear asymptote of $E^\mathrm{eff}_{00}$ towards the true ${E_{0}=0.1}$.
Meanwhile, overlaps $Z_{k1}$ for $\ket{\psi_1}$ are suppressed for the ground state but large for odd-$k$ states, including the first excited state.
This results in an approximate ``pseudo-plateau'' in $E^\mathrm{eff}_{11}$ near the true $E_1 = 0.2$ with slow but visible convergence towards its asymptotic value $E_0$. 
The ground-state overlap $Z_{01}$ for $\ket{\psi_1}$ is negative, while all other $Z_{ka}$ are positive.
This gives the observed non-monotonic behavior (i.e.~large oscillations) for the off-diagonal $E^\mathrm{eff}_{01} = E^\mathrm{eff}_{01}$ at early times and convergence towards $E_0$ from below.
\Cref{sec:noiseless:me} extends this example to resemble the case of different initial- and final-state quantum numbers for the matrix element demonstration.

As discussed in Refs.~\cite{Wagman:2024rid,Hackett:2024xnx}, applying Lanczos methods in the noiseless case requires high-precision arithmetic to avoid numerical instabilities; see Appendix~A of Ref.~\cite{Hackett:2024xnx} for a detailed discussion. 
As in that work, we use the \texttt{mpmath} Python package~\cite{mpmath}; 100 decimal digits of precision is sufficient for these demonstrations.
In practice, the same precision concerns do not apply in the noisy case as explored in \cref{sec:lattice}. 
More specifically, applications to noisy data below require Lanczos recursions to be performed in high precision while $\Tm$ eigensolves can be safely performed in double precision, apart from the case of thermal modes discussion in \cref{sec:thermal}.

In this and following sections, we compare block Lanczos with scalar Lanczos applied only to the diagonal elements of the correlator matrix.
As explored in Ref.~\cite{Wagman:2024rid}, scalar Lanczos may in principle be applied to the off-diagonal elements without issue for spectroscopy, but in this noiseless example it produces non-positive Ritz values that cannot be shown straightforwardly.
Properly treating overlap factors and matrix elements with off-diagonal Lanczos requires nontrivial extensions of the formalism presented in Ref.~\cite{Hackett:2024xnx}, which are left for future work.

\subsection{Spectrum}
\label{sec:noiseless:spectrum}

Computing the various quantities in \cref{sec:formalism},
it is straightforward to verify that all $R$ and $L$ quantities coincide 
for any symmetric convention $\bm{\beta}_j = \bm{\gamma}_j$
or any other convention that reduces to a symmetric one in the noiseless limit.
As in the scalar case, all physical quantities---Ritz values as well as overlap factor and matrix element estimators---are insensitive to choice of oblique Lanczos convention, but basis-dependent quantities like Krylov coefficients may vary.

\Cref{fig:noiseless-spectrum-block-vs-scalar} shows the Ritz values computed by block Lanczos, compared with those extracted by scalar Lanczos applied to the diagonal correlators $C_{00}$ and $C_{11}$.
The improvements offered by block Lanczos are immediately apparent.
Each block iteration produces $r=2$ new Ritz values, in contrast to scalar Lanczos, which yields only one.
Consequently, block Lanczos exactly solves the spectrum of this 16-state system exactly after $m = 16/r = 8$ steps, i.e.~twice as quickly as scalar Lanczos, which requires $m=16$ steps.
Roughly speaking, this may be characterized as block Lanczos achieving similar convergence to scalar Lanczos in $1/r$ steps, 
albeit
while consuming matrices of $r^2$ values per iteration.

More quantitatively, we observe that the low-lying states in the block Lanczos spectrum converges at least as well as the union of the two scalar Lanczos spectra.
As can be observed in \cref{fig:noiseless-spectrum-block-vs-scalar}, scalar Lanczos applied to $C_{00}$ quickly extracts $E_0$ and $E_2$, for which the corresponding overlaps are large, but does not give a Ritz value for the suppressed state $E_1$ until its 8th iteration.
The emergence of a Ritz value for $E_1$ is not associated with any disruption of convergence of its neighboring Ritz values towards $E_0$ or $E_2$.
Meanwhile, the lowest-lying Ritz value for scalar Lanczos applied to $C_{11}$ rapidly approaches $E_1$; due to the suppressed overlap associated with $E_0$, the lowest-lying Ritz value does not begin converging towards $E_0$ until $m \geq 5$.\footnote{We caution against interpreting these behaviors as ``pseudo-plateaus'' like those that arise in effective energies, as inspection of the residual bounds indicates true convergence to states but in a non-monotonic order; see Appendix~\ref{app:scalar-res-bounds}.}
In contrast, block Lanczos yields a sequence of Ritz values that quickly and smoothly converge to the true $E_0$ and $E_1$.
Similar comparisons hold for all other pairs of energies in the spectrum, as expected given the alternating pattern of overlap suppression in the example.
Further quantitative comparisons are presented in \cref{sec:bounds}, where they can be better understood in the context of the rigorous bounds on convergence presented therein.

Block Lanczos Ritz values appear to arise in co-moving pairs.
In this example, the first such pair appear\footnote{This association of Ritz values between different Lanczos iterations is purely visual and not quantitative.} to eventually converge to the ground and first excited state, while the second such pair apparently converge to the two highest-lying excited states.
This is expected behavior for Lanczos algorithms, which tend to first extract the most extremal eigenvalues~\cite{Parlett,Parlett:1990,Kuijlaars:2000,Garza-Vargas:2020}.

\subsection{Overlap factors}
\label{sec:noiseless:overlaps}

\begin{figure}
    \includegraphics[width=\linewidth]{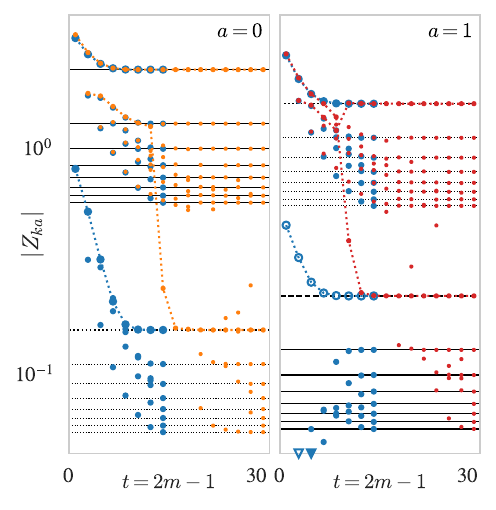}
    \caption{
        Overlap factors estimated in the noiseless example \cref{eq:noiseless:ex-def} by block Lanczos (large blue) versus scalar Lanczos applied to $C_{00}$ (small orange at left) and $C_{11}$ (small red at right), with $Z_{k0}$ at left and $Z_{k1}$ at right.
        Black horizontal lines represent the true overlaps $Z_{ka}$, with solid lines for even $k$, dotted for odd $k$, and a dashed line in the right panel for the absolute value of the single negative overlap $Z_{01}$. 
        Lines and symbols for the two lowest-lying states are bolded, and to guide the eye, dotted lines connect estimates corresponding to the two largest Ritz values in each extraction.
        No visual distinction is made between values for different $k$ otherwise, but note that, restricting to even or odd $k$, the true $Z_{ka}$ decrease monotonically in $k$.
        Hollow symbols represent negative values whose absolute value is shown instead; this includes all block estimates of the negative overlap $Z_{01}$.
        The triangles at the bottom of the right panel indicate block estimates outside the plot range.
        Block Lanczos solves for all overlaps exactly after 8 steps, while each scalar Lanczos requires 16 and finds the wrong sign for $Z_{01}$; see the discussion in the text.
    }
    \label{fig:noiseless-Zs}
\end{figure}

As presented in \cref{sec:formalism:overlaps}, block Lanczos provides an estimator for overlap factors $Z_{ka} = \braket{k}{\psi_a}$, generalizing the scalar construction.
As with the spectrum, block Lanczos solves for the full set of overlap factors exactly after $m=8$ steps, and scalar does after $m=16$ up to the sign ambiguity discussed further below.
\Cref{fig:noiseless-Zs} compares block and scalar extractions of the overlap factors in this example.
It is clear from the comparison that the relative sizes of the overlaps govern which order states are resolved in.
Scalar Lanczos applied to either diagonal correlator preferentially resolves the states with larger overlaps in that correlator, not yielding values for the suppressed states until the latter half of iterations.
Block Lanczos, which sees at least one interpolator with unsuppressed overlap with each state, shows no similar structure---as with the spectrum, block overlap estimates converge at least as well as the union of scalar estimates.
This indicates that block Lanczos is able to cleanly resolve spectra which must be accessed using multiple interpolators with nearly-disjoint support in the space of states.
This is exactly the circumstance that arises in lattice studies of multi-particle spectroscopy.

Importantly, while the overall phase of overlap factors is a matter of definition,\footnote{The true transfer matrix eigenstates are eigenvectors, which are defined only up to an overall phase. Conventionally, overlap factors for one interpolator are defined to be real to remove this ambiguity. Typically, this is sufficient to guarantee all overlaps are real---see the discussion of correlator reality in \cref{sec:lattice} and Appendix~\cref{app:real}.} relative signs between different overlaps with the same state are physical.
Block Lanczos is able to correctly extract these relative signs (in this case, between $Z_{00}$ and $Z_{01}$). In contrast, scalar Lanczos---which does not simultaneously process data involving $\ket{\psi_0}$ and $\ket{\psi_1}$---is fundamentally incapable of doing so.
In this example, when scalar Lanczos applied to $C_{11}$ solves this noiseless example exactly at $m=16$, our definitions necessarily return a positive $|Z_{01}|$.
The ability to resolve such signs represents a qualitative improvement in capabilities of block over scalar Lanczos.

\subsection{Matrix elements}
\label{sec:noiseless:me}

For the demonstration of matrix elements, we extend the example defined in \cref{eq:noiseless:ex-def} to resemble the case of an off-forward three-point function, with different initial- and final-state momenta.
Specifically, we additionally define a final-state two-point function and an off-diagonal three-point function
\begin{equation}\begin{aligned}
    C'_{ab}(t) &= \sum_k Z^{\prime *}_{ka} Z'_{kb} e^{-E'_k t},
    \\
    C^\mathrm{3pt}_{ab}(\sigma, \tau) &= \sum_{fi} Z^{\prime *}_{fa} J_{fi} Z^{\vphantom{\prime}}_{ib} e^{-E'_f \sigma - E_i \tau}
    ~,
\end{aligned}
\end{equation}
respectively, with parameters
\begin{equation}
    \begin{gathered}
        E_k = 0.1 (k+1) \,, \quad
        E'_k = \sqrt{E_k^2 + 0.1^2} \,, \\
        Z_{ka} = \frac{\tilde{Z}_{ka}}{\sqrt{2 E_k}} \,, \quad
        Z'_{ka} = \frac{\tilde{Z}_{ka}}{\sqrt{2 E'_k}} \,, \\
        J_{fi} = \sqrt{\frac{4 E'_0 E_0}{4 E'_f E_i}} \tilde{J}_{fi} \,,
    \end{gathered}
    \label{eq:noiseless:me-ex-def}
\end{equation}
where $\tilde{Z}_{ka}$ is as in \cref{eq:noiseless:ex-def}, $\tilde{J}_{00} = 1$, and all other elements of $\tilde{J}_{fi}$ are randomly drawn from a unit normal distribution.\footnote{The precise values used are provided in the attached file.}
The initial-state spectrum $E_k$ and overlaps $Z_{ak}$ are the same as \cref{eq:noiseless:ex-def}. The final-state spectrum $E'_k$ and overlaps $Z'_{ak}$ are defined to resemble those of the initial state in a boosted reference frame. The matrix elements $J_{fi}$ are defined to impose the expected energy scaling in $C^\mathrm{3pt}$ for single-particle states while maintaining $J_{00}=1$.
The block and scalar Lanczos treatments of the final-state spectrum are similar to the initial-state results presented in \cref{sec:noiseless:spectrum} and thus not shown.

\begin{figure}
    \includegraphics[width=\linewidth]{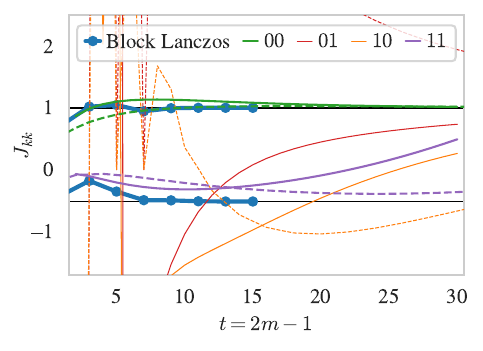}
    \caption{
        For the noiseless example \cref{eq:noiseless:me-ex-def}, estimates of diagonal matrix elements for the ground state $J_{00}$ (top) and first excited state $J_{11}$ as computed by block Lanczos (blue), as compared with the same summation (\cref{eq:summation-Jeff}, solid) and power iteration (\cref{eq:power-iter-Jeff}, dashed) estimators.
        The horizontal lines indicate the true values of $J_{00}=1$ and $J_{11}$.
        Block Lanczos solves for the true $J_{00}$ and $J_{11}$ exactly after $m=\betaT/4=8$ steps.
    }
    \label{fig:noiseless-ME-block-vs-jeffs}
\end{figure}

The example is sufficiently pathological that standard ratio plots are not useful to provide.
The data are displayed below in \cref{fig:noiseless-ME-block-vs-jeffs}, which  shows effective estimators for $J_{00}$ defined for the different elements of $C^\mathrm{3pt}_{ab}$ separately.
Specifically, we use the same two definitions as in Ref.~\cite{Hackett:2024xnx} following from the standard ratio used in the scalar case\footnote{As noted in Ref.~\cite{Hackett:2024xnx}, this is not the precise ratio prescribed by the power iteration method; we employ the standard ratio \cref{eq:std-ratio} for ease of comparison with other works.}
\begin{equation}\begin{aligned}
    R(\sigma, \tau) &= 
    \frac{ C^\mathrm{3pt}(\sigma, \tau) }{ C'(\sigma+\tau) }
    \sqrt{
    \frac{ C(\sigma) }{ C'(\sigma) }
    \frac{ C'(\sigma+\tau) }{ C(\sigma+\tau) }
    \frac{ C'(\tau) }{ C(\tau) }
    } ~ .
    \label{eq:std-ratio}
\end{aligned}\end{equation}
From this ratio, power iteration effective matrix elements can be constructed as
\begin{equation}
    J^\mathrm{PI}_{00}(t) = \begin{cases}
        R(\frac{t}{2}, \frac{t}{2} ), & t\text{ even} \\
        \frac{1}{2} \left[ R(\frac{t+1}{2}, \frac{t-1}{2}) + R(\frac{t-1}{2}, \frac{t+1}{2}) \right], & t\text{ odd} 
    \end{cases}
    \label{eq:power-iter-Jeff}
\end{equation}
and summation-method~\cite{Maiani:1987by,Dong:1997xr,Capitani:2012gj,Hackett:2023nkr} effective matrix elements as 
\begin{equation}\begin{aligned}
    \Sigma_{\Delta_\tau}(t_f) 
    &= \sum_{\tau=\Delta_\tau}^{t_f-\Delta_\tau} R(t_f-\tau, \tau),
    \\
    J^\Sigma_{00,\Delta_\tau}(t_f) &= \Sigma_{\Delta_\tau}(t_f+1) - \Sigma_{\Delta_\tau}(t_f)
    \\ &= J_{00} + (\text{excited states}) ~ .
\end{aligned}
\label{eq:summation-Jeff}
\end{equation}
Each may be applied to any individual element $a,b$ of a three-point correlator matrix $C^\mathrm{3pt}_{ab}(t)$.
As visible in \cref{fig:noiseless-ME-block-vs-jeffs}, while analysis of $C^\mathrm{3pt}_{00}$ provides estimators which converge near $J_{00}$ quickly, those for $C^\mathrm{3pt}_{11}$ neither converge to $J_{00}$---each initially ``anti-converges'' away from its asymptotic value---nor provide an accurate estimator of $J_{11}$ with which the overlap is larger.
The estimators for off-diagonal $C^\mathrm{3pt}_{ab}$ show large oscillations at early $t$ and poor convergence in the available range $0 \leq t < 32$.
Considered collectively, these estimators do not provide any convincing estimate of any matrix element.

\Cref{fig:noiseless-ME-block-vs-jeffs} compares the same estimators against the block Lanczos extractions of the diagonal matrix elements for the ground and first excited state, $J_{00}$ and $J_{11}$.
In contrast to the confused and incoherent behavior of the effective estimators, block Lanczos estimates converge quickly and reliably to the true values, providing a clear picture.
This improvement is just as observed for scalar Lanczos over effective estimators in Ref.~\cite{Hackett:2024xnx}.
Comparisons with improved GEVP estimators for $J$ in \cref{sec:gevp} are similar.
While we have not repeated the adversarial exercise of Ref.~\cite{Hackett:2024xnx} in the block case, we expect block Lanczos to inherit the same robustness against pathological cases observed for scalar Lanczos.

\begin{figure}
    \includegraphics[width=\linewidth]{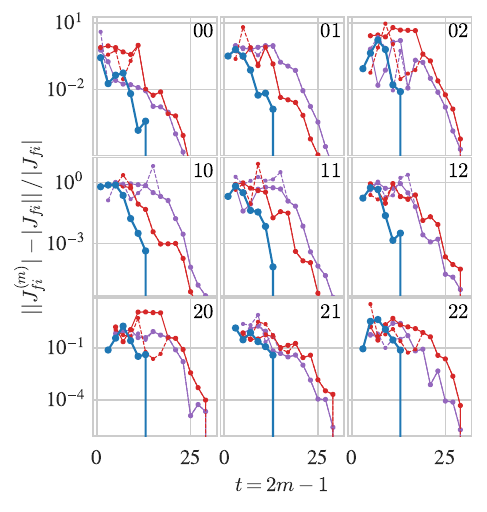}
    \caption{
        For the noiseless example \cref{eq:noiseless:me-ex-def}, relative errors in (absolute values of) estimates of operator matrix elements $J_{fi}$ involving the lowest-lying three states, as computed by block Lanczos (blue) and scalar Lanczos applied to $C_{00}$ (purple) and $C_{11}$ (red).
        Solid lines connect points for relative errors defined with respect to $i$th and $f$th largest Ritz values.
        To check for the effects of state misidentification, dashed lines connect points for the scalar results where the error is defined with respect to matrix elements for the states with true energies nearest to the initial- and final-state Ritz values. In most cases, these coincide with the other points.
        Block Lanczos solves for all matrix elements exactly after $m=\betaT/4=8$ steps.
        Scalar Lanczos solves after $m = \betaT/2 = 16$ steps, but in some cases is off by an overall sign, as discussed in the text.
    }
    \label{fig:noiseless-ME-block-vs-scalar}
\end{figure}

\cref{fig:noiseless-ME-block-vs-scalar} compares the convergence of block and scalar Lanczos estimates for low-lying diagonal and transition matrix elements.
The improvement is similar as with energies and overlaps: block Lanczos provides a better estimator than scalar Lanczos applied to either diagonal correlator, even allowing for state misidentification in scalar Lanczos where the spectrum is extracted non-monotonically.
However, the advantage is not as obvious as over the effective estimators.

A critical improvement over scalar Lanczos is not reflected in \cref{fig:noiseless-ME-block-vs-scalar}: block Lanczos allows self-consistent determination of the relative phases between transition matrix elements for states probed by different interpolators.
The formalism presented in both \cref{sec:formalism} and Ref.~\cite{Hackett:2024xnx} provides self-consistent determinations of relative signs between transition matrix elements.
In the scalar case, this means that the phases of operator matrix elements are defined relative to the overlap factors---for which it produces real and positive values by convention.
In this example, this results in scalar Lanczos applied to $C_{11}$ yielding ground-excited matrix elements off by a sign from the definitions in the problem.
In contrast, because block Lanczos is able to resolve the relative phases of overlaps as discussed above, it is thereby able to consistently determine relative phases of other transition matrix elements as well.
This may be useful for e.g.~lattice studies of electroweak transitions involving resonances~\cite{Briceno:2014uqa,Leskovec:2024pzb}.

\section{Convergence and bounds}
\label{sec:bounds}

The same formal frameworks use to quantify and bound convergence in scalar Lanczos can be extended to the block case as well.
In this section, we first review the block extension of Kaniel-Paige-Saad (KPS) convergence theory~\cite{Kaniel:1966,Paige:1971,Saad:1980} and discuss its implications for correlator matrix analyses.
We then derive the block generalization of the oblique Lanczos residual bounds originally presented in Ref.~\cite{Wagman:2024rid}.
In each case, we demonstrate the validity of the bounds using the noiseless examples of \cref{sec:noiseless}.

\subsection{KPS Bound}

In finite-dimensional linear algebra applications of the Lanczos algorithm,
the convergence of Ritz values and vectors to the eigenvalues and eigenvectors of 
a matrix is described by KPS convergence theory.
The KPS bound not only quantifies the asymptotic converge rates of Ritz values and vectors, it also provides two-sided bounds on differences between Ritz values and true eigenvalues after finitely many iterations.
The KPS bound for applications of scalar Lanczos to infinite-dimensional transfer matrices in LQCD is discussed in Refs.~\cite{Wagman:2024rid,Hackett:2024xnx}.

Two important limitations arise for LQCD applications of the KPS bound.
First, the KPS bound depends on the full spectrum of $T$ and cannot be directly computed only from the matrix elements of $\Tm$ available after finitely many steps of Lanczos.
Second, the KPS bound only applies to non-oblique Lanczos and therefore only to LQCD results for diagonal correlator matrices in the infinite-statistics limit.
For these reasons, the KPS bound should be thought of as a theoretical bound on the convergence rate of Lanczos applications to LQCD at infinite statistics, rather than as a directly computable measure of finite-iteration effects in a given calculation.
In the remainder of this subsection, $L/R$ superscripts will therefore be omitted.

The block Lanczos extension of Kaniel-Paige-Saad (KPS) convergence theory~\cite{Kaniel:1966,Paige:1971,Saad:1980} was developed by Saad and reported along with results for standard Lanczos in Ref.~\cite{Saad:1980},
\begin{equation}\label{eq:KPS}
    0 \leq \frac{ \lambda_n - \lambda_n^{(m)} }{ \lambda_n - \lambda_{\infty} } \leq \left[ \frac{ K_n^{(m)} \tan \theta_n}{ T_{m-n-1}(\Gamma_n^r)} \right]^2,
\end{equation}
where $r = \text{dim}[C_{ab}(0)]$ is the size of the initial Lanczos block,  the $T_k(x)$ are Chebyshev polynomials of the first kind defined by $T_k(\cos x) = \cos(k x)$,
\begin{equation}
\begin{split}
    \Gamma_n^r &\equiv 1 + \frac{2(\lambda_n - \lambda_{n+r})}{\lambda_{n+r} - \lambda_{\infty}} = 2 e^{E_{n+r}-E_n} - 1,
    \end{split}
    \label{eq:kps-gamma}
\end{equation}
and
\begin{equation}
    K_n^{(m)} \equiv \prod_{l=0}^{n-1} \frac{\lambda_l^{(m)} - \lambda_{\infty}}{\lambda_l^{(m)} - \lambda_n}, \hspace{20pt} n > 0,
    \label{eq:kps-K}
\end{equation}
with $K_0^{(m)} \equiv 1$ and $\lambda_{\infty}$ the smallest eigenvalue of $T$---for a bounded infinite-dimensional operator $T$ it must be that $\lambda_{\infty} = 0$.
The remaining ingredient in the KPS bound is the angle $\theta_n$ between the vector $\ket{n}$ and the subspace $\mathcal{K}_1$ spanned by the initial Lanczos vectors.
It can be expressed as 
\begin{equation}
    \tan^2\theta_n = ||\ket{n} - \ket{\hat{x}_n}||^2,
\end{equation}
where $\ket{ \hat{x}_n }$ is the vector whose orthogonal projection onto $\mathcal{K}_1$ is equal to $\ket{n}$. 
The explicit construction of this vector is discussed in Ref.~\cite{Saad:1980} and summarized in Appendix~\ref{app:KPS}, where it is shown that this can be expressed as
\begin{equation}
    \tan^2\theta_n = [X_{(n,r)}]^{-1}_{nn} - 1,
\end{equation}
where $X_{(n,r)}$ is the $r\times r$ block of the matrix
\begin{equation}
    [X]_{n', n''}
    \equiv \sum_{ab} Z_{n'a} C^{-1}_{ab}(0) Z^*_{n'' b},
\end{equation}
where $n',n'' \in [n,n+r-1]$.
For the ground state this simplifies to
\begin{equation}\label{eq:KPS0} \begin{split}
    0 \leq \frac{\lambda_0 - \lambda_0^{(m)}}{\lambda_0} &\leq  \frac{\tan^2\theta_0}{T_{m-1}(2 e^{\delta_r} - 1)^2},
    \end{split}
\end{equation}
where $\delta_r = E_r - E_0$ and, with $\bm{Z}$ the $r \times r$ matrix defined by $Z_{na}$ with $n \in [0,r-1]$, 
\begin{equation}
    \tan^2\theta_0 = \left[ \bm{Z} \bm{C}(0)^{-1} \bm{Z}^\dagger \right]^{-1}_{00} - 1.
\end{equation}
For large $m$, $T_{m}(x) \approx \frac{1}{2}(x + \sqrt{x^2 - 1})^{m}$, so this further simplifies to
\begin{equation}\label{eq:block_convergence}
    0 \leq \frac{\lambda_0 - \lambda_0^{(m)}}{\lambda_0 } \lesssim  4 \tan^2\theta_0  \times \begin{cases} e^{-2 (m-1) \delta_r }    &  \delta_r \gg 1 \\
    e^{-4(m-1)\sqrt{\delta_r}} &  \delta_r \ll 1 
    \end{cases} .
\end{equation}
For $r=1$ these results reduce to scalar Lanczos results, while for $r > 1$ block Lanczos converges exponentially faster than scalar Lanczos because $\delta_r > \delta_1$.

\begin{figure}
    \includegraphics[width=\linewidth]{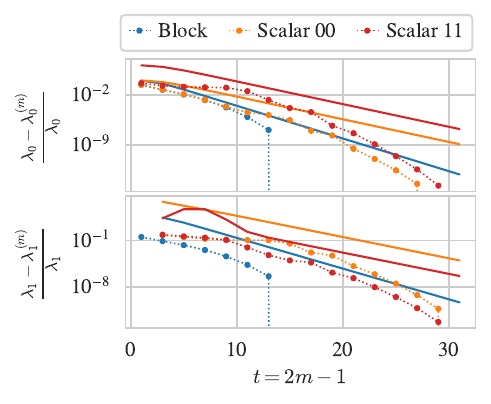}
    \caption{
        Convergence of block (blue) and scalar (orange, red) extractions of the two largest eigenvalues for the noiseless example of \cref{sec:noiseless} (\cref{eq:noiseless:ex-def}), alongside corresponding KPS bounds.
        Markers connected by dashed lines indicate the true relative error, while solid lines indicate the bound.
        The block KPS curve beyond $m = 8$ is an extrapolation computed by taking $K^{(m)}_n = K^{(8)}_n$.
    }
    \label{fig:noiseless-spectrum-kps}
\end{figure}

The convergence of block and scalar Lanczos and the corresponding KPS bounds are compared in  \cref{fig:noiseless-spectrum-kps}.
For the ground state, block Lanczos shows nearly identical convergence as for scalar Lanczos applied to $C_{00}$ until the last few iterations before solving the system exactly.
Convergence for scalar Lanczos applied to $C_{11}$ parallels that of $C_{00}$ after an initial pause due to the delay in resolving the true ground state.
\Cref{fig:noiseless-spectrum-kps}  further demonstrates the validity of the block and scalar KPS bounds in comparisons with relative errors of eigenvalue estimates computed for the noiseless example of \cref{sec:noiseless}.
Each bound is satisfied, although the bound is not well-saturated in any case and especially not for the excited state; it becomes increasingly less-saturated for higher states not shown.
More positively, observed convergence is substantially faster than guaranteed.
The same bounds predict block Lanczos will converge exponentially faster than scalar Lanczos.
This is clearly reflected in empirical convergence for the excited state, for which block Lanczos's improved extraction is especially apparent, as expected for a method which analyzes correlator matrices rather than individual correlators.

\subsection{Residual bounds}

The residual bound provides a distinct and complementary two-sided bound on the minimum distance between a Ritz value at finite $m$ and a true eigenvalue of $T$ for scalar Lanczos~\cite{Paige:1971,Parlett:1979,Parlett,Parlett:1995}.
Unlike the KPS bound, it can be calculated directly from matrix elements of $\Tm$ and therefore provides a practically computable bound on the size of excited-state effects at finite $m$.
However, also unlike the KPS bound, it is noteworthy that the residual bound applies to the difference between a given Ritz value and \emph{the closest true eigenvalue}, as opposed to a particular true eigenvalue such as $\lambda_0$. 
It was shown in Ref.~\cite{Wagman:2024rid} that the residual bound applies to oblique Lanczos assuming $T = T^\dagger$, even when $\Tm \neq [\Tm]^\dagger$.
We discuss the block generalization next.

The derivation proceeds by comparing the matrix-element expression for a Ritz-vector residual norm with its spectral expansion.
Right and left Ritz-vector residual norms are defined by 
\begin{equation}
    \begin{split}
        R^{R/L(m)}_k &\equiv \mbraket{y^{R/L(m)}_k}{ |T - \Tm|^2 }{y^{R/L(m)}_k} ,
    \end{split}
\end{equation}
and quantify the difference between the action of the exact transfer matrix $T$ and the Krylov-space approximation $\Tm$ on Ritz vectors.
For oblique block Lanczos, the right residual norms can be computed by noting that Eq.~\eqref{eq:Tdiff-lanczos} implies
\begin{equation}\label{eq:Tdiff}
\begin{aligned}[]
    [T - \Tm] \ket{y^{R(m)}_k} &= \mathcal{N}_k^{(m)} \sum_{ab} \ket{ v^R_{(m+1)b }} \gamma_{(m+1)ba} \omega^{(m)}_{mak} ,
\end{aligned}
\end{equation}
i.e., that the action of $T$ on a Ritz vector is a rescaling within $rm$-dimensional Krylov space by its Ritz value plus the addition of a term orthogonal to this space.
It is then straightforward to derive
\begin{equation}
\begin{split}
        R^{R(m)}_k &= |\mathcal{N}_k^{(m)}|^2 \sum_{abcd} \omega^{(m)*}_{mak}  \gamma_{(m+1)ba}^* \\
        &\hspace{10pt} \times \braket{v^R_{(m+1)b}  }{ v^R_{(m+1)c} } \gamma_{(m+1)cd} \omega^{(m)}_{mdk}.
\end{split}
\end{equation}
An analogous calculation for left residual norms gives
\begin{equation}
\begin{split}
        R^{L(m)}_k &= \frac{1}{|\mathcal{N}_k^{(m)}|^2} \sum_{abcd} (\omega^{-1})^{(m)}_{kma}  \beta_{(m+1)ab} \\
        &\hspace{10pt} \times \braket{v^L_{(m+1)b}  }{ v^L_{(m+1)c} } \beta_{(m+1)dc}^* (\omega^{-1})^{(m)*}_{kmd}.
\end{split}
\end{equation}
These expressions differ from the scalar Lanczos analogs only by the presence of block indices.

The derivation of the spectral representation of the residual bound in Ref.~\cite{Wagman:2024rid} applies here without modification, under the same assumption that $T$ is Hermitian---the Ritz vectors $\bigl| y_k^{R/L(m)} \bigr>$ can be expanded as linear combinations of true energy eigenstates regardless of whether they are obtained from scalar or block Lanczos.
This provides the inequalities
\begin{equation}
    \min_{\lambda \in \{\lambda_n\}} \left| \lambda^{(m)}_k - \lambda \right|^2 \leq  B^{R/L(m)}_k,
    \label{eq:block_bound}
\end{equation}
where
\begin{equation}
    B^{R/L(m)}_k \equiv 
     \left| \frac{ R^{R/L(m)} }{ \braket{y^{R/L(m)}_k} { y^{R/L(m)}_k} } \right|,
\label{eq:residual-B}
\end{equation}
can be conveniently expressed as
\begin{equation}
\begin{split}\label{eq:residual-BV}
        B^{R(m)}_k &= \left| \sum_{ab} \omega^{(m)*}_{mak} [ \bm{\gamma}_{(m+1)}^\dagger \bm{V}^{R(m)}_{k}  \bm{\gamma}_{(m+1)} ]_{ab} \omega^{(m)}_{mbk} \right|, \\
        B^{L(m)}_k &= \left| \sum_{ab}  (\omega^{-1})^{(m)}_{kma} [ \bm{\beta}_{(m+1)} \bm{V}^{L(m)}_{k} \bm{\beta}_{(m+1)}^\dagger ]_{ab} (\omega^{-1})^{(m)*}_{kmb} \right|,
\end{split}
\end{equation}
in terms of the matrices
\begin{equation}\begin{aligned}
    V^{R(m)}_{kab}
    &\equiv
   \frac{ |\mathcal{N}^{(m)}_k|^2  \braket{v^R_{(m+1)a}}{v^R_{(m+1)b}} }{ \braket{y^{R(m)}_k }{ y^{R(m)}_k} }
    \\ &= \frac{ \braket{v^R_{(m+1)a}}{v^R_{(m+1)b}} }
        { \sum_{ijcd} \omega^{(m)*}_{ick} \braket{v^R_{ic} }{ v^R_{jd} } \omega^{(m)}_{jdk} },
    \\
    V^{L(m)}_{k ab} &\equiv
       \frac{ \braket{v^L_{(m+1)a}}{v^L_{(m+1)b}} }{ |\mathcal{N}^{(m)}_k|^2 \braket{y^{L(m)}_k }{ y^{L(m)}_k} }
    \\ &= \frac{  \braket{v^L_{(m+1)a}}{v^L_{(m+1)b}} }
        { \sum_{ijcd} (\omega^{-1})^{(m)}_{kic} \braket{v^L_{ic} }{ v^L_{jd}} (\omega^{-1})^{(m)*}_{kjd} }
    ~ .
\end{aligned}
\label{eq:residual-V}
\end{equation}

\begin{figure}
    \includegraphics[width=\linewidth]{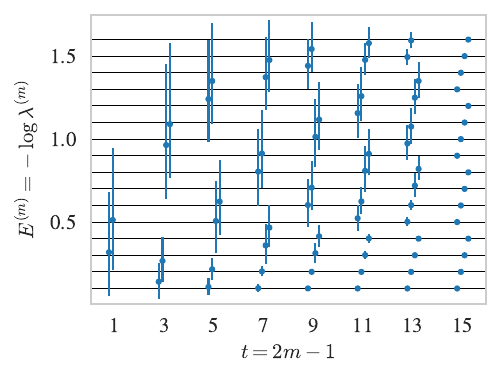}
    \caption{
        Demonstration of block Lanczos residual bounds for the noiseless example \cref{eq:noiseless:ex-def}.
        Error bars are not statistical, but rather represent the extent of values allowed by the residual bound \cref{eq:block_bound}.
        The window of allowed values is computed for Ritz values and mapped through the logarithm, and thus asymmetric for energies.
    }
    \label{fig:noiseless-spectrum-bounds}
\end{figure}

Given a symmetric choice of oblique convention $\bm{\beta}_j = \bm{\gamma}_j$, the matrices $V^{R/L(m)}_{kab}$ are equal to $\delta_{ab}$ in the noiseless limit where $\iket{v^R_j} = \iket{v^L_j}$ and the oblique algorithm reduces to the standard symmetric one.
However, they are non-trivial and must be computed explicitly in applications to noisy data where the oblique formalism is required.
This may be accomplished straightforwardly using \cref{eq:lanczos:wv} once Krylov coefficients have been computed.
Thus, two-sided Ritz value error bounds can be computed in practical applications for all $m$ where Lanczos results with $m+1$ iterations are available using Eqs.~\eqref{eq:block_bound}-\eqref{eq:residual-V} and Lanczos-vector norms in Eq.~\eqref{eq:lanczos:wv}, i.e., for all iterations but the last one, $m=N_t/2$.
These bounds hold for oblique block Lanczos with Hermitian $T$, regardless of whether $\Tm$ is Hermitian.
The explicit absolute values in \cref{eq:residual-B,eq:residual-V} are not required for noiseless applications but are helpful for regulating negative values that can arise in applications to noisy data.

\Cref{fig:noiseless-spectrum-bounds} demonstrates the validity of the residual bound for Ritz values extracted by applying block Lanczos to the noiseless example of \cref{sec:noiseless}; \cref{app:scalar-res-bounds} examines the residual bounds for scalar Lanczos applied to the same example.
In the figure, each Ritz value is shown with an error bar representing the extent of the window of values allowed by the bound.
For all points, the error bar crosses at least one line representing a true eigenvalue, indicating the bound is satisfied for all values.
A separate question is whether the bounds are sufficiently tight to be useful in practice.
As can be observed in \cref{fig:noiseless-spectrum-bounds}, as a Ritz value converges, its residual bound shrinks as well, with each Ritz value quickly becoming consistent with only a single true value.
For the lower-lying states, the bounds rapidly become smaller than visible on the plot.

\section{Noise}
\label{sec:lattice}

Block Lanczos applied to noisy correlator data exhibits an analogous version of the ``Lanczos phenomenon''~\cite{Cullum:1981,Cullum:1985} as the scalar case: some Ritz vectors consistently describe physical states that have all the properties expected at infinite statistics, while other Ritz vectors describe states that have manifestly unphysical properties and must arise from noise.
This section describes techniques for identifying and removing such spurious states from Lanczos results, including the construction of a simple physical picture that is equivalent to the Cullum-Willoughby (CW) test
for scalar Lanczos.
This picture, based on the identification of Ritz vectors that have pathologically small overlaps with all of the 
initial states ($\sim$ interpolating operators) considered, 
is used to define a physically motivated block Lanczos generalization of the CW test.
This is the first generalization of the CW test to block Lanczos, as far as we are aware.

Proof-of-principle results are demonstrated for the extraction of energies and overlap factors from a $2\times 2$ matrix of nucleon correlation functions.
Comparisons are made between block and scalar results for this dataset, as well as scalar Lanczos results from the higher-statistics dataset with identical physical parameters studied in Ref.~\cite{Hackett:2024xnx}.
Comparisons with GEVP are deferred to \cref{sec:gevp}.
We focus on exploring how block versus scalar Lanczos affects the spurious eigenvalue filtering and state identification that arise during spectroscopy; once Ritz coefficients have been computed for a given state, the subsequent evaluation of matrix elements for external currents only involves matrix multiplication of three-point functions and proceeds identically for block and scalar Lanczos.

\subsection{Problem setup \& data}
\label{sec:setup}

\begin{figure}
    \includegraphics[width=\linewidth]{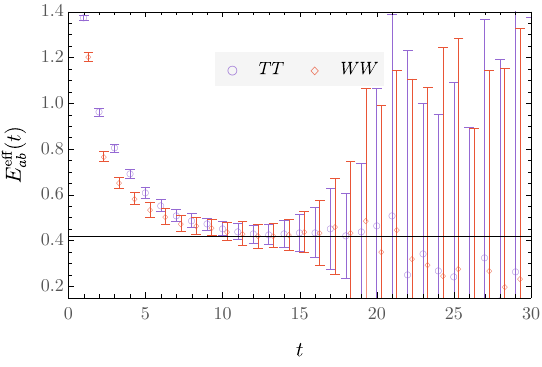}
    \caption{
      Effective energies $E^\mathrm{eff}_{ab} = -\ln C_{ab}(t) / C_{ab}(t-1)$ for the diagonal elements of the correlator matrix, $a=b=T$ (purple, $TT$) and $a=b=W$ (red, $WW$) with a horizontal offset for clarity.
      Recall $a=T$ indicates the interpolator with ``thin'' smearing and $a=W$ the ``wide'' one.
      Uncertainties are computed using bootstrap confidence intervals.
        The black line is the fit from Refs.~\cite{Hackett:2023rif,Hackett:2023nkr}, $0.4169(18)$; its uncertainties are not visible on the scale of the plot.
    }
    \label{fig:nuc-meffs}
\end{figure}

For the demonstration, we use data computed 
for the same ensemble with $a \approx 0.091~\mathrm{fm}$ and $m_{\pi} \approx 170~\mathrm{MeV}$~\cite{Park:2021ypf,Yoon:2016jzj,Mondal:2020ela} as used for demonstrations in Refs.~\cite{Wagman:2024rid,Hackett:2024xnx}. 
Configurations were generated by the JLab/LANL/MIT/WM groups~\cite{ensembles} using the tadpole-improved L\"uscher-Weisz gauge action~\cite{Luscher:1984xn} and $N_f = 2+1$ flavors of clover fermions~\cite{Sheikholeslami:1985ij} defined with stout smeared~\cite{Morningstar:2003gk} links on a $48^3 \times 96$ lattice volume.
Quark propagators and nucleon correlator matrices were computed by the NPLQCD Collaboration.

The data for the block Lanczos example are a $2 \times 2$ matrix of nucleon two-point functions projected to zero momentum, computed on $N_\mathrm{cfg}=80$ configurations independently from and on a (partially) different subensemble than the data used in Refs.~\cite{Wagman:2024rid,Hackett:2024xnx}.
The two interpolators differ only in the quark smearing, and are labeled by $a \in \{T,W\}$ where $T$ is for ``thin'' and $W$ for ``wide''.
Each is of the form
\begin{equation}
    \psi_a(x) = \epsilon^{ijk} [u_j^S(x)^T C \gamma_5 d_k^S(x)] u_i^S(x),
\end{equation}
where $C$ is the charge conjugation matrix, and $u^S(x)$ and $d^S(x)$ are up- and down-quark fields smeared using gauge-invariant Gaussian smearing, with radius $3.0$ for $\psi_T$ and $4.5$ for $\psi_W$.
The measurement for each configuration is obtained by evaluating the correlator
\begin{equation}
    C_{ab}(t; x_0) = \sum_{\vec{x}} \Tr \left[ \Gamma_{\pm}  \avg{ \psi_a(\vec{x}, t+t_0) \overline{\psi}_b(x_0) }  \right],
\end{equation}
where
\begin{equation}
    \Gamma_\pm = P_+ (1 \pm i \gamma_x \gamma_y) \text{   with   } P_+ = \frac{1}{2} (1+\gamma_t) ~ .
\end{equation}
on an $8^3$ grid of $N_{\rm src}=512$ source positions all on a single random timeslice, then summing over sources and averaging both signs of $\Gamma_\pm$, i.e.~over the up-up and down-down channels.
This source- and spin-averaged correlator matrix is denoted $\bar{C}_{ab}(t)$.
The $\psi_a$ will also sometimes be denoted $N^+_a$ below in order to distinguish them from interpolating operators $N^-_a$ obtained by replacing $P_+ = \frac{1}{2} (1+\gamma_t)$ with $P_- = \frac{1}{2} (1-\gamma_t)$ in all expressions above.

It is essential for the spurious state filtering methods described below to explicitly take the real and symmetric part of the correlator matrix, denoted
\begin{equation}
  C'_{ab}(t) = \frac{1}{2} \text{Re}\left[ \bar{C}_{ab}(t) + \bar{C}_{ba}(t) \right].
\end{equation}
As with the imaginary part of a scalar correlator, the imaginary and antisymmetric parts of the correlator matrix are zero in expectation, and we are free to set them to their known infinite-statistics values.
Hermiticity follows immediately for any correlator matrix built from the same source and sink operators, $C_{ab}(t) = \mbraket{\psi_a}{T^t}{\psi_b}$, from the underlying Hermiticity of the transfer matrix.
The stronger assumption of a real, symmetric correlator matrix is valid for any interpolating operators that are complex conjugated by a $CP$ transformation and a $2\pi$ rotation, as shown in Appendix~\ref{app:real}. This includes the interpolators used here as well as all commonly used single- and multi-hadron interpolating operators with plane-wave spatial wavefunctions built from products of (momentum-)smeared~\cite{Gusken:1989ad,Gusken:1989qx,Bali:2016lva} quark fields.

To compare overlap factors between different operators, it is convenient to work directly with interpolating operators $\hat{\psi}_a$ creating unit-normalized states such that ${\langle |\hat{\psi}_a|^2 \rangle} = 1$.
The corresponding correlator matrix built from unit-normalized interpolating operators is
\begin{equation}
  C_{ab}(t) = \frac{C'_{ab}(t)}{\sqrt{C'_{aa}(0) C'_{bb}(0)}},
\end{equation}
and satisfies $C_{aa}(0) = C_{bb}(0) = 1$.
This is not a necessary precursor to applying the Lanczos analysis, which may treat correlators with arbitrary norms, but instead simply to render the overlap factors for the different smearings of comparable magnitude.
This normalization is not included in the error propagation.\footnote{That is, it is computed from the mean of $C_{ab}(0)$ and this common value is applied for all bootstrap draws.}
This furthermore does not affect the value of the ZCW cut quantity $\DZCW$ defined and discussed in \cref{sec:CW} below.

We employ bootstrap resampling to study the effects of noise on the analysis and propagate uncertainties. In particular, we construct 200 bootstrap ensembles (bootstraps) by drawing 80 samples with replacement from the original dataset of 80 and apply Lanczos to the ensemble average within each bootstrap. We further study the effects of a nested bootstrap procedure in which these 200 (outer) bootstrap ensembles are resampled again, generating 200 (inner) bootstrap ensembles for each outer ensemble by again drawing 80 samples with replacement from the 80 (non-unique) samples of the outer ensemble. We then apply Lanczos to the ensemble average of each inner bootstrap. As discussed below, this enables the calculation of an outlier-robust bootstrap-median based estimator and its uncertainties.

For some tests below, we also use the same high-statistics dataset of $N_\mathrm{cfg}=1381$ configurations employed in Ref.~\cite{Hackett:2024xnx}, generated in the course of the studies in Refs.~\cite{Hackett:2023nkr,Hackett:2023rif}.
This scalar nucleon correlator is defined similarly to the $WW$ channel of the $2\times 2$ matrix, using the same interpolator $\psi_W$ with smearing radius 4.5.
Two-point functions are computed and summed over two offset $4^3 \times 8$ grids of 512 source positions with an overall random offset, for a total of 1024 positions per configuration.
It is similarly projected to zero momentum, averaged over both signs of $\Gamma_\pm$, and the real part is taken.
The correlator is normalized to equal unity at $t=0$, which is similarly not included in the error propagation.
Effective energies for the diagonal
entries of the correlator matrix are shown in Fig.~\ref{fig:nuc-meffs}.

\subsection{Hermitian subspace filtering}\label{sec:Hermitian}

As in the scalar case, a subset of block Lanczos outputs (states) are real Ritz values $\lambda^{(m)}_k = \lambda^{(m)*}_k$ with corresponding degenerate left/right Ritz vectors $\iket{y^{R(m)}_k} = \iket{y^{L(m)}_k}$.
These obey the physical properties that follow from Hermiticity of $T$ in the infinite-statistics limit and define a Hermitian subspace of Krylov space in the sense made precise below.
We denote\footnote{The definition of $\mathcal{H}$ necessarily differs for different $m$ and could more explicitly be denoted $\mathcal{H}^{(m)}$. We leave this $m$ dependence implicit to improve readability.} this subset of Ritz values and vectors by $k \in \mathcal{H}$ where $\mathcal{H} \subset \{0,1,2,\ldots,rm-1\}$.
The objective of Hermitian subspace filtering is to identify and retain the physically interpretable states in $\mathcal{H}$ and discard those in its complement $\overline{\mathcal{H}}$.

Sorting the Ritz values and vectors into sets labeled by $\mathcal{H}$ and $\overline{\mathcal{H}}$ provides 
the operator-level decomposition
\begin{equation}
    T^{(m)} = T^{(m)}_{\mathcal{H}} + T^{(m)}_{\overline{\mathcal{H}}},
\end{equation}
where $T^{(m)}_{\mathcal{H}} = T^{(m)\dagger}_{\mathcal{H}}$ acts on states in the Hermitian subspace as
\begin{equation}
    T^{(m)}_{\mathcal{H}} \equiv \sum_{k \in \mathcal{H}} \ket{y^{(m)}_k} \lambda^{(m)}_k \bra{y^{(m)}_k} ~ ,
\end{equation}
with $\lambda^{(m)}_k = \lambda^{(m)*}_k$ and $L/R$ labels omitted because $\iket{y^{R(m)}_k} = \iket{y^{L(m)}_k}$.
Distinctions between left and right quantities only arise for states in the non-Hermitian subspace where the action of $\Tm$ is governed by
\begin{equation}
    T^{(m)}_{\overline{\mathcal{H}}} \equiv \sum_{k \in \overline{\mathcal{H}}} \ket{y^{R(m)}_k} \lambda^{(m)}_k \bra{y^{L(m)}_k} ~ .
\end{equation}
The associated Hermitian and non-Hermitian subspaces of Krylov space can be explicitly defined as
\begin{equation}
\begin{split}
    \mathcal{K}^{(m)} &\equiv \text{span}\{ \bigl|y_k^{(m)} \bigr>, k \in \mathcal{H} \}, \\
    \overline{\mathcal{K}}^{L(m)} &\equiv \text{span}\{ \bigl|y_k^{L(m)} \bigr>, k \in \overline{\mathcal{H}} \}, \\
    \overline{\mathcal{K}}^{R(m)} &\equiv \text{span}\{ \bigl|y_k^{R(m)} \bigr>, k \in \overline{\mathcal{H}} \}.
    \end{split}
\end{equation}
Ritz vector orthogonality $\bigl< y_k^{L(m)} | y_l^{R(m)} \bigr> = \delta_{kl}$ is exact even for noisy applications because $\bigl< v_{ia}^L | v_{jb}^R \bigr> = \delta_{ij}\delta_{ab}$ follows identically from the recursion relations.
This means that these spaces provide an exact decomposition of Krylov space of the form
\begin{equation}
\begin{split}
    \mathcal{K}^{L(m)} &= \mathcal{K}^{(m)} \oplus \overline{\mathcal{K}}^{L(m)}, \\
    \mathcal{K}^{R(m)} &= \mathcal{K}^{(m)} \oplus \overline{\mathcal{K}}^{R(m)}.
    \end{split}
\end{equation}
We define all non-spurious states to include only elements of the Hermitian subspace $\mathcal{K}^{(m)} \subset ( \mathcal{K}^{L(m)} \cap \mathcal{K}^{R(m)} )$.

The features of the remaining states with labels $\overline{\mathcal{H}}$ clearly identify them as noise artifacts (or at least noise-contaminated).
This can be seen by noting their role in the correlator decomposition (cf.~\cref{sec:formalism:corr-decomp}),
\begin{equation}
    C_{ab}(t) = \sum_k Z^{R(m)*}_{ka} Z^{L(m)}_{kb} [\lambda^{(m)}_k]^t,
\end{equation}
which holds exactly for all $t \leq 2m-1$.
Reproducing noisy data generically requires states with complex and/or negative Ritz values that make oscillatory contributions, since noise does not preserve the existence of a convex spectral representation.
Complex-eigenvalue states necessarily come in conjugate pairs when $\bm{C}(t)$ is real.
Other states have real $\lambda^{(m)}_k > 0$ but distinct $L/R$ overlaps $\bigl< y^{R/L(m)}_k | \psi_a \bigr>$, allowing them to make unphysical negative contributions to diagonal correlators.

One may also separately filter on the positivity of Ritz values, i.e.~discard any $\lambda^{(m)}_k < 0$, on the grounds that in the infinite-statistics limit the underlying transfer matrix is positive-definite.
However, it remains unclear whether this is the best option in practical analyses of noisy data.
While such states make oscillatory contributions that are clearly associated with noise, it is not unreasonable for a small eigenvalue to fluctuate negative, and clipping the distribution at zero may complicate error analysis.
The results below do not filter explicitly on positivity and instead rely on subsequent filtering by the Cullum-Willoughby test to define which are spurious.

The practical Hermitian subspace filtering prescription proceeds similarly as in the scalar case.
States with complex Ritz values may be immediately sorted into $\overline{\mathcal{H}}$, defining reality at working precision as
\begin{equation}\label{eq:real-at-finite-prec}
  \left| \frac{\text{Im}[\lambda_k^{(m)}]}{ \lambda_k^{(m)} } \right| = \left| \sin \arg \lambda_k^{(m)} \right| < \eRe,
\end{equation}
where $\eRe \sim 10^{-8}$ is often appropriate.\footnote{When eigensolves are performed with floating precision arithmetic, lowering $\varepsilon^{\rm float}$ to significantly smaller values than $10^{-8}$ leads to all Ritz values being labeled as outside the Hermitian subspace. If eigensolves are performed in higher precision, then $\varepsilon^{\rm float}$ can be lowered accordingly without changing the number of states in the Hermitian subspace. This suggests that these states have imaginary parts arising solely from the use of floating point precision at some stages of the calculation, motivating the notation $\varepsilon^{\rm float}$.}
Further removal of unphysical $k$ where right and left Ritz vectors do not coincide
may be accomplished with the block version of the ``norm trick'' introduced in Ref.~\cite{Hackett:2024xnx}, by attempting to compute $|\mathcal{N}^{(m)}_k|^2$ per \cref{eq:norm_calc} as
\begin{equation}
    \frac{ \sum_b (\omega^{-1})^{(m)}_{k1b}  \gamma_{1ba}  }{ 
    \left[ \sum_c \beta_{1ac} \omega^{(m)}_{1ck} \right]^* } ,
    \label{eq:lattice:norm-trick}
\end{equation}
yielding for each state $k$ a set of $r$ values indexed by $a$.
As in the scalar case, equality with real $|\mathcal{N}^{(m)}_k|^2 > 0$ requires that these values all be real and positive; states are thus sorted into $\overline{\mathcal{H}}$ for which $\text{arg}[N_{ka}^{(m)}] > \eRe$ for any $a$.
As advertised in \cref{sec:formalism:overlaps}, in the block case we also impose an additional criterion---that the normalization of a Ritz vector $\iket{y^{(m)}_k}$ can be computed consistently from its overlap with any $\ket{\psi_a}$. In practice, this amounts to requiring that the value computed by \cref{eq:lattice:norm-trick} is numerically identical for all $a$ at the level of $\eRe$.
We note that this last condition fails to hold for \emph{any} state if the correlator matrix is not taken to be real, even if it is Hermitian.
It is essential to apply oblique block Lanczos to the real, symmetric part of a noisy correlator matrix $\text{Re}[\bm{C}(t) + \bm{C}(t)^T]/2$ rather than the complex, Hermitian part $(\bm{C}(t) + \bm{C}(t)^\dagger)/2$ in order for this part of the Hermitian subspace filter to be valid.

Following this filtering, we find that left and right Ritz coefficients of the surviving states coincide within numerical precision, i.e.,
\begin{equation}
    \sum_b \gamma^{-1}_{1ab} P^{R(m)}_{ktb} = \sum_b P^{L(m)*}_{ktb} \beta^{-1}_{1ba} ~ ,
\end{equation}
where the factors of $\bm{\beta}^{-1}_1$ and $\bm{\gamma}^{-1}_1$ give a convention-independent definition.
This is sufficient to guarantee that all left and right estimators of physical quantities will be identical.

\subsection{The Cullum-Willoughy test}
\label{sec:CW}

In the context of numerical linear algebra, procedures like the Cullum-Willoughby (CW) test~\cite{Cullum:1981,Cullum:1985} and selective reorthogonalization~\cite{Parlett:1979} are motivated by the observation that numerical noise (i.e.~errors from finite-precision arithmetic) causes a loss of (bi)orthogonality in the Lanczos vectors after many iterations.
In practice, this results in an artificial expansion of Krylov space outside what would be obtained in the noiseless case and thus the seeding of spurious Ritz vectors which must be diagnosed and removed.
Intuitively speaking, the effect of noise is to mix new information into the Lanczos vectors which is independent of the initial vectors $\ket{v_1^R}$ and $\bra{v_1^L}$.
The observation motivates the CW test, which diagnoses spuriosity by examining sensitivity of Ritz values to the removal of the initial vectors from the Krylov space.
As discussed in detail below, this same observation motivates an alternative ``ZCW test'' that diagnoses spuriosity by directly computing the overlaps of Ritz vectors with the initial vectors.

\begin{figure}
    \includegraphics[width=\linewidth]{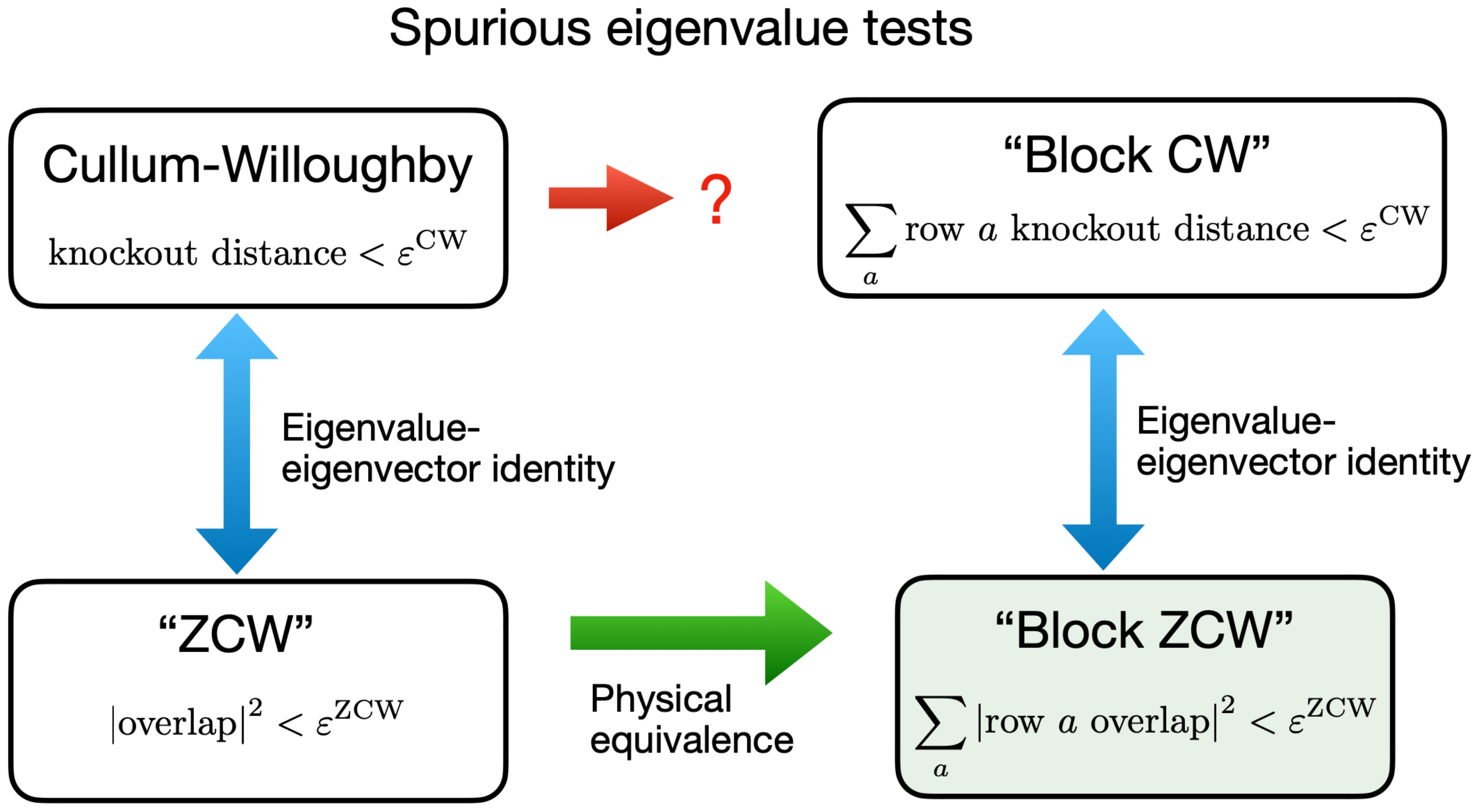}
    \caption{
       Schematic depiction of the line of argumentation used to construct block generalizations of the CW test in \cref{sec:CW}.
    }
    \label{fig:CW_diagram}
\end{figure}

\Cref{fig:CW_diagram} schematically illustrates the line of argumentation of this subsection, whose goal is to provide a block generalization of CW filtering.
As discussed in detail below, the original CW test is challenging to generalize to the block case directly; in contrast, the ZCW test generalizes naturally on physical grounds.
Separately, we find the CW and ZCW tests are closely related by the eigenvalue-eigenvector identity~\cite{Denton:2019pka} and functionally equivalent in practice.
This finally allows an indirect derivation of block CW, although we find the resulting prescription is practically equivalent to block ZCW and more expensive.

\subsubsection{The CW test}

As presented in its original formulation~\cite{Cullum:1981,Cullum:1985} and employed in Refs.~\cite{Wagman:2024rid,Hackett:2024xnx}, the CW test compares the Ritz values with the eigenvalues $\widetilde{\lambda}^{(m)}_k$ of the matrix $\CWm$ constructed by removing (``knocking out'') the first row and column from $T^{(m)}$.
The matrix $\CWm$ corresponds to the projection of $T$ to a Krylov space with the initial states removed.
Thus, by the reasoning above, Ritz values insensitive to this removal must have been seeded by noise and can be identified as spurious, where ``insensitive'' is defined as
\begin{equation}
   \DCW \equiv \min_l \left| \lambda^{(m)}_k - \widetilde{\lambda}^{(m)}_l \right|
    < \eCW,
    \label{eq:CW-test}
\end{equation}
i.e., whether any eigenvalue of $\CWm$ is equal to a given Ritz value within a chosen threshold $\eCW$.
In numerical applications where noise is due to finite-precision arithmetic, $\eCW$ may be taken to be machine precision.
Where noise is statistical, as in analysis of lattice correlator data, $\eCW$ must be set accordingly larger.
Refs.~\cite{Wagman:2024rid,Hackett:2024xnx} propose different adaptive procedures to set this threshold.
Below, we propose a simpler alternative to these procedures which follows naturally from the reformulation of the CW test in more physically intuitive terms.

As already advertised, our new physical perspective presents a straightforward path to a block generalization of the CW test not offered by its original motivating argument.
The original argument~\cite{Cullum:1981,Cullum:1985} relies on an identity relating the (scalar) residual bound (\cref{eq:residual-B} with $r=1$) to the characteristic polynomials of $\widetilde{T}$ and $T^{(m-1)}$,
defined as $\widetilde{a}_m(\mu) \equiv \det(\CWm - \mu \mathbbm{1})$ and  ${a_{m-1}(\mu) \equiv \det(T^{(m-1)} - \mu \mathbbm{1})}$:
\begin{equation}\label{eq:CW_resid}
    B_k^{(m)} = \frac{\prod_{j=2}^{m+1} |\beta_j|^2}{\widetilde{a}_m\left(\lambda_k^{(m)}\right) a_{m-1}'\left(\lambda_k^{(m)}\right) }, \hspace{15pt} (r=1),
\end{equation}
where $a_m'(\mu)$ denotes the derivative of $a_m(\mu)$.
This identity was derived for symmetric Lanczos; we have numerically verified that it remains true for oblique Lanczos applications to noisy LQCD data.
This identity shows that $B_k^{(m)}$ has a pole whenever $\widetilde{a}_m(\lambda_k^{(m)}) = 0$, 
i.e.~whenever $\lambda^{(m)}_k$ is also an eigenvalue of $\CWm$ and thus $\DCW=0$.
This directly associates $\DCW=0$ with non-convergence of $\lambda^{(m)}_k$ to a true eigenvalue, as indicated by a diverging residual bound.
As noted in Eq.~\eqref{eq:CW_resid}, this identity is only valid for scalar Lanczos ($r=1$) and we are not aware of a simple generalization that is valid for block Lanczos with $r>1$.
In the scalar case, it is valid to identify Ritz values with spuriously small $\DCW$ and Ritz values with spuriously large residuals; however, this this perspective does not generalize straightforwardly to the block case.

\subsubsection{CW in terms of overlaps (ZCW)}

The original motivation for the CW test---independence of the starting vectors as a characteristic property of spurious Ritz vectors---suggests an alternative and more literal procedure: 
directly compute and examine the overlaps of Ritz vectors with the starting states $\sim \braket{v_1}{y_k}$, with a small (i.e.~zero up to noise) overlap indicating a spurious state.
These are simply the usual overlap factor estimators $Z^{(m)}$, up to normalization (and, in the block case, choice of basis, as discussed further below).
Specifically, restricting initially to scalar Lanczos, we consider the normalized overlap factor product as our test quantity,
\begin{equation}\label{eq:DZCW_scalar}\begin{aligned}
    \DZCW 
    &= \left| \braket{v^L_1}{y^{R(m)}_k} \braket{y^{L(m)}_k}{v^R_1} \right|
    \\&= \left| \frac{Z^{R(m)*}_k Z^{L(m)}_k}{C(0)} \right|
    = \left| \omega^{(m)}_{1k} (\omega^{-1})^{(m)}_{k1} \right| \,,
\end{aligned}\end{equation}
and define the ZCW test for spuriosity as
\begin{equation}\label{eq:ZCW-test}
    \DZCW < \varepsilon^\mathrm{ZCW} ~ .
\end{equation}
How to choose the cut $\varepsilon^\mathrm{ZCW}$ is discussed in the next subsection.

\begin{figure}
    \includegraphics[width=\linewidth]{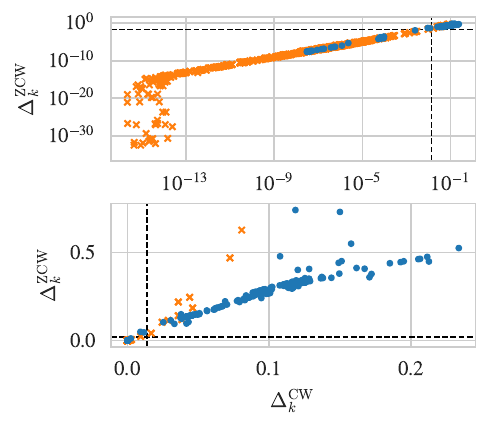}
    \caption{
        Comparison of $\DCW$ and $\DZCW$ for scalar Lanczos results using the high-statistics nucleon correlator analysis of Ref.~\cite{Hackett:2024xnx}.  
        Each marker corresponds to a pair of values for some $m,k$, for all $m \in [2,48]$ and all $k$, as computed on the central-value correlator.
        No graphical distinction is made between different $m$ and $k$.
        Blue dots (orange Xs) are states inside (outside) the Hermitian subspace.
        The top and bottom panels show the same data with log and linear axes.
        The vertical line indicates the cut $\eCW \approx 1.4 \times 10^{-2}$ selected using the bootstrap histogram procedure of Ref.~\cite{Wagman:2024rid}, while the horizontal line indicates the cut $\eZCW \approx 2.2 \times 10^{-2}$ selected by \cref{eq:eZCW}.
    }
    \label{fig:nuc-vy-vs-dCW}
\end{figure}

\Cref{fig:nuc-vy-vs-dCW} illustrates a practical comparison of the CW and ZCW tests using results for $\DCW$ and $\DZCW$ computed for the high-statistics scalar nucleon correlator of Ref.~\cite{Hackett:2024xnx}.
The relation between $\DZCW$ and $\DCW$ is remarkably linear for the range of $\DCW$ corresponding to spurious eigenvalues when $\eCW$ is defined using the bootstrap histogram method from Ref.~\cite{Wagman:2024rid}.
This demonstrates that an appropriate choice of $\eZCW$ will lead to ZCW test results that are completely equivalent to those from the CW test with this choice of $\eCW$.

The close relationship between the CW and ZCW tests may be understood in terms of the ``eigenvalue-eigenvector identity'', first derived by Jacobi in 1834~\cite{Jacobi:1834} and re-discovered many times since~\cite{Denton:2019pka}.
For $\CWm$ specifically,\footnote{
Similarly to how Cullum \& Willoughby use Eq.~\eqref{eq:CW_resid} to relate $B_k^{(m)}$ to eigenvalues of a minor of $\Tm$ (i.e.,~$\CWm$), the general eigenvalue-eigenvector identity relates products of eigenvector elements $\omega^{(m)}_{ik} (\omega^{-1})^{(m)}_{kj}$ to eigenvalues of the minor of $\Tm$ with row $i$ and column $j$ knocked out.
} it reads
\begin{equation}
    \prod_{\widetilde{l}} \left[ \lambda^{(m)}_k - \widetilde{\lambda}^{(m)}_{\widetilde{l}} \right]
    = \omega^{(m)}_{1k} (\omega^{-1})^{(m)}_{k1} \prod_{l \neq k} \left[ \lambda^{(m)}_k - \lambda^{(m)}_l \right] ~ .
\end{equation}
For any state, $\DCW$ appears in the product on the LHS.
Dividing the RHS by the remaining terms and recognizing $\DZCW$ allows $\DCW$ to be expressed as
\begin{equation}\label{eq:CW_ZCW}
\begin{split}
    \DCW 
    &= \DZCW \left( \frac{\prod_{l \neq k} \left| \lambda^{(m)}_k - \lambda^{(m)}_l \right|}{\prod_{\widetilde{l} \neq \widetilde{c}} \left| \lambda^{(m)}_k - \widetilde{\lambda}^{(m)}_{\widetilde{l}} \right|} \right),
    \end{split}
\end{equation}
where $\widetilde{c} = \text{argmin}_{\widetilde{l}} \left| \lambda^{(m)}_k - \widetilde{\lambda}^{(m)}_{\widetilde{l}} \right|$ indexes the closest eigenvalue of $\CWm$ to $\lambda^{(m)}_k$.

Equation~\eqref{eq:CW_ZCW} implies that if $\DZCW$ is small then so must be $\DCW$, and vice versa, and thus that they are equivalent quantities for filtering.
The key feature of Eq.~\eqref{eq:CW_ZCW} is that the factor in parentheses corresponding to $\DCW/\DZCW$ is necessarily non-zero and finite.
To see this, note first that the eigenvalues of any tridiagonal matrix are non-degenerate~\cite{Parlett}.
This means $\lambda^{(m)}_l \neq \lambda^{(m)}_k$ for $l \neq k$ and therefore the numerator $\prod_{l \neq k} \left| \lambda^{(m)}_k - \lambda^{(m)}_l \right|$ has no zeros.
It is possible to have $\lambda^{(m)}_k = \widetilde{\lambda}^{(m)}_{\widetilde{l}}$, but only for one choice of $\widetilde{l}$ because $\CWm_{ij}$ is tridiagonal.
This means that it is only possible to have $\lambda^{(m)}_k = \widetilde{\lambda}^{(m)}_{\widetilde{l}}$ when $\widetilde{l} = \widetilde{c}$,
so the denominator---defined as a product over $\widetilde{l} \neq \widetilde{c}$, which excludes the one possible zero---has no zeroes.
The finiteness of the numerator and denominator follow immediately from the fact that they are finite products of finite factors.
It follows that $\DCW = 0$ can occur if and only if $\DZCW = 0$.
Since both $\DCW$ and $\DZCW$ are smooth functions of $\lambda_k^{(m)}$, Ritz values within a sufficiently small neighborhood of $\DCW = 0$ are in one-to-one correspondence with the Ritz values within some small neighborhood of $\DZCW = 0$.
Therefore, the CW test $\DCW < \eCW$ for sufficiently small $\eCW$ is formally equivalent to 
a ZCW test $\DZCW < \eZCW$ for some $\eZCW$.

\subsubsection{Physical interpretation and choice of $\eZCW$}

Rephrasing the CW test in terms of overlaps suggests a clear physical (re)interpretation.
Specifically, filtering with the CW test corresponds to restricting to the physical subspace of states to those with the correct quantum numbers.
In practice, for any interpolating operator, the eigenexpansions of the initial states $\iket{v^{L/R}_1}$ have support over all states with the desired quantum numbers, and noiseless time evolution preserves quantum numbers.
Thus, any Ritz vector $\iket{y^{L/R(m)}_k}$ corresponding to a physical state must have a finite overlap with the initial state, i.e.~$\Delta^\mathrm{ZCW} > 0$. By the same logic, a state with zero overlap cannot have the correct quantum numbers.
Such states are necessarily noise artifacts arising from artificial expansion of Krylov space by noise, do not admit a physical interpretation, and should be disregarded.\footnote{In some cases, such as sectors where finite-volume analogs of multi-particle scattering states are present at low energies, there can be physical states that are approximately orthogonal due e.g.~to approximate symmetries that emerge at low energies. Noise contributions then implicitly include contributions from any approximately orthogonal physical states whose overlaps are too small to be resolved from zero at a given level of statistics. }

In practice, the overlaps and thus $\DZCW$ can only be measured noisily, and the formal spuriosity criterion $\bigl< \psi | y^{R(m)}_k \bigr> = 0$ must be replaced by the practical ZCW test, $\DZCW < \eZCW$.
However, physical intuition about overlaps can be used to guide the selection of $\eZCW$.
For physical states, $\DZCW$ corresponds to the absolute, normalized squared overlap factor, such that $0 \leq \DZCW \leq 1$.
A simple choice of $\eZCW = 10^{-2}$ or $10^{-3}$ is sufficient for analyses of typical datasets where available precision allows extraction of states with $\DZCW \sim O(0.1)$.
Estimates for any physical state with overlaps above the cut will converge in the limit of infinite statistics, while physical states below the cut will not be extracted.
However, small-overlap states are difficult to resolve from noise; studies targeting such states 
will necessarily require higher statistical precision, allowing smaller values of $\eZCW$ to be chosen.

Although simple and practical, a fixed choice of $\eZCW$ recovers only a subset of the full spectrum of Ritz values in the infinite-statistics limit.
It is thus desirable to define a procedure with a better approach to this limit.
The same line of physical reasoning provides a simple and natural definition.
First, note that results with small $m$ are least impacted by noise and for at least some range of $m$ will have all Ritz vectors in the Hermitian subspace and all Ritz values satisfying $0<\lambda_k^{(m)}<1$.
Define $m_{\mathcal{H}}$ to be the largest iteration where 
this holds,
that is
\begin{equation}\label{eq:mH}
    m_{\mathcal{H}} \equiv  \max \{ m\ |\ [  k\in \mathcal{H}\ \& \ 0<\lambda_k^{(m)}<1 ] \  \forall k \}.
\end{equation}
The minimum value of $\Delta_k^{\mathrm{ZCW}(m_{\mathcal{H}})}$ provides a natural cutoff for the smallness of overlap factors that can be unambiguously associated with physical states.
For subsequent iterations $m > m_{\mathcal{H}}$, Ritz values can be safely labeled as spurious if the overlap factors are ``much smaller'' than this minimum value.
This provides the cutoff $\eZCW$ needed to define the ZCW test can then be defined as
\begin{equation}\label{eq:eZCW}
    \eZCW \equiv \frac{1}{F^{\mathrm{ZCW}}} \min_{k\in \{1,\ldots,m_{\mathcal{H}}\} } \Delta_k^{\mathrm{ZCW}(m_{\mathcal{H}})}.
\end{equation}
For sufficiently noisy data, spurious states can arise as early as $m=2$ leading to $m_{\mathcal{H}} = 1$; 
 this results in $m_{\mathcal{H}} = 1$ where by definition $\Delta_k^{\mathrm{ZCW}(1)} = 1$.
For the $m_{\mathcal{H}} = 1$ case in particular it is preferable to impose a simple cut $\eZCW = 10^{-2}$ or $10^{-3}$ since the data does not provide additional information on the sizes of physical overlaps.

The only hyperparameter that enters \cref{eq:eZCW} is the factor $F^{\mathrm{ZCW}}$, which simply quantifies the ``much'' in ``much smaller.''
All numerical results below apply this procedure with a default value of $F^{\mathrm{ZCW}} = 10$; we find results to be broadly insensitive to other choices in the range $F^{\mathrm{ZCW}} \in [2,20]$.
When applying bootstrap resampling, $m_{\mathcal{H}}$ and $\eZCW$ can be computed independently for each bootstrap sample to provide idiomatic bootstrap estimators.

One important edge case that must be considered separately is the appearance of exponentially growing modes associated with thermal effects on Euclidean lattices with finite temporal extent.
Exponentially growing modes have effective overlaps suppressed by $O(e^{-\beta E / 2})$, and for practical analyses where this is much smaller than $\eZCW$ their identification requires a separate thermal ZCW test as discussed in Sec.~\ref{sec:thermal} below.

\subsubsection{Block ZCW}

The physical perspective of ZCW as filtering on the overlaps between Ritz vectors and the initial state generalizes immediately to block Lanczos.
Specifically, Eq.~\eqref{eq:DZCW_scalar} generalizes naturally to define a separate test quantity for each initial state $\ket{v_{1a}}$,
\begin{equation}\begin{aligned}\label{eq:DZCW_def}
    \Delta^{\mathrm{ZCW}(m)}_{ka}
    &\equiv \left| \braket{v^L_{1a}}{y^{R(m)}_k} \braket{y^{L(m)}_k}{v^R_{1a}} \right| \\
    &= \left| \omega^{(m)}_{1ak} (\omega^{-1})^{(m)}_{k1a} \right| \,.
\end{aligned}\end{equation}
The same logic as above---states that have vanishingly small overlap with the initial state cannot arise from physical transfer matrix evolution and must be due to noise---applies in the block case to Ritz vectors that have vanishingly small overlap with \emph{all} interpolating operators present.
This suggests a particular construction of the ``block ZCW test''
\begin{equation}\begin{aligned}
    \eZCW > \DZCW &\equiv \left| \sum_a \omega^{(m)}_{1ak} (\omega^{-1})^{(m)}_{k1a} \right| 
    \\
    & = \left|  \sum_{ab} Z^{R(m)*}_{ka} [\mathbf{C}(0)^{-1}]_{ab} Z^{L(m)}_{k b} \right| ~,
    \label{eq:block-ZCW-test}
\end{aligned}\end{equation}
where the sum over $a$ encodes that all overlaps must be small and provides a choice independent of block oblique convention, as made clear by the re-expression in terms of convention-independent quantities in the second equality.
The exact same prescription for choosing a physically motivated $\eZCW$ using Eq.~\eqref{eq:mH}-\eqref{eq:eZCW} can be used for the block case. 
No special treatment is required for oscillating modes in applications to staggered fermions; see Appendix A of Ref.~\cite{Wagman:2024rid}.

\subsubsection{Block CW}

\begin{figure}
    \includegraphics[width=\linewidth]{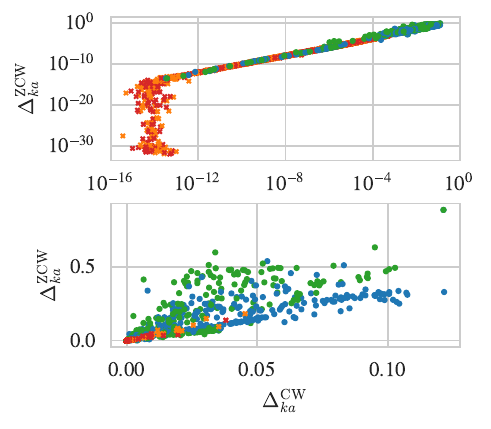}
    \caption{
     Comparison of $\Delta^{\mathrm{ZCW}(m)}_{ka}$ and $\Delta^{\mathrm{CW}(m)}_{ka}$ for block Lanczos results, computed on the central-value correlator of the $2 \times 2$ nucleon example.
     Each marker corresponds to a pair of values for some $m,k,a$, for all smearings $a \in \{T,W\}$ and $k$ for all $m \in [2,48]$.
     No graphical distinction is made between different $m$ and $k$.
     Blue and orange markers indicate results for $a=T$, while green and red are for $a=W$.
     Blue and green dots (orange and red Xs) are states inside (outside) the Hermitian subspace.
     The top and bottom panels show the same data with log and linear axes.
    }
    \label{fig:nuc-CW-vs-ZCW-a}
\end{figure}

\begin{figure}
    \includegraphics[width=\linewidth]{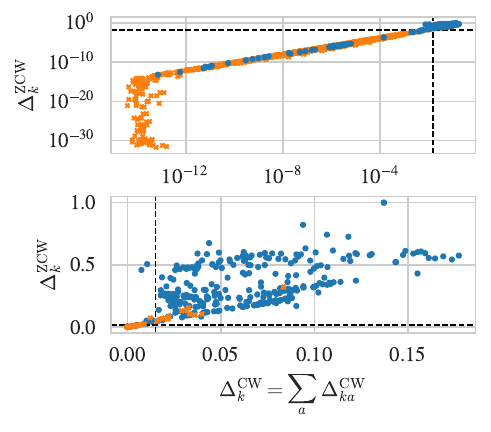}
    \caption{
     Comparison of $\DZCW$ and $\DCW$ for block Lanczos results.
     Details are as in \cref{fig:nuc-vy-vs-dCW}, but for the block generalizations \cref{eq:block-ZCW-test} and \cref{eq:block-CW-sum-a}.
     The data is the same as shown in \cref{fig:nuc-CW-vs-ZCW-a} but summed over $a$.
     The vertical line indicates the cut $\eCW \approx 1.5 \times 10^{-2}$ selected using the bootstrap histogram procedure of Ref.~\cite{Wagman:2024rid}, while the horizontal line indicates the cut $\eZCW \approx 1.7 \times 10^{-2}$ selected by \cref{eq:eZCW}.
    }
    \label{fig:nuc-CW-vs-ZCW-sum}
\end{figure}

The eigenvalue-eigenvector identity can be applied to $\Tm$ exactly as in the scalar case to give
\begin{equation}\label{eq:block-eval-evec}
    \prod_{\widetilde{l}} \left[ \lambda^{(m)}_k - \widetilde{\lambda}^{a(m)}_{\widetilde{l}} \right]
    = \omega^{(m)}_{1ak} (\omega^{-1})^{(m)}_{k1a} \prod_{l \neq k} \left[ \lambda^{(m)}_k - \lambda^{(m)}_l \right] ~,
\end{equation}
where $\widetilde{\lambda}^{a(m)}_{\widetilde{l}}$ denotes the $\widetilde{l}$-th eigenvalues of the matrix obtained by removing the $a$-th row and column of $\Tm$.
Similar reasoning as in the scalar case gives the $a$-dependent relation
\begin{equation}
    \Delta^{\mathrm{CW}(m)}_{ka} = \Delta^{\mathrm{ZCW}(m)}_{ka} 
    \left( \frac{\prod_{l \neq k} \left| \lambda^{(m)}_k - \lambda^{(m)}_l \right|}{\prod_{\widetilde{l} \neq \widetilde{c}} \left| \lambda^{(m)}_k - \widetilde{\lambda}^{a(m)}_{\widetilde{l}} \right|} \right),
\end{equation}
with $\widetilde{c} = \text{argmin}_{\widetilde{l}} \left| \lambda^{(m)}_k - \widetilde{\lambda}^{a(m)}_{\widetilde{l}} \right|$ and
\begin{equation}
    \Delta^{\mathrm{CW}(m)}_{ka} \equiv \min_{\widetilde{l}} \left| \lambda^{(m)}_k - \widetilde{\lambda}^{a(m)}_{\widetilde{l}} \right|~.
\end{equation}
The same reasoning leading to \cref{eq:block-ZCW-test} motivates considering
\begin{equation}\label{eq:block-CW-sum-a}
    \DCW \equiv \sum_a \min_l \left| \lambda^{(m)}_k - \widetilde{\lambda}^{a(m)}_l \right|,
\end{equation}
i.e.~the sum of the minimum distances between a Ritz value and the eigenvalues obtained after knocking out the first $r$ rows and columns.
This provides a quantity that vanishes if and only if $\DZCW$ vanishes and so provides a physically motivated starting point for a ``block CW'' test $\DCW < \eCW$ based on the sum of minimum knockout distances.
Numerical tests on the $2 \times 2$ nucleon correlator matrix shown in \cref{fig:nuc-CW-vs-ZCW-a,fig:nuc-CW-vs-ZCW-sum} illustrate clear proportionality between $\Delta^{\mathrm{CW}(m)}_{ka}$ and $\Delta^{\mathrm{ZCW}(m)}_{ka}$, as well the same quantities \cref{eq:block-ZCW-test,eq:block-CW-sum-a} summed over $a$.
Unlike the scalar case, proportionality breaks down for Ritz values where both are $O(1)$.

Several other plausible extensions of the CW test could be imagined, for example based on knocking out the first block of rows/columns collectively.
Ones we have explored tend to provide roughly similar levels of spurious eigenvalue filtering as the block ZCW / block CW tests defined above.
Unsurprisingly, they show slightly worse correspondence with $\DZCW$ than this definition.

At a practical level, the block ZCW test provides almost identical spurious eigenvalue filtering as the block CW test as seen in \cref{fig:nuc-CW-vs-ZCW-sum}.
The block ZCW test is computationally more efficient, since it only requires calculation of $\omega^{(m)}_{1ak} (\omega^{-1})^{(m)}_{k1a}$, which must already be computed to evaluate residual bounds, overlap factors, and/or matrix elements.
In contrast, the block CW test requires $r$ additional eigensolves at every iteration.
Eigensolves of $\Tm$ are one of the dominant computational costs of Lanczos analyses, and so the need to perform $r+1$ times as many amounts to a significant overall increase.

It is interesting to note that, as visible in both the scalar and block cases in \cref{fig:nuc-vy-vs-dCW,fig:nuc-CW-vs-ZCW-a,fig:nuc-CW-vs-ZCW-sum}, states in and out of the Hermitian subspace are well-separated such that---although there is no reason not to use the Hermitian subspace test, which introduces no ambiguities or additional hyperparameters---in principle, an appropriate choice of $\eCW$ or $\eZCW$ is a sufficient state filtering scheme by itself.

\subsubsection{Results of filtering}

\begin{figure}
    \includegraphics[width=\linewidth]{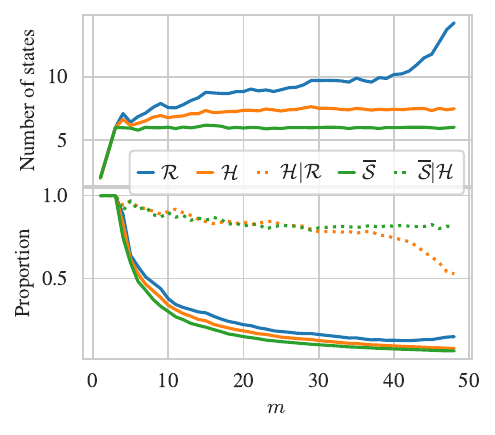}
    \caption{
      Census of states surviving the various stages of spuriosity filtering, computed as a mean over $N_{\rm boot}=200$ (outer) bootstraps on the $2 \times 2$ nucleon correlator matrix example.
        In the legend, $\mathcal{R}$ denotes states with numerically real Ritz values per \cref{eq:real-at-finite-prec},
        $\mathcal{H}$ denotes states in the Hermitian subspace,
        and $\overline{\mathcal{S}}$ denotes non-spurious states which survive both Hermitian subspace and block ZCW filtering.
        $\mathcal{H} | \mathcal{R}$ indicates the fraction of states which are in the Hermitian subspace, given their Ritz values are real.
        $\overline{\mathcal{S}} | \mathcal{H}$ indicates the fraction of states which are non-spurious, given they are in the Hermitian subspace.
    }
    \label{fig:census}
\end{figure}

Our final prescription for spurious state identification used below is application of the Hermitian subspace test, as described in \cref{sec:Hermitian}, followed by application of the ZCW test, \cref{eq:block-ZCW-test}.
\Cref{fig:census} summarizes the results.
After the initial few iterations, there are a gradually increasing number of real Ritz values, but typically $\sim 8-10$.
Roughly $80\%$ of real Ritz values correspond to states in the Hermitian subspace, with a decreasing fraction at late $m$, and a similar fraction of Hermitian states survive additional ZCW filtering.
The average number of states in the Hermitian subspace is nearly constant at $\approx 7$ after $m \sim 5-10$.
More sharply, the total number of states identified as physical is nearly constant at $\approx 6$ after $m=4$.
However, these figures represent averages, and will vary from bootstrap to bootstrap.

\subsection{Assessing state identification}

Providing estimates and uncertainties for energy levels requires state identification, i.e., labeling the Ritz values within each bootstrap as the ground state $n=0$, first excited state $n=1$, etc.
In this work, we restrict to the maximally simple ``filter and sort'' method where, after filtering using Hermitian subspace and ZCW tests, we associate the largest surviving eigenvalue with the ground state, the second-largest with the first excited state, etc.
However, the cautious analysis below finds some ambiguities in this scheme, likely due to the low statistics used for the demonstration.
State identification is significantly more challenging for states beyond the lowest-lying two, i.e.~those with $n \geq r$, and future work on improved state identification methods may be useful for improving the precision of excited-state energy determinations.

\begin{figure}[t]
    \includegraphics[width=\linewidth]{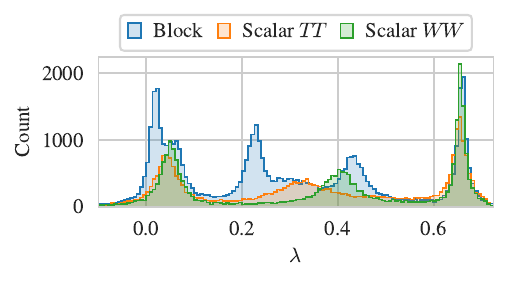}
    \caption{
        For the nucleon correlator matrix, distributions of Ritz values for all $m \geq 6$ over $N_{\rm boot}=200$ (outer) bootstraps as extracted by block Lanczos or scalar Lanczos applied to either diagonal correlator, after removing spurious eigenvalues with Hermitian subspace and ZCW tests.
    }
    \label{fig:nuc-eigs-block-vs-scalar}
\end{figure}

However, to begin, we can demonstrate the advantage of block over scalar Lanczos without the need to discuss state identification.
\Cref{fig:nuc-eigs-block-vs-scalar} shows histograms of the overall distributions of Ritz values, comparing those extracted by block Lanczos with those found by scalar Lanczos applied to the diagonal correlators $C_{TT}$ or $C_{WW}$.
The histograms are restricted to Ritz values that survive spurious state filtering and to $m \geq 6$ to remove pre-convergence behavior, but then all remaining values for all $m,k$ over all bootstraps are binned together.
All three procedures produce similar-shaped peaks for the ground state at right, as well as a mass of Ritz values near $\lambda = 0$ corresponding to very high-energy states.

\begin{figure}
    \includegraphics[width=\linewidth]{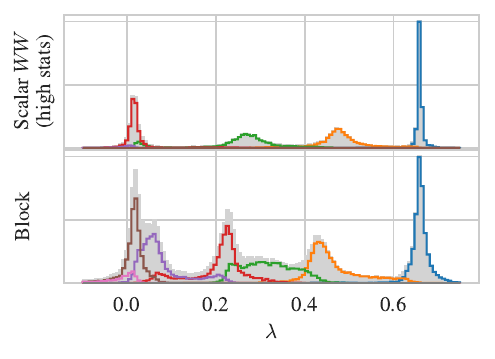}
    \caption{
       Distributions of Ritz values for all $m \geq 6$ over $N_{\rm boot}$ (outer) bootstraps after removing spurious eigenvalues with Hermitian subspace and ZCW tests, for scalar Lanczos on the high-statistics dataset of Ref.~\cite{Hackett:2024xnx} (top) and block Lanczos on the $2 \times 2$ nucleon correlator example (bottom).
        In each panel, the gray filled histogram indicates all Ritz values, while colored histograms indicate associations of Ritz values with different states, with state identification made simply by sorting Ritz values after filtering.
        The colored histograms sum to the gray ones. 
    }
    \label{fig:nuc-eigs-states}
\end{figure}

The important difference lies in the intermediate part of the spectrum: block Lanczos yields two clearly distinct peaks, while each scalar Lanczos analysis yields only one.
Qualitatively, this suggests exactly the improvement in ability to distinguish nearby states expected when analyzing correlator matrices instead of scalar ones.
Complicating this interpretation is that, as found in Ref.~\cite{Hackett:2024xnx}, when scalar Lanczos is applied to a higher-statistics measurement of $C_{WW}$, it similarly finds two states in the intermediate regime (as reproduced in \cref{fig:nuc-eigs-states}).
This demonstrates that scalar Lanczos with sufficiently large statistics can sometimes achieve similar energy resolution as block Lanczos.

\Cref{fig:nuc-eigs-states} assesses the performance of filter-and-sort state labeling by decomposing the total Ritz value histograms into separate histograms for each state label.
The figure shows results for both the block Lanczos analysis, as well as a scalar Lanczos analysis of the high-statistics $C_{WW}$ data for comparison.
Interestingly, despite the much lower statistics, the block Lanczos analysis appears to resolve an intermediate state not seen by the high-statistics scalar analysis.
The peak for this state is wide and thus not apparent in the total histograms of \cref{fig:nuc-eigs-block-vs-scalar}.

The reduced statistics in the block example cause difficulties with state identification.
In the high-statistics setting, each of the four peaks is clearly associated with a single state, up to some minor confusion between the second and third excited state (i.e., the small green sub-peak under the mostly-red leftmost peak).
In the block example, the validity of the labeling is less obvious.
In several instances, the labeling for a single state includes density from multiple peaks (e.g.~red, especially). 
While unambiguous peak-state association in such a plot is not necessary, it does raise suspicion.

Comparison with the noiseless demonstration of \cref{sec:noiseless} provides a mechanistic picture of the issue: the spectrum is not extracted in monotonic order, and rather states with large overlaps tend to be resolved earlier.
The apparent mislabelings can be explained as a signal-dependent version of this same effect, which gives rise to dislocations in the indexing of the spectrum due to intermediate states being resolved in some bootstraps but not others.
These labeling ambiguities motivate our use of outlier-robust estimators based on bootstrap medians  discussed in the next subsection.

Before proceeding, we caution against overinterpreting histograms like those of \cref{fig:nuc-eigs-block-vs-scalar,fig:nuc-eigs-states}.
These are intended only to give a simple, global view of the distribution of Ritz values extracted by Lanczos analyses.
They do not admit a clean statistical interpretation in terms of uncertainties or confidence about the location of true eigenvalues.
However, exploring state-identification-agnostic constructions which may admit such interpretations could be an interesting topic for future work.

\subsection{Bootstrap median with nested bootstrap confidence intervals}
\label{sec:median}

The state-identification challenges above underscore the fact that noise affects the spectra of Ritz values obtained from Lanczos in non-trivial ways, even after 
spurious state filtering.
Moreover, filtering is imperfect in practice, introducing additional difficulties.
The combination of bootstrap resampling and outlier-robust estimators provides a natural means to work past these difficulties, giving reliable results as long as the imperfect filtering and labeling are not \emph{too} imperfect.
In particular, we find that the median of Ritz values over bootstraps provides a more reliable energy estimator than the sample mean.
This can be intuitively understood from the fact that state misidentification is a finite-statistics artifact that can lead to large effects in comparison with statistical fluctuations of a particular Ritz value;  outlier-robust estimators that remove these large effects can be expected to provide more accurate, as well as more precise, estimators at finite statistics.

The statistical uncertainties of the bootstrap median may be computed by bootstrap resampling and then repeating the entire analysis, including a second (nested) step of bootstrap resampling and then taking the median.
To calculate uncertainties, outer and inner bootstrap ensembles are constructed as detailed in \cref{sec:setup}.
To define a Ritz-value estimator for a given outer ensemble, compute (filtered and labeled) Ritz values for each of the $N_\mathrm{boot}$ inner ensembles resampled from the outer one, then take the median over the inner estimates.
Uncertainties are computed simply from the variance over outer bootstraps: for any quantity $X$ with bootstrap samples $X^{(b)}$ the uncertainty $\delta X$ is defined as~\cite{Efron:1982,DiCiccio:1996,Davison:1997}
\begin{equation}\label{eq:var}
    \begin{split}
    \delta X^2 &= \frac{1}{N_\mathrm{boot}} \sum_b (X^{(b)})^2 - \left( \frac{1}{N_\mathrm{boot}} \sum_b X^{(b)} \right)^2 ~ .
    \end{split}
\end{equation}
This formula provides an asymptotically unbiased estimator for the variance of any random variable; in particular we apply it to the case where the random variable $X$ is 
\begin{equation}\label{eq:bootmed}
E_n^{(m)} \equiv -\ln \text{median}_{b \in \mathbb{S}_n^{(m)}} \lambda_n^{(b,m)}.
\end{equation}
where $\mathbb{S}_n^{(m)}$ is the subset of bootstrap draws $b$ that contain at least $n$ non-spurious Ritz values for iteration $m$---i.e., the median over draws where an estimate is available.
The bootstrap samples $X^{(b)}$ are then
\begin{equation}
E_n^{(b,m)} \equiv -\ln \text{median}_{b' \in \mathbb{S}_n^{(b,m)}} \lambda_n^{(b,b',m)},
\end{equation}
where $\mathbb{S}_n^{(b,m)}$ is the subset of 
inner bootstrap draws $b'$ from the outer ensemble $b$
that contain at least $n$ non-spurious Ritz values for iteration $m$,
from which the bootstrap variance $\delta E_n^{(m)}$ can be computed using Eq.~\eqref{eq:var}.
Note that in noisy data applications below, $E_n^{(m)}$ will always refer to the bootstrap median estimator, \cref{eq:bootmed}, rather than the sample 
mean estimator $-\ln \lambda_n^{(m)}$ unless explicitly specified.

To define the overall number of non-spurious states $N_{\rm max}^{(m)}$ below, we first define $N_{\rm max}^{(b,m)}$ for each outer bootstrap as the maximum $n$ for which at least $n$ non-spurious Ritz values exist in 95\% of the $\{\lambda_n^{(b,b',m)}\}$, or in other words the largest $n$ for which $\mathbb{S}_n^{(b,m)}$ contains at least $0.95 N_{\rm boot}$ elements.\footnote{The choice of 95\% is arbitrary; any choice $\gtrsim 50\%$ will lead to identical results in this case. Note that demanding 100\% inclusion would lead to pathological behavior in the $N_{\rm boot} \rightarrow \infty$ limit since outliers due to (spurious or non-spurious) state misidentification occur with non-zero probability.}
Applying Eq.~\eqref{eq:var} requires that $E_n^{(b,m)}$ is defined for each outer bootstrap,\footnote{The pathological behavior of the $N_{\rm boot} \rightarrow \infty$ limit for inner bootstraps with a 100\% acceptance cut does not apply here because the inner bootstrap median is already an outlier-robust quantity.}
and therefore $N_{\rm max}^{(m)} = \min N_{\rm max}^{(b,m)}$ is the maximum number of states for which consistent error quantification is possible.

Both $E_n^{(m)}$ and $\delta E_n^{(m)}$ are asymptotically unbiased but include finite-statistics bias suppressed by $O(1/N)$.
Bias correction factors that reduce this bias to $O(1/N^2)$ are well known~\cite{Young:2014}.
However, as $N\rightarrow \infty$ the correlation function for $t \leq 2m-1$ will approach a convex form and (assuming numerical precision is also taken to infinity appropriately) there will be no spurious eigenvalues at iteration $m$.
Outlier-robust estimators will eventually become unimportant in the asymptotic regime where $N\rightarrow \infty$ at fixed $m$.
Iterations where the variance of bootstrap-median estimators is significantly smaller than that of sample-mean estimators are therefore in a qualitatively different regime than the asymptotic regime, and it is not obvious that $1/N$ bias correction will produce a more accurate estimator.
Since there are additional $1/\sqrt{N_{\rm boot}}$ statistical uncertainties introduced by the bias correction to $E_n^{(m)}$ and taking $N_{\rm boot}$ arbitrarily large becomes computationally expensive for nested median estimators, we expect $E_n^{(m)}$ to be a preferable estimator to the bias-corrected analog $2E_n^{(m)} - \frac{1}{N_{\rm boot}} \sum_b E_n^{(b,m)}$.

\begin{figure}
    \includegraphics[width=\linewidth]{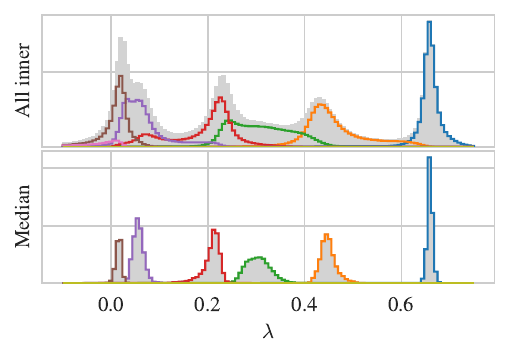}
    \caption{
      Distributions of Ritz values estimated over all $N_\mathrm{boot}^2$ inner bootstraps (top), and of the $N_{\rm boot}$ outer bootstrap samples of the median estimator $E_n^{(b,m)}$ constructed from them (bottom), as extracted by block Lanczos applied to the $2 \times 2$ nucleon correlator matrix, for all $m \geq 6$ after removing spurious eigenvalues with Hermitian subspace and ZCW tests.
      The colored histograms sum to the gray ones. 
    }
    \label{fig:nuc-eigs-pre-post-med}
\end{figure}

\begin{figure*}
    \includegraphics[width=\linewidth]{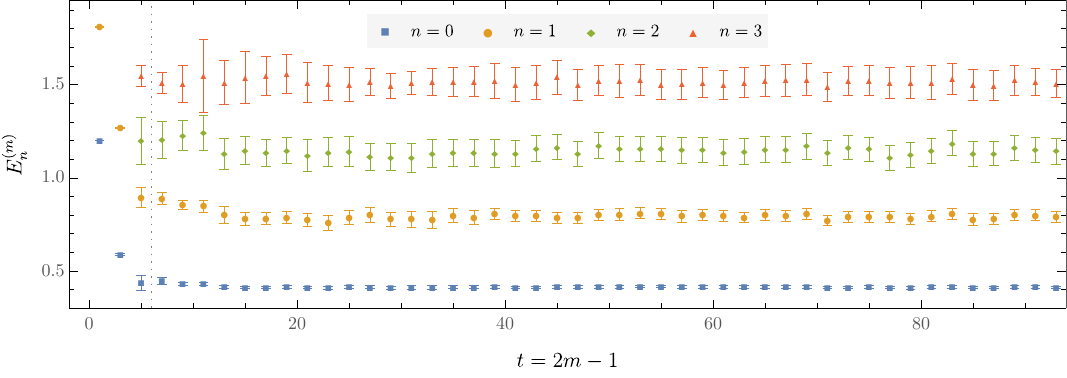}
    \caption{
        Lanczos energy estimators for the lowest-lying four states in the nucleon spectrum  extracted using block Lanczos.
        Errors are computed for $E_n^{(m)}$ using bootstrap-median estimators as described in the text.
        Dotted vertical lines are placed just after $m_{\mathcal{H}}$ as defined in \cref{eq:mH}, the last iteration when all Ritz values/vectors satisfy all the physical constraints that arise in the infinite-statistics limit, which is used to determine $\eZCW$ for the ZCW test.
        For $m=1$ there are only two Ritz values; for $m=2$ the higher pair of energies are much above the plot range.
    }
    \label{fig:block-nuc-spectrum}
\end{figure*}

\Cref{fig:nuc-eigs-pre-post-med} illustrates how this scheme regulates the issues with state identification noted in the previous subsection.
Note that the median is only defined \emph{after} state labeling, and thus the median estimator necessarily depends on the precise choice of state identification scheme.
The distributions of Ritz values for states $n \in \{1,2,3\}$ are clearly more well-separated in the bootstrap median case and show better resolved peaks.
These peaks qualitatively resemble log-normal distributions that would correspond to Gaussian $E_n^{(b,m)}$.
As detailed in \cref{app:ci},  bootstrap median Lanczos energy estimators are much closer to Gaussian distributed than sample-mean versions of the estimators; however, $E_0^{(b,m)}$ and $E_3^{(b,m)}$ still have observable departures from Gaussianity as quantified by the Kolmogorov-Smirnov and Shapiro-Wilk tests.

\Cref{eq:var} provides an asymptotically unbiased estimate of the variance of any random variable; however, $s \delta E_n^{(m)}$ can only be interpreted as an $s \sigma$ confidence interval under the stronger assumption that the $E_n^{(b,m)}$ are Gaussian distributed.
Given these observed departures from Gaussianity, it is important to test how accurately $s \delta E_n^{(m)}$ approximates $s\sigma$ confidence intervals.
These Gaussian approximations are compared with direct confidence-interval calculations using empirical bootstrap confidence intervals in \cref{app:ci}.
Despite the departures from Gaussianity visible in Shapiro-Wilk test results for $E_0$, results for $n\in \{0,1\}$ at the largest $m$ show that Gaussian confidence interval estimates $s \delta E_n^{(m)}$ are within 10\% of the empirical bootstrap confidence intervals for $s \sigma$ with $s \in \{1,2,3\}$.
For $n\in \{2,3\}$ the differences are somewhat larger but still less than 20\% in all cases.
This quantifies the systematic uncertainties associated with identifying $s \sigma \approx s \delta E_n^{(m)}$.

\begin{figure}
    \includegraphics[width=\linewidth]{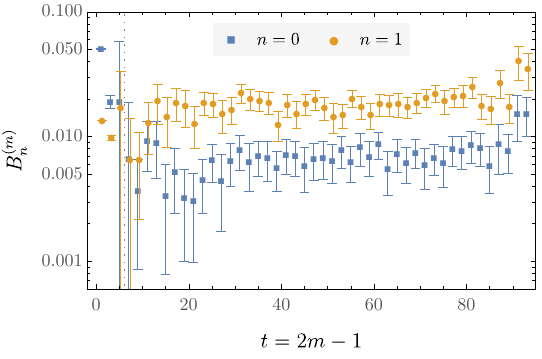} \\
    \includegraphics[width=\linewidth]{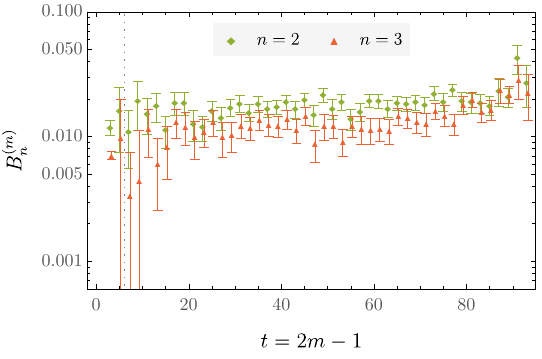}
    \caption{
        Residual bounds corresponding to the energy estimators shown in \cref{fig:block-nuc-spectrum}. 
        Note the uncertainties are defined symmetrically but appear asymmetric due to the logarithmic $y$ axes.
    }
    \label{fig:block-nuc-resdiauls}
\end{figure}

\Cref{fig:block-nuc-spectrum} shows the spectrum as extracted by the block analysis using the Hermitian subspace test, ZCW test, and bootstrap-median estimators.
Energies $E_n^{(m)}$ of the cleanly identified Ritz values, $n= \{0,\ldots,3\}$, converge after 5-10 iterations and then fluctuate about stable values with roughly constant uncertainties.
The corresponding residual bounds $B_n^{(m)}$ are shown in \Cref{fig:block-nuc-resdiauls}.

\begin{figure}
    \includegraphics[width=\linewidth]{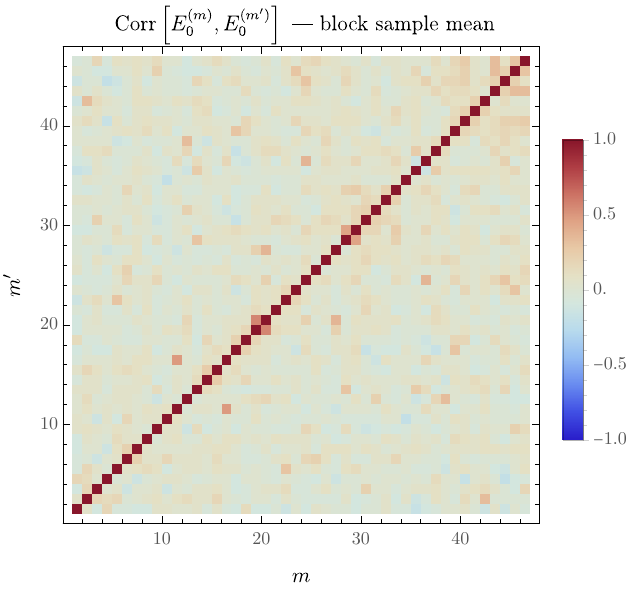}\\
    \includegraphics[width=\linewidth]{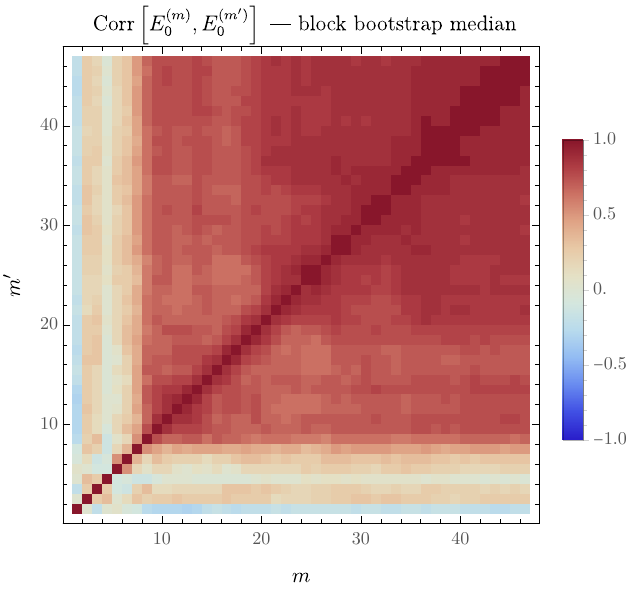}
    \caption{
        Correlations between Lanczos energy estimators at different iteration counts for determinations using sample-mean Ritz values (top) and bootstrap-median Ritz values with idiomatic bootstrap uncertainties (bottom). The bootstrap-median case corresponds to the (Pearson) correlation matrix of the ``outer bootstrap'' samples of the ``inner bootstrap'' medians. 
    }
    \label{fig:block-nuc-correlations}
\end{figure}

Using bootstrap-median estimators allows simply taking the estimates at the last iteration where residual bounds can be computed, $m_\mathrm{max} = N_t/2-1$, as the final output of the analysis.
As observed in the initial application of bootstrap-median estimators to scalar Lanczos results in Ref.~\cite{Wagman:2024rid}, correlations between Ritz values at large $m$ emerge for bootstrap-median estimators that are obscured by additional uncorrelated noise present in sample-mean estimators.\footnote{These differences between sample mean and nested median estimators only appear after 10s of iterations, when many spurious eigenvalues are present. In contrast, the uncertainties of the effective mass are effectively identical when calculated using bootstrap median and sample-mean estimators.}
This holds similarly for block Lanczos analyses; as shown in \cref{fig:block-nuc-correlations}, large correlations ($\gtrsim 0.7$) between energy estimators at large $m$ arise for block Lanczos results using bootstrap-median estimators but not for results using sample-mean estimators.
These correlations signal that most of the useful statistical information found in block Lanczos analysis of $\mathbf{C}(t)$ considered over all $m$ is captured by energy estimators at any single large value of $m$.
There is therefore little gain in precision from e.g.~fitting the large-$m$ points to a constant, which moreover comes at the cost of rigor.
Such fitting processes require estimating the covariance matrix of many strongly correlated data points, which requires statistical techniques to regulate numerical ill-posedness that often introduce additional subjective analysis choices and/or hyperparameters.

\begin{figure}
    \includegraphics[width=\linewidth]{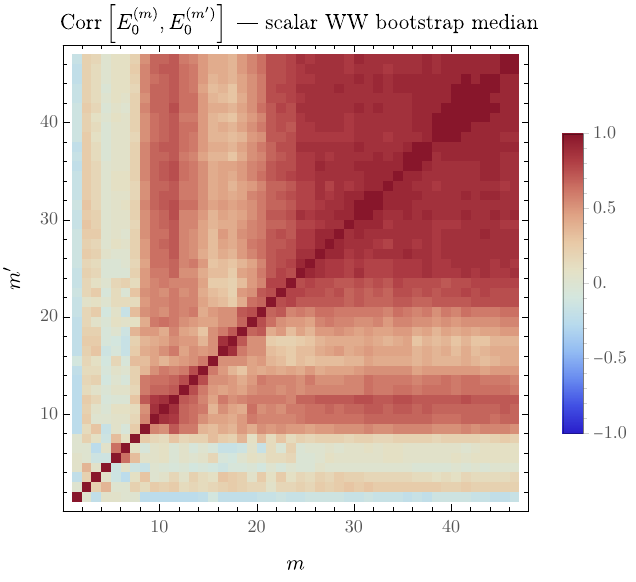}\\
    \includegraphics[width=\linewidth]{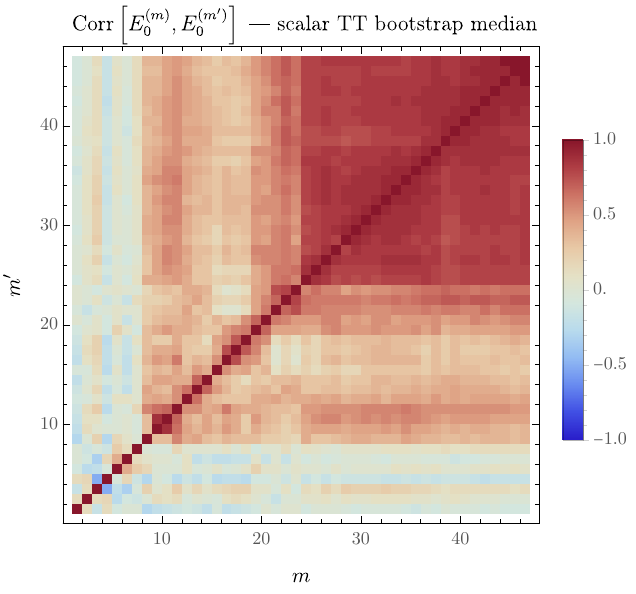}
    \caption{
        Correlations between scalar Lanczos energy estimators using bootstrap-median Ritz values estimators; details are as in \cref{fig:block-nuc-correlations}.
    }
    \label{fig:scalar-nuc-correlations}
\end{figure}

It is interesting to note that, as apparent from comparison of \cref{fig:block-nuc-correlations} and \cref{fig:scalar-nuc-correlations}, correlations $\gtrsim 0.7$ appear at smaller $m \gtrsim 8$ in the block case than for scalar Lanczos applied to either diagonal correlation function, which do not stably display similar correlations until $m \gtrsim 20$ for $WW$ and $m \gtrsim 25$ for $TT$.
The faster approach of $E_0^{(m)}$ to values strongly correlated with the large-$m$ asymptotic result could be related to the faster convergence of block than scalar Lanczos for noiseless data.
However, from this example alone it is not clear whether this behavior is generic or what features of the Lanczos algorithm (e.g.~loss of biorthogonality after convergence of particular Ritz vectors~\cite{Paige:1971,Parlett}) give rise to these differences.

\begin{table}[t]
    \centering
    \begin{ruledtabular}
    \begin{tabular}{cllll}
    Lanczos Analysis & $E_0$ & $E_1$ & $E_2$ & $E_3$ \\ \hline \vspace{0.2em}
    Block  & 
        $0.4122(72)$ & $0.792(30)$ & $1.14(7)$ & $1.50(8)$ \\
    Scalar $TT$  & 
        $0.4240(91)$ & $0.937(67)$ & $-$ & $-$\\
    Scalar $WW$  & 
        $0.4217(69)$ & $0.891(38)$ & $-$ & $-$\\
    High-stats $WW$ &
        $0.4170(22)$ & $0.735(25)$ & $1.31(7)$ & $-$ \\
    High-stats $WW$~\cite{Hackett:2024xnx} &
        $0.4175(17)$ & $0.736(34)$ & $1.30(8)$& $-$
    \end{tabular}
    \end{ruledtabular}
    \caption{
        Comparison of $E_0^{(m_{\rm max})}$ and $E_1^{(m_{\rm max})}$ from block and scalar Lanczos on the $2\times 2$ matrix of $\{ \psi_W, \psi_T\}$ operators described in Sec.~\ref{sec:setup}, its diagonals, and the high-statistics superset of this data analyzed in Ref.~\cite{Hackett:2024xnx}. The row labeled ``High-stats $WW$~\cite{Hackett:2024xnx}'' quotes fits results to the estimators without nested bootstrap used in that work.
    }
    \label{tab:E-comps}
\end{table}

\begin{table}[]
    \centering
    \begin{ruledtabular}
    \begin{tabular}{cllll}
    Lanczos analysis & $Z_{0W}$ & $Z_{0T}$ & $Z_{1W}$ & $Z_{1T}$ \\ \hline \vspace{0.2em}
    Block  & 
        $0.366(15)$ & $0.238(11)$ & $0.535(21)$ & $0.437(23)$ \\
    Scalar $TT$  & 
        $-$ & $0.261(15)$ & $-$ & $0.642(58)$ \\
    Scalar $WW$  & 
        $0.392(16)$ & $-$ & $0.650(13)$ & $-$ \\
    High-stats $WW$ &
        $0.375(4)$ & $-$ & $0.486(15)$ & $-$  \\
   High-stats $WW$~\cite{Hackett:2024xnx} &
       $0.376(7)$ & $-$ & $0.482(29)$ & $-$
    \end{tabular}
    \end{ruledtabular}
    \caption{
        The corresponding overlap factors, details as in Table~\ref{tab:E-comps}. Note that Ref.~\cite{Hackett:2024xnx} uses non-unit-normalized interpolating operators whose overlaps differ from those shown here by a normalization factor $\sqrt{C_{WW}(0)} \approx 6.2801 \times 10^{-4}$.
    }
    \label{tab:Zs}
\end{table}

\begin{figure}
    \includegraphics[width=\linewidth]{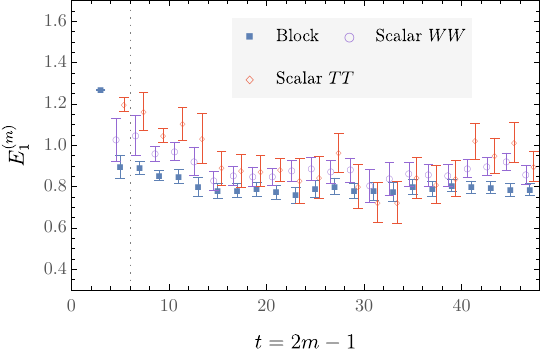}\\
    \includegraphics[width=\linewidth]{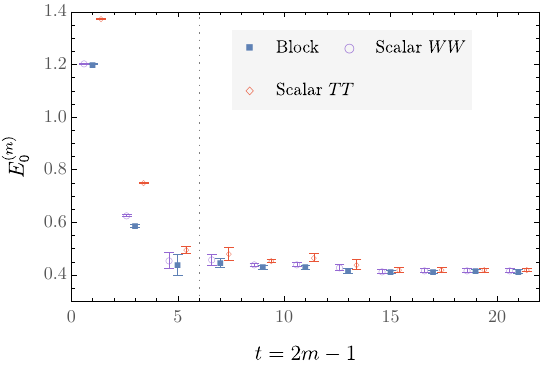}
    \caption{
        Comparisons of block Lanczos energy estimators with scalar Lanczos estimators for the corresponding diagonal correlators. Details are as in \cref{fig:block-nuc-spectrum}.
    }
    \label{fig:nuc-Es-vs-scalar-Es}
\end{figure}

\Cref{fig:nuc-Es-vs-scalar-Es} compares block Lanczos energy estimators with their scalar Lanczos counterparts for both of the diagonal correlators.
The strictly faster converge of block than scalar Lanczos is manifest but not large, particularly for the $\psi_W$
operator with larger ground-state overlap. The block Lanczos and $WW$ scalar Lanczos energies both converge within uncertainties for the ground state at $m=3$ ($t=5$), while the $TT$ correlation function converges within uncertainties at around $m=7$ ($t=13$).
Results and uncertainties for the ground state are very similar between block and both scalar analyses.

More significant differences between block and scalar Lanczos are seen in the determination of $E_1$.
Here, block estimates have significantly smaller uncertainties than either $WW$ or $TT$ scalar estimates.
Block Lanczos converges more quickly and stably to a 2-3$\sigma$ lower value than the $E_1$ determined from either $TT$ or $WW$ scalar Lanczos analyses.

As already discussed, the last iteration where residual bounds can be computed, 
$m_\mathrm{max} = \betaT /2 - 1$,
provides an estimator $E_n^{(m_{\rm max})}$ that uses the maximum amount of information in the correlation function to constrain the spectrum. 
\Cref{tab:E-comps,tab:Zs} compare these estimates with results from the scalar Lanczos analysis of the high-statistics $WW$ correlator, showing statistical consistency between all methods for $E_0$.
It is noteworthy that the $E_1$ obtained from block Lanczos is in 1$\sigma$ agreement with the high-statistics $E_1$ result.

The power of block Lanczos to resolve higher-energy excited states is visible in determinations of $E_2$ and $E_3$, both of which have comparable precision to high-statistics scalar results for $E_2$.
It is noteworthy that the high-statistics $E_2$ results are in between the two block results and consistent with each within 
2$\sigma$ (and the corresponding residual bounds overlap at 1$\sigma$, see Sec.~\ref{sec:gevp} below), but it is difficult to conclude from this data alone which determination is more accurate.

\begin{figure}
    \includegraphics[width=\linewidth]{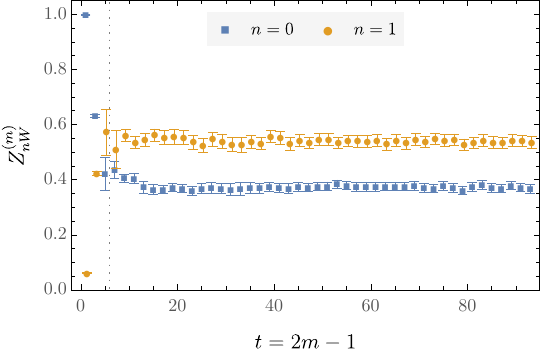}\\
    \includegraphics[width=\linewidth]{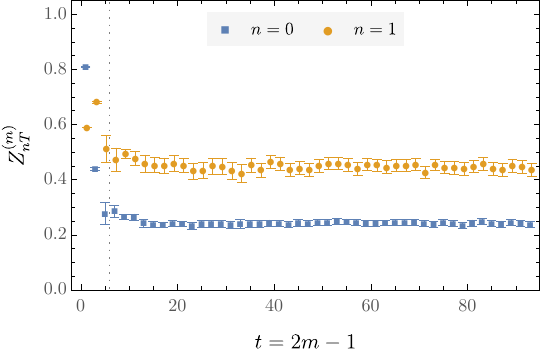}
    \caption{
        For the nucleon correlator matrix example, overlap factors $Z^{(m)}_{na}$ as estimated by $m$ steps of block Lanczos for states $n \in \{0,1\}$ and smearings labeled by $a \in \{W,T\}$.
        Uncertainties are computed using bootstrap-median estimators with idiomatic bootstrap confidence intervals.
    }
    \label{fig:nuc-Zs}
\end{figure}

As presented in \cref{sec:formalism:overlaps}, block Lanczos provides an estimator for the overlap factors.
\Cref{fig:nuc-Zs} shows the resulting estimates, $Z^{(m)}_{na}$,\footnote{For states in the Hermitian subspace, the $R$ and $L$ estimators necessarily become identical, thus no distinction is necessary.} for each interpolator $\psi_W$ and $\psi_T$
and the lowest two states $n\in\{0,1\}$.
As with the spectrum, the overlap estimators converge rapidly and provide clean signals with no signal-to-noise ratio (SNR) degradation with increasing $m$.
Predictably, the signal is cleaner for the ground-state overlaps than for the excited state.

\begin{figure}
    \includegraphics[width=\linewidth]{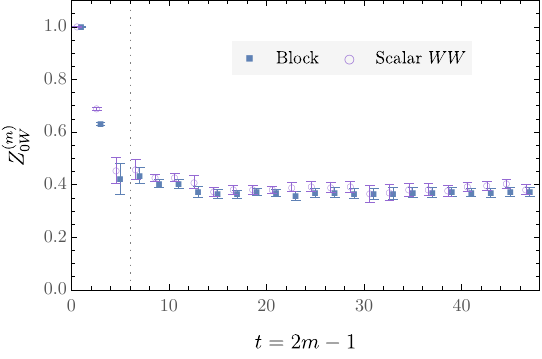}\\
    \includegraphics[width=\linewidth]{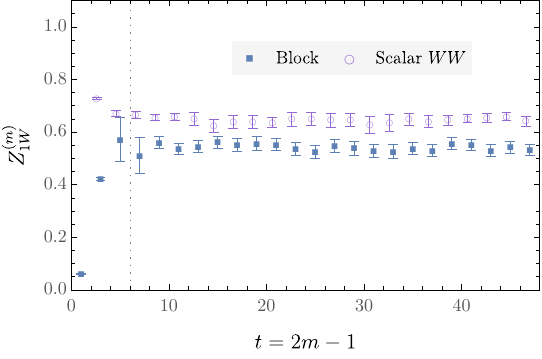}
    \caption{
        Comparisons of the block Lanczos overlap factors for the ``wide'' $\psi_W$ interpolator with those obtained using scalar Lanczos, details as in \cref{fig:nuc-Zs}.
    }
    \label{fig:nuc-Zs-vs-scalar-0}
\end{figure}

\begin{figure}
    \includegraphics[width=\linewidth]{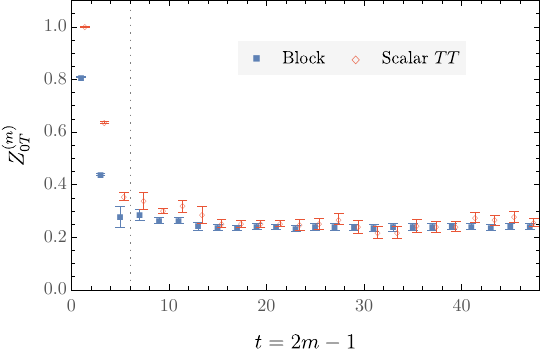}\\
    \includegraphics[width=\linewidth]{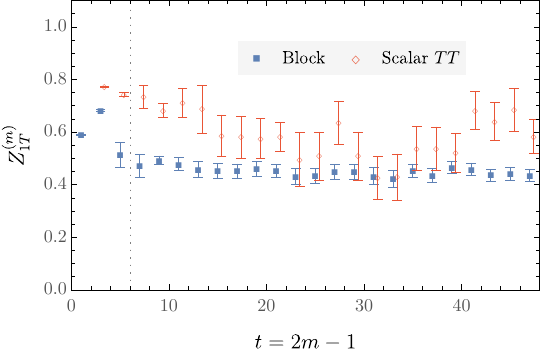}
    \caption{
        Comparisons of the block Lanczos overlap factors for the ``thin'' $\psi_T$ interpolator with those obtained using scalar Lanczos, details as in \cref{fig:nuc-Zs}.
    }
    \label{fig:nuc-Zs-vs-scalar-1}
\end{figure}

As with the spectrum, the block Lanczos extraction of the overlaps demonstrates clear advantages over those made by scalar Lanczos applied to the diagonal correlators $C_{WW}$ and $C_{TT}$, as compared in \cref{fig:nuc-Zs-vs-scalar-0,fig:nuc-Zs-vs-scalar-1}.
While the block and scalar estimators for the ground-state overlaps $Z_{0a}$ are similar in value and uncertainty, the excited-state estimates differ substantially.
Each scalar estimate for $Z_{1a}$ is shifted upwards versus its block equivalent; this is likely associated with the downward shift of the corresponding peaks in the scalar-extracted Ritz value histograms of \cref{fig:nuc-eigs-block-vs-scalar} versus the block-extracted one.
The scalar estimator for $Z_{1T}$ is significantly noisier than the block one, which may again be associated with the relatively broader (orange) peak in \cref{fig:nuc-eigs-block-vs-scalar}.

To summarize, in comparison with scalar Lanczos, block appears most advantageous for excited-state energy and overlap determinations.
These advantages---in terms of both signal-to-noise and control of excited-state effects---are clear but do not qualitatively change the results for the lowest-energy states.
However, it is important to note that the two interpolators $\psi_T$ and $\psi_W$ used in these example are very similar, differing only by the choice of quark smearing radius.
This is dissimilar to the most important use-case for correlator matrix data, wherein qualitatively different operators are used to probe different single- and multi-particle channels.
Comparisons between block and scalar analyses in this setting are a critical topic for future work.
Separately but no less importantly, as explored in \cref{sec:gevp} below, the advantages offered by block Lanczos over GEVP are qualitative and readily apparent even in this example.

\subsection{Thermal states}
\label{sec:thermal}

The ZCW test proposed in \cref{sec:CW} and employed to produce the results above is based on the assumption that states with sufficiently small overlaps with all $\ket{\psi_a}$ must be spurious noise artifacts.
However, as already noted, there is a physically motivated counter-example to this assumption: thermal effects lead to a modification of the spectral representation
\begin{equation}
    C(t) = \sum_{n=0}^\infty \left[ |Z_n|^2 e^{-E_n t} + |Z_n^T|^2 e^{-E_n^T (\betaT - t)} \right], 
\end{equation}
where the $E_n^T$ are the energies associated with states that have time-reversed quantum numbers compared to those of the $\ket{\psi_a}$.
For the (positive-parity) nucleon example, $E_n^T$ corresponds to the energy spectrum of negative-parity nucleon states---i.e.~states with baryon-number 2, isospin 1/2, and $G_1^-$ cubic transformation properties, describing negative-parity spin-1/2 finite-volume states.
Rearranging this expression gives
\begin{equation}
    C(t) = \sum_{n=0}^\infty \left[ |Z_n|^2 e^{-E_n t} + |\mathcal{Z}_n^T|^2 e^{E_n^T t} \right],
\end{equation}
defining ``apparent overlaps'' $\mathcal{Z}_n^T \equiv e^{-E_n^T \betaT /2} Z_n^T$.
Thermal states therefore enter the spectral expansion of $C(t)$ as growing exponentials with very small overlap factors $\mathcal{Z}_n^T \sim O(e^{-\betaT E/2})$.

As discussed in Ref.~\cite{Wagman:2024rid}, applying Lanczos to a thermal field theory correlator $C(t)$ provides approximations to both the transfer-matrix eigenvalues $\lambda_n = e^{-E_n}$ and also the thermal modes $\lambda_n^T = e^{E_n^T}$.
Thermal states can be expected to have physically reasonable overlaps $Z_n^T > \eZCW$ with the time-reversed initial state, but contribute to the correlator with apparent overlaps $\mathcal{Z}_n^T \ll 1$ likely much smaller than $\eZCW$ for realistic lattice sizes and statistical precision.
Nevertheless, Lanczos should in principle be able to compute accurate approximations to these modes from correlation functions at large $t$, where their contributions grow exponentially and statistically precise signals dominated by these modes are visible.

Accessing these exponentially growing modes requires an adapted ``thermal ZCW test'' analogous to Eq.~\eqref{eq:ZCW-test}, which defines states as non-spurious if
\begin{equation}
   1 < [ \lambda_k^{(m)} ]^{\betaT} \DZCW < \eZCW.
\end{equation}
Here, scaling $\DZCW$ by $[ \lambda_k^{(m)} ]^{\betaT} = e^{\betaT E_k^{(m)}}$ is equivalent to applying the ZCW test to the physical overlaps $Z_n^T$ rather than the apparent overlaps $\mathcal{Z}_n^T$.

To study the practical utility of this prescription,
we apply the thermal ZCW test to the Ritz values obtained from analyzing the high-statistics nucleon correlator from Ref.~\cite{Hackett:2024xnx}.
For iterations $m < 44$, no Ritz values pass the thermal ZCW test.
For $m \in [44, 46]$, a single Ritz value passes the thermal ZCW test.
For $m = m_{\rm max} = 47$, two Ritz values pass the ZCW test with energies $0.632(7)$ and $1.128(23)$.%
\footnote{In practice, finding any thermal modes passing both the ZCW test and the Hermitian subspace test with $\eRe = 10^{-8}$ requires performing $\Tm$ eigensolves using high-precision arithematic (in addition to the Lanczos recursions). In contrast, all analyses of non-thermal modes presented herein use double-precision eigensolves and find negligible differences with results obtained using high-precision eigensolves.}
These states, which can be interpreted as part of the spectrum of negative-parity nucleon states, have thermally suppressed overlap factors 
which after rescaling by $e^{E_n \betaT/2}$ predict physical overlaps of $0.277(7)$ and $0.545(4)$.

These energies and overlap factors can be compared with the results for negative-parity states obtained by analyzing the time-reversed correlator $C^\mathrm{rev}(t) \equiv C(N_t-t)$ with $C(N_t) \equiv C(0)$ using the (non-thermal) ZCW test.
As illustrated in \cref{tab:E-comps-thermal,tab:Zs-thermal}, the negative-parity energies and overlap factors extracted from thermal and non-thermal modes are in excellent agreement.
It is noteworthy that almost identical precision is achieved using the time-reversed correlator where the negative-parity modes are visible for small $m$ and the non-time-reversed correlated where negative-parity modes are only visible for the last few iterations.

The thermal ZCW test can also be applied to the Ritz values from the time-reversed (and thus negative-parity) correlator analysis.
In this case thermal modes correspond to positive-parity states, and should therefore be consistent with the precise determinations of the positive-parity spectrum from the non-reversed correlator presented above.
This is indeed the case; as shown in \cref{tab:E-comps-thermal,tab:Zs-thermal} the energy and overlap results determined from thermal effects on negative-parity correlators are in 1$\sigma$ agreement with the standard determination of the positive-parity spectrum.
Again, the uncertainties of the thermal and non-thermal determinations are almost identical.

The nearly identical precision of results from reversed and non-reversed correlators is unsurprising from the point of view that, with $m=m_{\rm max}$, the Krylov spaces appearing in both cases should have almost the same physical content only differing by the inclusion of states for $t$ very close to either end of the correlator.
Since the Ritz values are uniquely defined, optimal eigenvalue approximations in Krylov space~\cite{Parlett}, the same Ritz values ought to be determined from reversed and non-reversed correlators.
The fact that this expectation is borne out in practice even in the presence of noise suggests that the spurious eigenvalue filtering described enough is sufficient to provide a physical subspace of states that obey the expected properties of Krylov-space estimators. 

Note that because the ZCW test is applied as a postprocessing step after all Ritz values are calculated, the non-thermal modes are completely unaffected by whether thermal modes are discarded as spurious or included as a separate category of physical Ritz values via the thermal ZCW test.
This means that the non-thermal version of the ZCW test can be safely applied in finite-temperature LQCD calculations and will not bias results for non-thermal modes; it will simply result in thermal modes being labeled as spurious rather than physical.
All numerical results outside of this subsection apply the non-thermal version of the ZCW test to avoid the need for high-precision eigensolves.

\begin{table}[]
    \centering
    \begin{ruledtabular}
    \begin{tabular}{ccll}
    State Parity & Spurious eval test & $E_0$ & $E_1$ \\ \hline \vspace{0.2em}
    $+$ & ZCW &  $0.4170(22)$ & $0.735(25)$ \\
    $+$ & Thermal ZCW & $0.4186(23)$ & $0.756(32)$ \\\\
    $-$ & ZCW &  $0.638(10)$ & $1.164(37)$ \\
    $-$ & Thermal ZCW & $0.632(7)$ & $1.128(24)$ \\
    \end{tabular}
    \end{ruledtabular}
    \caption{
        Comparison of $E_0^{(m_{\rm max})}$ and $E_1^{(m_{\rm max})}$ determined with the usual ZCW test with those determined from exponentially growing modes using the thermal ZCW test. The parity of the state corresponds to the (opposite of the) interpolating-operator parity for the ZCW (thermal ZCW) test.
        The correlator analyzed to produce the results of the second and third rows is the (time-)reverse of the correlator analyzed for the first and fourth rows, $C^\mathrm{rev}(t) = C(N_t - t)$.
    }
    \label{tab:E-comps-thermal}
\end{table}

\begin{table}[]
    \centering
    \begin{ruledtabular}
    \begin{tabular}{ccll}
    State Parity & Spurious eval test & $Z_{0W}$ & $Z_{1W}$ \\ \hline \vspace{0.2em}
    $+$ & ZCW &  $0.375(4)$ & $0.486(15)$ \\
    $+$ & Thermal ZCW &      $0.379(5)$ & $0.509(21)$ \\\\
    $-$ & ZCW &      $0.284(9)$ & $0.542(6)$ \\
    $-$ & Thermal ZCW &      $0.277(7)$ & $0.545(4)$
    \end{tabular}
    \end{ruledtabular}
    \caption{
        The corresponding overlap factors, details as in Table~\ref{tab:E-comps-thermal}.
    }
    \label{tab:Zs-thermal}
\end{table}

Although a detailed study of nucleon resonances requires extraction of a range of finite-volume energy levels including those overlapping with multi-particle (e.g.~$N\pi$) interpolating operators, it is interesting to note that in the positive parity sector $E_1^+ / E_0^- = 1.76(6)$ is nearly consistent with the experimental ratio $M_{N^*(1710)} / M_N \approx 1.823$, while $E_0^- / E_0^+ = 1.53(3)$ is reasonably close to the experimental ratio $M_{N^*(1535)} / M_N \approx 1.636$.
The deviations from the physical ratios may easily be accounted for by the slightly heavier quark masses used here ($m_{\pi} \approx 170~\mathrm{MeV}$) than in nature and finite lattice-spacing effects.
This underscores the fact that Lanczos converges to genuine energy levels in the spectrum; however, they are not necessarily the lowest energy levels.
In particular, it is noteworthy that this Lanczos analysis does not reveal an energy level close to the expected position of the positive-parity Roper resonance, $M_{N^*(1440)} / M_N \approx 1.535$.
Further, there are no energy levels obtained by this Lanczos analysis that are close to the energies of non-interacting {$P$-wave} $N\pi$ and $S$-wave $N\pi\pi$ states, which for this volume give $E_{N\pi} / M_N \gtrsim 1.26$ and $E_{N\pi\pi} / M_N \gtrsim 1.24$.
Block Lanczos analysis of correlator matrices involving $N$, $N\pi$, and $N\pi\pi$ are likely to be required in order to obtain a complete picture of the low-energy spectrum at finite statistics.

\section{Comparison with GEVP}
\label{sec:gevp}

The block Lanczos formalism and numerical results above can be compared with
the state-of-the-art method to date for analyzing LQCD correlator matrices $C_{ab}(t)$, which starts by considering the right\footnote{See Appendix~\ref{app:leftGEVP} for discussion of the corresponding left GEVP.} GEVP~\cite{Luscher:1990ck},
\begin{equation}\label{eq:gevp}
    \sum_b C_{ab}(t) G_{bk}(t, t_0) = \lambda_k(t,t_0) \sum_b C_{ab}(t_0) G_{bk}(t, t_0),
\end{equation}
where $\lambda_k(t,t_0)$ and $G_{bk}(t,t_0)$  are the $k$-th generalized eigenvalues and eigenvectors, respectively.
For Hermitian $\bm{C}(t) = \bm{C}(t)^\dagger$, the $\lambda_k(t,t_0)$ provide upper bounds on the largest eigenvalue $\lambda_0$ of $T$~\cite{Fox:1981xz,Michael:1982gb,Luscher:1990ck}.
Further, if the $\lambda_k(t,t_0)$ are ordered such that $\lambda_0(t,t_0) \geq \ldots \geq \lambda_{r-1}(t,t_0)$, then Cauchy's interlacing theorem guarantees that there are at least $k$ transfer matrix eigenvalues satisfying $\lambda_0 \geq \ldots \geq \lambda_k \geq \lambda_k(t,t_0)$ for all $k \in \{0,\ldots,r-1\}$~\cite{Fleming:2023zml}.
That is, $\lambda_k(t,t_0)$ provides a lower bound on the $k$th true eigenvalue $\lambda_k$.
Moreover, $\lambda_k(t,t_0)$ and $G_{bk}(t,t_0)$ may be used to construct estimators for energies, overlap factors, and matrix elements as described below.

\subsection{Block Lanczos generalizes GEVP}
\label{sec:gevp:vs-block}

Before proceeding, we may demonstrate immediately that GEVP coincides with a single step of block Lanczos.
Noting that, in numerical applications, correlator matrices $\bm{C}(t)$ are always invertible for all $t$ permits reworking the GEVP into an equivalent (non-generalized) eigenproblem,
\begin{equation}
    \sum_{b} [\bm{C}(t_0)^{-1} \bm{C}(t)]_{ab} G_{bk}(t, t_0) = G_{ak}(t, t_0) \lambda_k(t, t_0) ~ .
\end{equation}
Thus, the GEVP eigenvalues and eigenvectors are simply those of $\bm{C}(t_0)^{-1} \bm{C}(t)$.
Meanwhile, the Ritz values after a single step of block Lanczos are obtained by diagonalizing Eq.~\eqref{eq:alpha1},
\begin{equation}
   \bm{T}^{(1)}_{11} = \bm{\alpha}_1 = \bm{\beta}^{-1}_1  \bm{C}(1)  \bm{\gamma}^{-1}_1.
\end{equation}
Eigenvalues are invariant under conjugation by an arbitrary matrix and its inverse; conjugating by 
$\bm{\gamma}_1$ and using $\bm{C}(0) = \bm{\beta}_1 \bm{\gamma}_1$ per \cref{eq:C0_bg} gives
\begin{equation}\label{eq:block_GEVP}
\begin{aligned}
    \bm{\gamma}^{-1}_1  
    \bm{T}^{(1)}_{11}
     \bm{\gamma}_1
    &= 
   \bm{\gamma}^{-1}_1  \bm{\beta}^{-1}_1  \bm{C}(1)\\
   &= \bm{C}(0)^{-1} \bm{C}(1)  ~ .
\end{aligned}
\end{equation}
Therefore the Ritz values from one step of block Lanczos and the GEVP eigenvalues for $t_0=0$ and $t=1$ are both equal to the eigenvalues of $\bm{C}(0)^{-1} \bm{C}(1)$ and thus identical to each other,
\begin{equation}
     \lambda_k(1,0) = \lambda_k^{(1)}.
\end{equation}
Alternately, one may simply observe that a valid choice of oblique convention is $\bm{\beta}_1 = \bm{C}(0)$ and $\bm{\gamma}_1 = \bm{1}$, for which $\bm{T}^{(1)}_{11} = \bm{C}(0)^{-1} \bm{C}(1)$ identically.

The coincidence applies not only for eigenvalues, but also eigenvectors and the problem in abstract.
Rearranging  Eq.~\eqref{eq:block_GEVP} gives $T^{(1)}_{1a1b} = [\bm{\gamma}_1 \bm{C}(0)^{-1} \bm{C}(1) \bm{\gamma}_1^{-1}]_{ab}$ which, in conjunction with the eigenvector relation
$\sum_b T^{(1)}_{1a1b} \omega^{(1)}_{k1b} = \omega^{(1)}_{1ak} \lambda^{(1)}_k$ for $m=1$, gives
\begin{equation}
 \sum_b [\bm{\gamma}_1 \bm{C}(0)^{-1} \bm{C}(1) \bm{\gamma}_1^{-1}]_{ab} \omega^{(1)}_{1bk} = \omega^{(1)}_{1ak} \lambda^{(1)}_k.
\end{equation}
Left multiplying both sides by $\bm{C}(0)\bm{\gamma}^{-1}_1$ yields the equivalent GEVP,
\begin{equation}
   \sum_{bc} C_{ab}(1) \left( \gamma_{1bc}^{-1} \omega^{(1)}_{1ck} \right) = \lambda^{(1)}_k \sum_{bc}  C_{ab}(0) \left( \gamma_{1bc}^{-1} \omega^{(1)}_{1ck} \right) .
\end{equation}
This is precisely the form of Eq.~\eqref{eq:gevp}, allowing the GEVP eigenvectors to be identified as
\begin{equation}
    G_{bk}(1,0) = \sum_c  \gamma_{1bc}^{-1}\omega^{(1)}_{1ck} ~ .
\end{equation}

The exact relation between the two methods holds in full generality, not only for $t_0=0$ and $t_d=1$: a single step of block Lanczos applied to the reindexed correlation function
\begin{equation}
    \widetilde{C}_{ab}(t) = C_{ab}(t_0 + t \, t_d),
\end{equation}
will result in an eigenproblem for $\widetilde{\bm{C}}(0) = \bm{C}(t_0)$ and $\widetilde{\bm{C}}(1) = \bm{C}(t_d)$ equivalent to the GEVP for any $t_0, t_d$. 
The Ritz values $\widetilde{\lambda}_k^{(1)}$, eigenvectors $\widetilde{\omega}^{(1)}_{1bk}$, and $\widetilde{\bm{\gamma}}_1^{-1}$
obtained from this reindexed correlation function are therefore related to the GEVP eigenvalues and eigenvectors by
\begin{equation}\label{eq:GEVP=block1}
\begin{split}
    \lambda_k(t_d,t_0) &= \widetilde{\lambda}_k^{(1)}, \\
    G_{ak}(t_d,t_0) &=  \sum_b  \widetilde{\gamma}_{1ab}^{-1} \widetilde{\omega}^{(1)}_{1bk}.
    \end{split}
\end{equation}
Formally, this may be understood as applying a single step of block Lanczos to estimate the eigenvalues of the operator $T^{t_d-t_0}$ with redefined initial states. For even $t_0 \neq 0$, the initial states may be taken as $\bra{\chi_a} \rightarrow \bra{\chi_a} T^{t_0/2}$ and $\ket{\psi_b} \rightarrow T^{t_0/2}\ket{\psi_b}$.
For odd $t_0$, it must instead be viewed as applying a single step of oblique block Lanczos with distinct initial and final states, e.g.\ $\bra{\chi_a} \rightarrow \bra{\chi_a} T^{(t_0-1)/2}$ and $\ket{\psi_b} \rightarrow T^{(t_0+1)/2}\ket{\psi_b}$.

Block Lanczos can therefore be viewed as a generalization of GEVP that uses an $m$-times larger-dimensional Krylov space to approximate the transfer matrix after $m$ iterations.
This is analogous to how scalar Lanczos generalizes the power-iteration algorithm (i.e.~standard effective estimators for energies, overlaps, and matrix elements~\cite{Wagman:2024rid,Hackett:2024xnx}).
Since both GEVP and block Lanczos construct optimal Krylov-space approximations to the transfer-matrix eigensystem by explicitly diagonalizing their respective approximations of the transfer matrix, the larger Krylov space explored by block Lanczos necessarily leads to faster convergence than GEVP.
Indeed, for systems with small gaps (in lattice units) such that $E_r - E_0 \ll 1$, where achieving fast convergence is most important, block Lanczos converges to the ground state at a rate of $e^{-2t\sqrt{E_r - E_0}}$ after $m$ iterations  with $t = 2m-1$ according to Eq.~\eqref{eq:block_convergence}, while GEVP converges at an exponentially slower rate of $e^{-t (E_r - E_0)}$~\cite{Luscher:1990ck,Blossier:2009kd}.

\subsection{Numerical comparisons}

To demonstrate that block Lanczos not only extends but improves upon GEVP methods, we compare the two in applications to both the noiseless mock-data examples of \cref{sec:noiseless} and the noisy nucleon example of \cref{sec:lattice}.
Comparisons are complicated by the many different variations in precise definitions of GEVP estimators for energies, overlap factors, and matrix elements.
For the sake of demonstration, we define and apply two different sets of estimators---specifically, one ``moving pivot'' scheme and another with a ``fixed pivot'', as defined precisely below.
In the noiseless applications, in extractions of the spectrum, overlap factors, and matrix elements, we demonstrate the improved convergence properties of block Lanczos over both GEVP and scalar Lanczos.
In determinations of the spectrum and overlap factors in the noisy case, we demonstrate not only improved convergence but also an advantage in SNR properties, extending the success of scalar Lanczos to the block case as well.
While different variations on the GEVP analyses may perform slightly differently, the demonstrated advantages of block Lanczos should be expected to hold generically.

\subsubsection{GEVP definitions}

Rather than using GEVP eigenvalues directly, it is common practice to define an effective energy estimator~\cite{Luscher:1990ck,Blossier:2009kd}, 
\begin{equation}
    E_k(t, t_0) \equiv -\ln\left( \frac{\lambda_k(t, t_0)}{\lambda_k(t-1, t_0)} \right),
    \label{eq:gevp-meff}
\end{equation}
resembling but with distinct properties from the usual effective energy, as discussed further below.
Different ``moving pivot'' (MP) schemes to choose $t_0$ as a function of $t$ are employed in practice.
For large $t$, choosing\footnote{Where $\lfloor \cdot \rfloor$ indicates the integer floor operation, i.e.~rounding down to the nearest integer.} $t_0 = \lfloor t/2 \rfloor$ to define
\begin{equation}
    E^{\rm MP}_k(t) \equiv E_k(t, \lfloor t/2 \rfloor)
    \label{eq:gevp-mp-meff}
\end{equation}
guarantees that $E^\mathrm{MP}_0(t)$ converges to $E_0$ exponentially quickly with excited-state contamination suppressed by $\mathcal{O}(e^{-t(E_r - E_0)})$~\cite{Luscher:1990ck,Blossier:2009kd}.
At finite $t$, the inequalities above guarantee that there exist $k$ energies with $E_0 \leq \ldots \leq E_k \leq E_k^{\rm MP}(t)$ for all $k \in \{0,\ldots,r-1\}$.
This allows GEVP solutions to provide rigorous one-sided bounds on LQCD energy levels as emphasized in Refs.~\cite{Amarasinghe:2021lqa,Fleming:2023zml,Detmold:2024iwz}.

A downside of the moving-pivot approach is that 
$\lambda_k(t, t_0)$
does not possess an exact spectral representation as a sum of exponentials
even for fixed $t_0$, 
because the relative contributions of different energy eigenstates to $\lambda_k(t,t_0)$ depend on both $t$ and $t_0$.
This can sometimes complicate the analysis of excited-state effects.
A widely used approach for circumventing this issue are ``fixed pivot'' (FP) schemes, wherein one defines fixed-pivot correlation functions~\cite{Fox:1981xz,Michael:1982gb,Luscher:1990ck} 
\begin{equation}
\label{eq:principal_correlators}
\hat{C}_{k}(t;t_d,t_0) \equiv \left< \Psi_{k}(t;t_d,t_0)\Psi_{k}^{\dagger}(0;t_d,t_0) \right>,
\end{equation}
using optimized interpolators 
\begin{equation}
  \Psi_{k}(t; t_d,t_0) \equiv \sum_a \psi_a(t) G^*_{ak}(t_d, t_0)
  \label{eq:gevp-interp}
\end{equation}
defined from the generalized eigenvectors or ``GEVP weights'' $G_{ak}(t_d, t_0)$ at pivot time $t_0$ and ``reference time'' $t_d$.
Such fixed-pivot correlation functions possess convex spectral representations as functions of $t$~\cite{Bulava:2010yg,Bulava:2016mks},
\begin{equation}
  \hat{C}_{k}(t;t_d,t_0) = \sum_{\mathsf{n}=0}^\infty |\hat{Z}_{kn}(t_d,t_0)|^2e^{-t E_{n}} \ ,
  \label{eq:fp-gevp-convex-spectrum}
\end{equation}
where $\hat{Z}_{kn}(t_d,t_0) = \mbraket{n}{\Psi_k(0; t_d,t_0)^\dagger}{\Omega}$.
These can be used to define an effective energy
\begin{equation}
    E_k^{\rm FP}(t; t_d,t_0) \equiv -\ln\left( \frac{\hat{C}_{k}(t; t_d,t_0)}{\hat{C}_{k}(t-1; t_d,t_0)} \right),
    \label{eq:fp-gevp-meff}
\end{equation}
which shares all the properties of the usual effective energy defined on a scalar correlator due to \cref{eq:fp-gevp-convex-spectrum}.

Explicit effective estimators can also be defined for overlap factors by generalizing simpler constructions from scalar correlator analyses~\cite{Bulava:2016mks}.
For example, the moving-pivot effective overlap employed here is
\begin{equation}
    Z^{\mathrm{MP}*}_{ka}(t, t_0) \equiv \frac{ 
        \sum_b C_{ab}(t) \, G_{bk}(t,t_0) 
    }{ 
        \lambda_k(t,t_0)^{ \frac{1}{2} t / (t-t_0) }
        \sqrt{\hat{C}_k(t;t,t_0)} 
    }
    \label{eq:gevp-mp-Z}
\end{equation}
where $\hat{C}_k(t;t,t_0)$ is as in \cref{eq:principal_correlators}; 
we take $t_0 = \lfloor t/2 \rfloor$ in all demonstrations below.
We use an analogous definition for the fixed-pivot estimator, 
\begin{equation}
    Z^{\mathrm{FP}*}_{ka}(t; t_d,t_0) \equiv \frac{ \sum_b C_{ab}(t) \, G_{bk}(t_d,t_0) }{ e^{-\frac{1}{2} t E_k^{\rm FP}(t; t_0,t_d)} \sqrt{\hat{C}_k(t;t_d,t_0)} }
    \label{eq:gevp-fp-Z}
\end{equation}
where $E_k^{\rm FP}(t; t_0,t_d)$ is the effective energy estimator of \cref{eq:fp-gevp-meff}.

Explicit GEVP estimators can be defined for operator matrix elements $J_{fi} = \mbraket{f'}{J}{i}$ in terms of the three-point function matrix $\bm{C}^\mathrm{3pt}(\sigma, \tau)$ and two-point matrices $\bm{C}(t)$ and $\bm{C}'(t)$ with initial- and final-state quantum numbers, respectively.
In demonstrations below, we use the moving-pivot effective estimator~\cite{Blossier:2009kd}
\begin{equation}
    J^\mathrm{MP}_{fi}(t, m) = \sum_{ab} G^{\prime *}_{af}(t,2 m) \, C^\mathrm{3pt}_{ab}(m, m) \, G_{bi}(t, 2 m)
    \label{eq:gevp-mp-Jeff}
\end{equation}
where $G$ and $G'$ have been computed from $\bm{C}$ and $\bm{C}'$, respectively.
This expression directly generalizes the power-iteration estimator for the effective matrix element derived in Ref.~\cite{Hackett:2024xnx}.
Note that only a single point of three-point function data at $\sigma=\tau=m$ is incorporated, and varying $t$ only changes how $G$ and $G'$ are derived from the two-point data.
An analog of the $t_0 = \lfloor t/2 \rfloor$ MP scheme can be obtained by taking $m = \lfloor t/4 \rfloor$.

Fixed-pivot approaches are more commonly employed in GEVP matrix element analyses.
Using GEVP interpolators as in \cref{eq:gevp-interp}, one may define~\cite{Dragos:2016rtx}
\begin{equation}\begin{aligned}
    \hat{C}^\mathrm{3pt}_{fi}(\sigma, \tau; t_d, t_0) 
    &\equiv \left< \Psi'_{f}(\sigma+\tau;t_d,t_0) J(\tau) \Psi_{i}^{\dagger}(0;t_d,t_0) \right>
    \\
    &= \sum_{ab} G^{\prime *}_{af}(t_d,t_0) \, C^\mathrm{3pt}_{ab}(\sigma, \tau) \, G_{ib}(t_d, t_0) ~ .
\end{aligned}
\label{eq:gevp-fp-3pt}
\end{equation}
As with the FP two-point function, the elements of the resulting FP three-point matrix admit good spectral expansions and thus can be analyzed as with any other three-point function.
While this is often accomplished by fitting, to reduce the complexity of comparisons we instead restrict to explicit effective estimators applied to the individual correlators in $\hat{C}_{fi}(t)$.
Specifically, we employ summation (\cref{eq:summation-Jeff}) and power-iteration (\cref{eq:power-iter-Jeff}) effective matrix elements as in \cref{sec:noiseless:me}.

Some additional GEVP definitions are required for analysis of noisy data.
Rather than filtering and treating the resulting partial data missingness as in a Lanczos analysis, we take the more standard approach of fully discarding GEVP estimates for $t_0, t_d$ where any complex eigenvalues arise.
Separately, the FP scheme requires some choice of $t_0$ and $t_d$ to define the pivot; here, we use $t_0=12$ and $t_d=19$ as selected using the automated procedure of Ref.~\cite{Detmold:2024iwz}.\footnote{
Contributions to $G_{ak}(t_d,t_0)$ from states with energies above $E_r$ are exponentially suppressed for large $t_0$ and $t_d$.
For large enough $t_0$ and $t_d$ where these contributions are negligible, $E_k^{\rm FP}(t,t_0,t_d))$ will be identical to $E_k^{\rm MP}(t)$ up to statistical uncertainties.
In this regime, the positive features of both moving- and fixed-pivot GEVP effective mass estimators, including monotonicity with $t$ and the interlacing theorem, apply to either definition.
However, statistical noise increases with $t_0$ and $t_d$ and results will become unreliable if $t_0$ and $t_d$ are taken too large at finite statistics.
The algorithm of Ref.~\cite{Detmold:2024iwz} selects the largest $t_0$ and $t_d$ where there is statistical agreement between fixed- and moving-pivot estimators.
}
The FP scheme is defined taking a common pivot matrix $G_{ak}$ across all bootstraps, computed from the central-value expectation of $\bm{C}(t)$---i.e., the change of operator basis is not included in the error propagation.
The same definitions are used below for overlaps and matrix elements.

\subsubsection{Spectrum}

\begin{figure}
    \includegraphics[width=\linewidth]{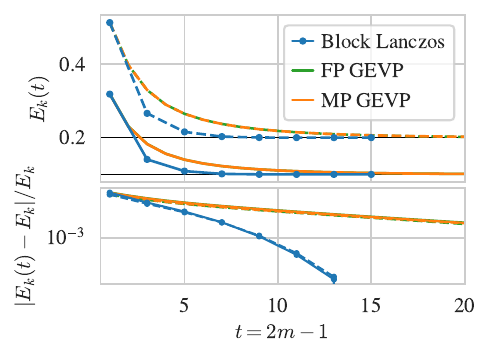}
    \caption{
        The lowest two energies in the spectrum of the noiseless example \cref{eq:noiseless:ex-def} (top) and their relative errors versus the true value (bottom) as extracted by 
        block Lanczos (blue), 
        GEVP with a moving pivot at $t_0 = \lfloor t/2 \rfloor$ (orange),
        and GEVP with a fixed pivot at $t_0=5$ and $t_d=10$ (green).
        For each $k$, the two GEVP estimates closely coincide at all $t$.
        In the bottom panel, the relative errors for both $k=0$ and 1 collapse into two curves, one for block Lanczos and one for the GEVP estimators.
        Block Lanczos solves for the true eigenvalues exactly after $m=\betaT/4=8$ steps. 
    }
    \label{fig:noiseless-spectrum-block-vs-gevp}
\end{figure}

We first compare block Lanczos with GEVP in application to the noiseless mock-data example from \cref{sec:noiseless}, \cref{eq:noiseless:ex-def}.
The faster convergence of block Lanczos over GEVP extractions of the spectrum is unambiguous, as shown in \cref{fig:noiseless-spectrum-block-vs-gevp}, which compares block Lanczos with both the moving-pivot scheme with $t_0 = \lfloor t_d/2 \rfloor$ as well as a fixed-pivot GEVP with $t_0=5$ and $t_d=10$ (note FP results are only weakly sensitive to the choice of $t_0$ and $t_d$ in the noiseless case).
With $m=1$, block Lanczos and both GEVP schemes give identical results as expected,\footnote{For GEVP estimators defined as effective energies, \cref{eq:gevp-mp-meff,eq:fp-gevp-meff}, this holds for the MP definition because $\lambda(0,0)=1$. This alignment of definitions is intentional, to reflect the deeper underlying coincidence between GEVP eigenvalues and Ritz values discussed above. For FP schemes the alignment is not exact, see noisy results below.}
and for a few iterations, GEVP and block Lanczos provide comparable extractions.
However, with increasing iterations, block Lanczos converges exponentially faster, as predicted by comparison of the block KPS bound of \cref{sec:bounds} leading to $e^{-2t\sqrt{E_r - E_0}}$ convergence in comparison to the $e^{-t(E_r-E_0)}$ convergence provided by GEVP~\cite{Luscher:1990ck,Blossier:2009kd}.
Besides improving convergence on the lowest $r=2$ states shown in \cref{fig:noiseless-spectrum-block-vs-gevp}, block Lanczos yields additional energy estimates for $(m-1)r$ higher states that are unconstrained by GEVP, as explored in \cref{sec:noiseless}.

\begin{figure}
    \includegraphics[width=\linewidth]{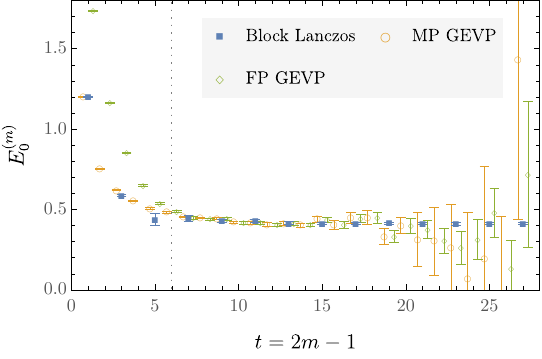}\\
    \includegraphics[width=\linewidth]{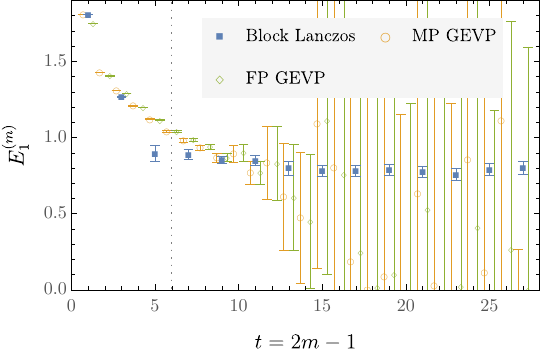}
    \caption{
        Energy estimates for the two lowest-lying states in the nucleon spectrum as extracted by block Lanczos (blue), GEVP with a moving pivot $t_0 = \lfloor t/2 \rfloor$ as in Ref.~\cite{Blossier:2009kd} (orange), and GEVP with a fixed pivot with $t_0=11$ and $t_d=18$ chosen using the algorithm of Ref.~\cite{Detmold:2024iwz} (green).
        Bootstrap-median estimators are used for block Lanczos. In all cases, uncertainties are computed using idiomatic bootstrap confidence intervals.
    }
    \label{fig:nuc-block-vs-gevp}
\end{figure}

Proceeding to the noisy nucleon example, \cref{fig:nuc-block-vs-gevp} compares the block Lanczos extraction of the ground and first excited states with estimates made with two different GEVP schemes: the FP scheme noted above and the moving-pivot (MP) scheme with $t_0 = \lfloor t / 2 \rfloor$.
The improved convergence and signal-to-noise properties of block Lanczos over either GEVP scheme are apparent.
At $t=1$, the block Lanczos and MP GEVP estimators are (statistically) identical; the FP estimators as defined here differ nontrivially in the noisy case.
However, examining $t>1$, energy estimates by block Lanczos reach their steady-state values earlier than those of either GEVP scheme; this is especially clear for the excited-state energy $E_1$.
At early times before convergence, the GEVP estimates are more precise than the Lanczos ones, similar as observed in analogous scalar correlator analyses~\cite{Wagman:2024rid,Hackett:2024xnx}.
However, signal-to-noise for the Lanczos extractions remains constant (up to fluctuations) after the estimates have converged, while SNR for both GEVP schemes decreases exponentially.
The Lanczos estimate of $E_0$ is consistent with the plateaus observed in each GEVP estimator, but for $E_1$, SNR degrades in the GEVP estimates before any convincing plateau is achieved.

\begin{figure}
    \includegraphics[width=\linewidth]{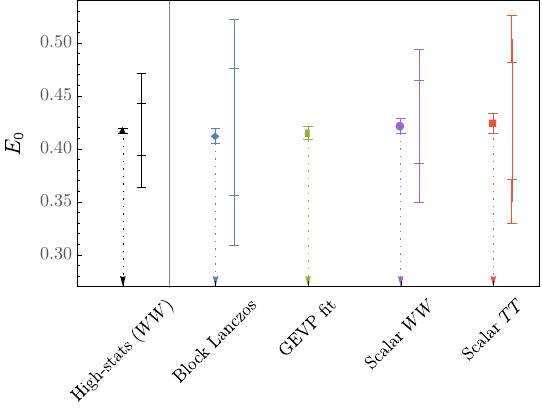}\\
    \includegraphics[width=\linewidth]{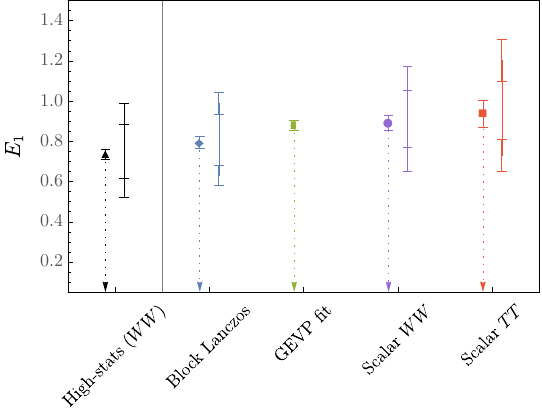}
    \caption{
        Block Lanczos results for $E_0^{(m_{\rm max})}$ (top) and $E_1^{(m_{\rm max})}$ (bottom) with $m_{\rm max} = 47$, shown in \cref{fig:block-nuc-spectrum}, are compared with FP GEVP fits performed as described in Refs.~\cite{NPLQCD:2020ozd,Amarasinghe:2021lqa,Detmold:2024iwz} as well as scalar Lanczos results from the corresponding diagonal correlator matrix entries, higher-statistics scalar Lanczos analysis. Results with bootstrap uncertainties are shown as points with error bars in all cases. Dotted arrows indicate variational bounds. Residual bounds for Lanczos results are shown as intervals whose upper and lower edges are marked by error bars showing bootstrap uncertainties on $-\ln( \lambda_k \pm \sqrt{B_k} )$.
    }
    \label{fig:nuc-block-vs-gevp-fit}
\end{figure}

\begin{figure}
    \includegraphics[width=\linewidth]{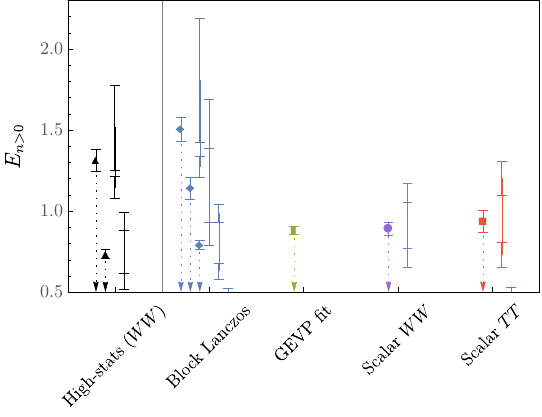}
    \caption{
        Results as in \cref{fig:nuc-block-vs-gevp-fit} shown for $E_2$ and $E_3$ where available along with $E_1$.
    }
    \label{fig:nuc-block-vs-gevp-fit-more}
\end{figure}

Block Lanczos results from the last iteration where residual bounds can be computed, $m_{\rm max} = 47$, are compared to multi-state fits to FP GEVP results as well as other scalar Lanczos results  in \cref{fig:nuc-block-vs-gevp-fit,fig:nuc-block-vs-gevp-fit-more}.
The only block Lanczos hyperparameter is $F^{\mathrm{ZCW}} = 10$, parameterizing how much smaller overlaps must be than ones at small $m$ to be labeled spurious.
This amounts to an enormous practical reduction in complexity versus the fits to the FP correlation functions, which are performed using the strategy summarized in Fig. 16 of Ref.~\cite{NPLQCD:2020ozd}: 
in brief, fits are performed for all possible $t_{\rm min} \in [2, t_{\rm max} - 6]$ where $t_{\rm max}$ is chosen as the largest $t$ where the SNR of $E_k^{\rm FP}(t)$ is greater than 0.2.
Linear shrinkage~\cite{stein1956,Ledoit:2004} is used to regulate numerical instabilities in covariance matrix inversion.
One-, two-, three-, ...~state fits are performed until adding states does not lower the Akaike information criterion~\cite{AkaikeAIC} (AIC) by more than a threshold of $0.5 N_{\rm dof}$.
Various checks on the consistency of numerical optimization detailed in Ref.~\cite{NPLQCD:2020ozd} are then performed.
A weighted average of the AIC-preferred fits for each $t_{\rm min}$ passing these checks is then taken, where the weights are proportional to the $p$-value divided by the variance of each fit~\cite{Rinaldi:2019thf}.
Hyperparameters enter this scheme through the numerical values specified above and others detailed in  Ref.~\cite{NPLQCD:2020ozd}, as well as the hyperparameters for choosing $t_0$ and $t_d$ summarized in Fig. 16 of Ref.~\cite{Detmold:2024iwz} that here lead to $t_0=11$ and $t_d=18$.
For $n=0$, the fits with weights above 0.1 that result from this procedure are one-state fits with $t_{\rm min} \in [9,12]$ and two-state fits for $t_{\rm min} = 4$.
For $n=1$, the only fits with weights above 0.1 are two-state fits with $t_{\rm min} \in [2,3]$.

Note that this GEVP analysis prescription is far from unique.
The GEVP formalism admits many other possible analyses of the same data, all of which ideally should give statistically compatible results.
\Cref{tab:GEVP_E0_fits,tab:GEVP_E1_fits} tabulates multi-state fit results for various $t_{\rm min}$ and numbers of states included in the fit.

\begin{table}[]
    \centering
    \begin{ruledtabular}
    \begin{tabular}{cclSc}
       $N_{\rm states}$ & $t_{\rm min}$ & $E_0$ & {$\chi^2 / N_{\rm{dof}}$} & $N_{\rm{dof}}$ \\ \hline \vspace{0.2em}
   $3$ &  $2$ & $0.403(12)$ & 1.0 & 13 \\
   $3$ &  $3$ & $0.4221(50)$ & 1.7 & 12 \\
   $3$ &  $5$ & $0.4025(93)$ & 1.3 & 10 \\
   $3$ &  $8$ & $0.4139(60)$ & 1.1 & 7 \\
   $3$ &  $9$ & $0.4157(40)$ & 1.4 & 6 \\
   $3$ & $10$ & $0.4143(42)$ & 1.6 & 5 \\
   $3$ & $11$ & $0.4138(71)$ & 1.9 & 4 \\
   $3$ & $12$ & $0.416(10)$ & 2.5 & 3 \\
   $3$ & $13$ & $0.42(21)$  & 3.7 & 2 \\
   $3$ & $14$ & $0.422(12)$   & 7.1 & 1 \\
   $2$ &  $2$ & $0.4293(30)$ & 2.0 & 15 \\
   $2$ &  $4$ & $0.4146(59)$ & 1.2 & 13 \\
   $2$ &  $5$ & $0.4448(26)$ & 7.2 & 12 \\
   $2$ &  $6$ & $0.4331(50)$ & 3.4 & 11 \\
   $2$ &  $7$ & $0.410(11)$ & 1.0 & 10 \\
   $2$ &  $8$ & $0.4233(49)$ & 2.3 & 9 \\
   $2$ &  $9$ & $0.4157(40)$ & 1.0 & 8 \\
   $2$ & $10$ & $0.4143(54)$ & 1.1 & 7 \\
   $2$ & $11$ & $0.4139(62)$ & 1.3 & 6 \\
   $2$ & $12$ & $0.4159(69)$ & 1.5 & 5 \\
   $2$ & $13$ & $0.4173(94)$  & 1.9 & 4 \\
   $2$ & $14$ & $0.42(21)$   & 2.4 & 3 \\
   $2$ & $15$ & $0.41(21)$   & 3.1 & 2 \\
   $1$ &  $2$ & $0.8147(84)$ & 260 & 17 \\
   $1$ &  $3$ & $0.5213(51)$ & 110 & 16 \\
   $1$ &  $4$ & $0.4658(30)$ & 21 & 15 \\
   $1$ &  $5$ & $0.4448(26)$ & 6.2 & 14 \\
   $1$ &  $6$ & $0.4331(30)$ & 2.9 & 13 \\
   $1$ &  $7$ & $0.4289(34)$ & 2.7 & 12 \\
   $1$ &  $8$ & $0.4233(36)$ & 1.9 & 11 \\
   $1$ &  $9$ & $0.4157(40)$ & 0.83 & 10 \\
   $1$ & $10$ & $0.4143(42)$ & 0.87 & 9 \\
   $1$ & $11$ & $0.4138(61)$ & 0.97 & 8 \\
   $1$ & $12$ & $0.4158(69)$ & 1.1 & 7 \\
   $1$ & $13$ & $0.417(10)$  & 1.2 & 6 \\
   $1$ & $14$ & $0.422(11)$  & 1.4 & 5 \\
   $1$ & $15$ & $0.411(16)$  & 1.5 & 4 \\
    \end{tabular}
    \end{ruledtabular}
    \caption{
        Multi-state fit results for the $n=0$ FP GEVP correlator with $t_{\rm max} = 19$; see the main text for details. Fits where $\chi^2$ is not decreased by increasing $N_{\rm states}$, indicating optimizer convergence failures, are omitted.
    }
    \label{tab:GEVP_E0_fits}
\end{table}

\begin{table}[]
    \centering
    \begin{ruledtabular}
    \begin{tabular}{cclSc}
    $N_{\rm states}$ & $t_{\rm min}$ & $E_1$ & {$\chi^2 / N_{\rm{dof}}$} & $N_{\rm dof}$ \\ \hline \vspace{0.2em}
    $3$ & $2$ & $0.86(10)$   & 0.60 & 3 \\
    $3$ & $3$ & $0.862(25)$  & 0.92 & 2 \\
    $2$ & $2$ & $0.892(12)$  & 1.1 & 5 \\
    $2$ & $3$ & $0.862(20)$  & 0.46 & 4 \\
    $2$ & $4$ & $0.84(14)$   & 0.44 & 3 \\
    $2$ & $5$ & $1.01(10)$   & 36 & 2 \\
    $1$ & $2$ & $1.2561(17)$ & 500 & 7 \\
    $1$ & $3$ & $1.1608(23)$ & 170 & 6 \\
    $1$ & $4$ & $1.0852(33)$ & 57 & 5 \\
    $1$ & $5$ & $1.0131(56)$ & 18 & 4 \\
    \end{tabular}
    \end{ruledtabular}
    \caption{
       Multi-state fit results for the $n=1$ FP GEVP correlator with $t_{\rm max} = 9$; see the main text and \cref{tab:GEVP_E0_fits} for details.
    }
    \label{tab:GEVP_E1_fits}
\end{table}

A weighted average of GEVP fit results performed as described above gives $E_0 = 0.4150(59)$. This is in excellent agreement with all Lanczos results, and all analyses of the same data give comparable uncertainty estimates.
For $E_1$, a weighted average of GEVP fit results gives 0.879(25), which differs from the high-statistics scalar Lanczos result 0.735(25) by 4$\sigma$, while the block Lanczos result, 0.792(32), shows only 1.4$\sigma$ tension (see \cref{tab:E-comps}).
Other choices of GEVP fit hyperparameters lead to similar tensions; for instance increasing the noise tolerance to $0.5$ leads to $t_{\rm max} = 11$ instead of $t_{\rm max}= 9$, which  changes the weighted average fit for $E_1$ to 0.879(21).
Scalar Lanczos results for the diagonal correlation functions also show 2-3$\sigma$ tensions in comparison with high-statistics results.
However, the residual bounds $-\ln( \lambda_1 \pm \sqrt{B_1} )$ provided by scalar Lanczos results\footnote{It is clear from \cref{fig:block-nuc-resdiauls} that $B_k^{(m)}$ shows more non-trivial $m$ dependence than $E_k^{(m)}$. For simplicity, we have computed the widths of the residual bounds using the $m$ where the central value of the bootstrap median estimator $B_k^{(m)}$ is minimized. The $m$-dependence of residual bounds, which must be a monotonic decrease for symmetric (block) Lanczos~\cite{Parlett} but can be non-trivial for oblique (block) Lanczos, will be investigated in more detail in future work. It will also be interesting to study whether the increase in the widths of the residual bounds between $r=1$ and $r=2$ seen in \cref{fig:nuc-block-vs-gevp-fit} appears in other noisy examples.
}
are in 1$\sigma$ agreement with the corresponding high-statistics results.
This residual-bound consistency shows that each result could be converging to the same true energy eigenvalue but with significant excited-state effects still present in one or more determinations.
Agreement at the level of Ritz value statistical uncertainties between block Lanczos and high-statistics scalar Lanczos suggests that the low-statistics scalar Lanczos results for $E_1$ still include statistically significant excited-state effects that are effectively removed by block Lanczos (or by going to higher statistics where they can be resolved more easily).
With only GEVP results for guidance, any interpretation of $E_1$ as more than a variational upper bound is therefore likely to be incorrect at 4$\sigma$ in this example. 

The residual bounds for $n\in \{2,3\}$ from block Lanczos both overlap the residual bounds for $n=2$ from the high-statistics scalar analysis, making it difficult to conclude whether block has resolved two distinct physical states or encountered an artificial degeneracy due to the appearance of multiple eigenvalues, which can generically occur for block Lanczos states with $n \geq r$~\cite{Parlett}.

\subsubsection{Overlap factors}

\begin{figure}
    \includegraphics[width=\linewidth]{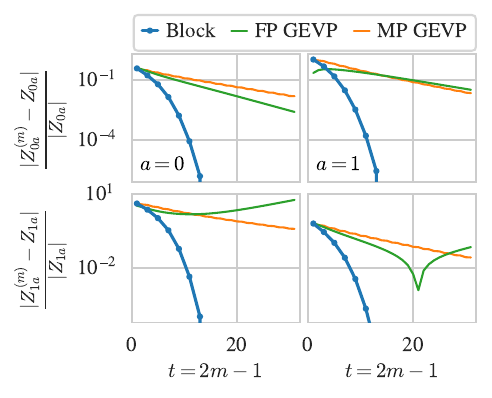}
    \caption{
        Relative error in ground-state overlap factors estimated in the noiseless example \cref{eq:noiseless:ex-def} by
        block Lanczos (blue),
        the moving-pivot GEVP estimator \cref{eq:gevp-mp-Z} with $t_0 = \lfloor t/2 \rfloor$ (orange),
        and the fixed-pivot GEVP estimator \cref{eq:gevp-fp-Z} with $t_0=5$ and $t_d = 10$ (green).
        Block Lanczos solves for the true overlap factors exactly at $m = \betaT/4 = 8$ steps.
        The sign convention $Z_{k0} > 0$ has been enforced by hand for the GEVP estimate of $Z_{01}$.
    }
    \label{fig:noiseless-Zs-vs-GEVP}
\end{figure}

For the noiseless example of \cref{sec:noiseless},
\Cref{fig:noiseless-Zs-vs-GEVP} compares convergence of different Lanczos and GEVP estimators of overlaps with $\psi_T$ and $\psi_W$ for the lowest-lying two states.
As with the spectrum, while GEVP provides a comparable-quality estimate at early times, block Lanczos estimators converge exponentially more rapidly.
In practice, it is necessary to manually enforce the sign convention $Z_{k0} > 0$ on the GEVP estimates. 

\begin{figure}
    \includegraphics[width=\linewidth]{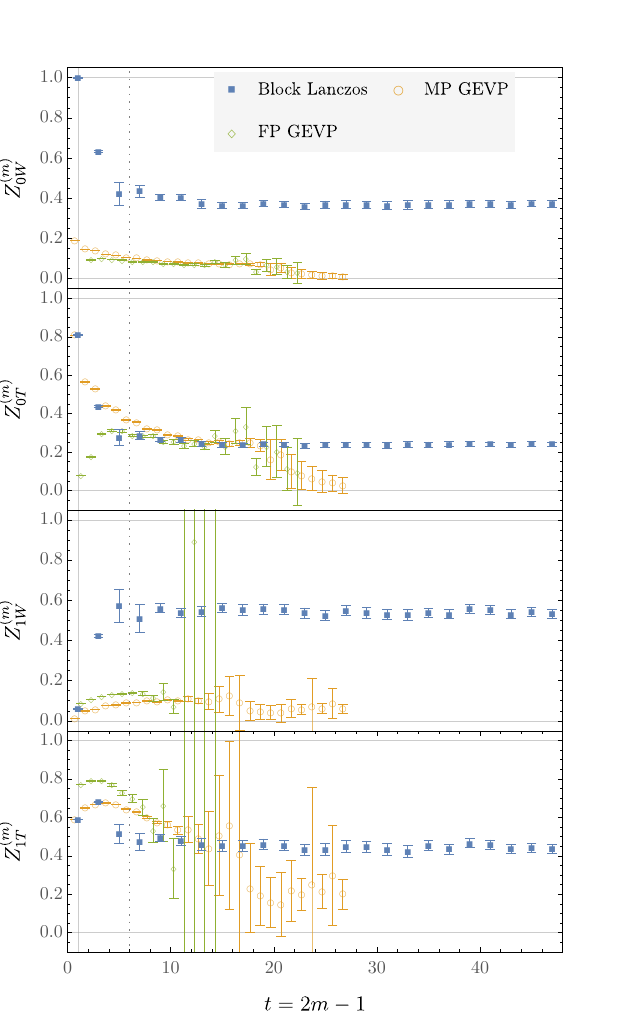}
    \caption{
        For the nucleon correlator matrix example, overlap factors as computed by block Lanczos (blue) and by GEVP with a moving pivot at $t_0 = \lfloor t/2 \rfloor$  as in Ref.~\cite{Blossier:2009kd} (orange), and GEVP with a fixed pivot with $t_0=11$ and $t_d=18$ chosen using the algorithm of Ref.~\cite{Detmold:2024iwz} (green). Bootstrap-median estimators are used for block Lanczos. In all cases, uncertainties are computed using idiomatic bootstrap confidence intervals.
    }
    \label{fig:nuc-Zs-block-vs-gevp}
\end{figure}

For the noisy nucleon data of \cref{sec:lattice}, \cref{fig:nuc-Zs-block-vs-gevp} compares block Lanczos overlap estimators with the GEVP estimators, demonstrating an even more clear advantage than for the spectrum.
Note that we find that the sign convention $Z_{k0} > 0$ must be enforced by hand for the GEVP results, achieved by applying an overall sign to $Z_{ka}$ at each $t$ and within each bootstrap.
For all GEVP overlaps, the signal breaks down more rapidly than for the spectrum, with SNR quickly degrading before each estimator begins producing complex values and thus becoming unreliable.
While the FP GEVP produces consistent results for the ground state overlaps $Z_{0a}$, the MP estimator does not provide a clear plateau and passes through the value indicated by the Lanczos estimator in each case.
For the excited state overlaps $Z_{1a}$, neither GEVP scheme produces any convincing plateau before signal is lost; the FP GEVP scheme for $Z_{1a}$ presents a brief but potentially deceptive pseudo-plateau.
In contrast, block Lanczos provides stable results with no SNR decay and none of the pathological behaviors observed for the GEVP estimators.

\subsubsection{Matrix elements}

\begin{figure}
    \includegraphics[width=\linewidth]{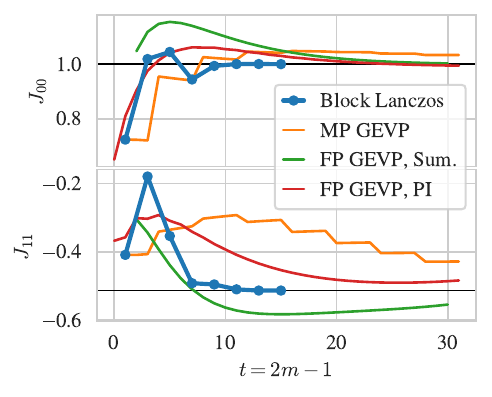}
    \caption{
        For the off-diagonal noiseless example \cref{eq:noiseless:me-ex-def}, estimates of diagonal matrix elements for the ground state $J_{00}$ (top) and first excited state $J_{11}$ as computed by block Lanczos (blue), by GEVP with a moving pivot at $t_0 = \lfloor t/2 \rfloor$ per \cref{eq:gevp-mp-Jeff}, and using the summation (\cref{eq:summation-Jeff}, green) and power iteration (\cref{eq:power-iter-Jeff}, red) estimators applied to the diagonals of a GEVP three-point constructed per \cref{eq:gevp-fp-3pt} for a fixed pivot at $t_0=5$ and $t_d=10$.
        Block Lanczos solves for the true $J_{00}$ exactly after $m=\betaT/4=8$ steps.
    }
    \label{fig:noiseless-ME-block-vs-gevp}
\end{figure}

For the noiseless mock-data example of \cref{sec:noiseless}, \cref{fig:noiseless-ME-block-vs-gevp} compares convergence of block Lanczos estimates of the diagonal matrix elements $J_{00}$ and $J_{11}$ with the moving-pivot GEVP estimator \cref{eq:gevp-mp-Jeff}, as well as different estimators constructed from the fixed-pivot GEVP three-point function \cref{eq:gevp-fp-3pt}.
As expected, the moving-pivot GEVP estimator \cref{eq:gevp-mp-Jeff} with $m = \lfloor t/4 \rfloor$ coincides with the block Lanczos one at $t=1$.
Block Lanczos provides a qualitatively improved extraction of each.
For either state, the initial Lanczos estimate changes non-smoothly---a feature advertising non-convergence---then converges quickly to the true value.
In contrast, the GEVP estimates vary smoothly\footnote{Up to expected even-odd effects in the MP estimator due to $t_0 = \lfloor t/2 \rfloor$, and in the power-iteration FP estimator due to the piecewise definition in \cref{eq:power-iter-Jeff}.} and converge poorly.
While the FP estimators for $J_{00}$ provide a good estimate by the final step, the MP estimator shows potentially deceptive behavior: the estimate changes only slowly while still visibly offset from the true value.
For the excited-state $J_{11}$, the MP estimator clearly does not converge before the end of the correlator. The two FP estimators drift near the true $J_{11}$ but will not provide a reliable estimate; in the FP scheme each will eventually converge to $J_{00}$, which behavior is suggested by the upwards turn in each by the end of the correlator.


\section{Conclusions}
\label{sec:conclusions}

We present a new method to extract energies, overlaps, and operator matrix elements from 
correlator matrix data based on an oblique block Lanczos algorithm and a new physically motivated test for spurious state filtering.
The method extends and improves upon GEVP methods---the present standard and state of the art---in control of excited-state effects, systematic uncertainty quantification, and signal-to-noise.

Reformulating the Cullum-Willoughby test in terms of a cut on overlap factors provides a fully physical picture of spurious state filtering.
Noise results in an apparent expansion of Krylov space outside the part of Hilbert space containing the states of physical interest, and the spurious states are a consequence of this expansion.
Following this motivation, the ZCW test identifies spurious states as those with ``spuriously small'' overlap factors, with a physically motivated threshold for ``small''.
After filtering, the remaining non-spurious Ritz values and vectors provide reliable estimates for some eigenstates of the physical transfer matrix.
The resulting estimators converge to the true spectrum faster than either GEVP or scalar Lanczos estimators, have directly calculable two-sided systematic uncertainty bounds, and have constant signal-to-noise at large iteration count. 

For spectroscopy applications, block Lanczos analysis is immediately applicable to existing datasets and can be adopted immediately.
As in the scalar case, the data requirements to use block Lanczos to extract operator matrix elements require evenly-spaced three-point data at small sink times.
This is in tension with the common strategy for sequential-source calculations where data are generated for only a few intermediate sink times where excited-state effects are reduced.
Data generation strategies should be adjusted.
This is especially true where correlator matrices are employed, as they are typically used in matrix element analyses only when excited-state effects (or properties) are a concern; the badly contaminated but well-resolved measurements at early sink times that are not typically computed are exactly those that Lanczos uses to control excited-state effects.

As demonstrated, filtering states, sorting on Ritz values, and using an idiomatically bootstrapped median estimator to define values and uncertainties provides a simple and effective analysis.
However, state identification (i.e., associating measurements between bootstraps to define the set to be averaged over) still poses some outstanding challenges.
In \cref{sec:lattice}, we observed that simply sorting measurements on the Ritz values led to some obvious misidentifications which complicate the analysis of relatively high-energy excited states.
Although ground-state energies and overlaps appear to be more robust, and there are indications that this effect is less severe at high statistics, it is nevertheless critical to check for such misidentification before blindly trusting the output of a Lanczos analysis.
Better schemes for state identification which can provide more reliable excited-state extractions at finite statistics are an important topic for further study.

The key improvements offered by Lanczos over previous methods are spurious state filtering and the ability to construct rigorous, computable two-sided bounds, rather than just the estimators derived in \cref{sec:formalism} absent this additional machinery.
As discussed above and in Refs.~\cite{Wagman:2024rid,Hackett:2024xnx,Ostmeyer:2024qgu,Chakraborty:2024scw}, the unfiltered Lanczos estimators of \cref{sec:formalism} provide numerically identical outputs to other methods in certain cases and with appropriately aligned definitions.
This includes not only standard effective estimators and GEVP in one-step limits, but also Prony's method\footnote{Equivalence between unfiltered Ritz values and Prony solutions, and the Prony generalized eigenvalue method (PGEVM)~\cite{Fischer:2020bgv} with particular parameters, was first noted in Ref.~\cite{Wagman:2024rid} and proven analytically in Refs.~\cite{Ostmeyer:2024qgu,Chakraborty:2024scw}.
In contrast to the claims of Refs.~\cite{Ostmeyer:2024qgu}, the generalization of PGEVM to correlator matrices discussed in Ref.~\cite{Fischer:2020bgv} is not equivalent to block Lanczos; see \cref{app:PGEVM}.} and its block generalization as introduced in Ref.~\cite{Fleming:2023zml}. 
Further connections between these conceptually distinct but numerically coincident methods---and opportunities for improvements suggested by these connections---will be explored in future work~\cite{future}.

Besides correlator matrix analyses, the block Lanczos spurious eigenvalue filtering algorithms defined in this work may also be useful for the more traditional application of computing eigenvalues and eigenvectors of numerical matrices.
To the best of our knowledge, this work represents the first construction of the ZCW test and its block generalization, as well as the block CW test obtained by consideration of the eigenvalue-eigenvector identity.
(Z)CW tests provide alternatives to selective reorthogonalization---the present standard approach to using Lanczos methods with finite-precision arithmetic---which do not require storage and use of converged Ritz vectors.
These may now be used with block Lanczos, which could provide practical advantages in some applications.

\begin{acknowledgments}
We thank Anthony Grebe, Ryan Abbott, Will Detmold, George Fleming, Christoph Lehner, Johann Ostmeyer, Robert Perry, Dimitra Pefkou, and Fernando Romero-L\'opez for stimulating discussions and helpful comments.
We thank Robert Edwards, Rajan Gupta, Balint Jo{\'o}, Kostas Orginos, and the NPLQCD collaboration for generating the gauge-field ensembles used in this study.
Quark propagators and nucleon correlator matrices were computed by the NPLQCD Collaboration.

The calculations were performed using an allocation from the Innovative and Novel Computational Impact on Theory and Experiment (INCITE) program using the resources of the Oak Ridge Leadership Computing Facility located in the Oak Ridge National Laboratory, which is supported by the Office of Science of the Department of Energy under Contract DE-AC05-00OR22725.
This research also used resources of the National Energy Research Scientific Computing Center (NERSC), a U.S. Department of Energy Office of Science User Facility located at Lawrence Berkeley National Laboratory, operated under Contract No. DE-AC02-05CH11231. We acknowledge USQCD computing allocations and PRACE for awarding access to Marconi100 at CINECA, Italy.
This research used resources of the National Energy Research Scientific Computing Center (NERSC), a U.S. Department of Energy Office of Science User Facility operated under Contract No.~DE-AC02-05CH11231.
This research used facilities of the USQCD Collaboration, which are funded by the Office of Science of the U.S. Department of Energy.
The Chroma~\cite{Edwards:2004sx}, QUDA~\cite{Clark:2009wm,Babich:2011np,Clark:2016rdz}, QDP-JIT~\cite{6877336}, and Chromaform~\cite{chromaform} software libraries were used to generate the data in this work. Numerical analysis was performed using NumPy~\cite{harris2020array}, SciPy~\cite{2020SciPy-NMeth}, pandas~\cite{jeff_reback_2020_3715232,mckinney-proc-scipy-2010}, lsqfit~\cite{peter_lepage_2020_4037174}, gvar~\cite{peter_lepage_2020_4290884}, mpmath~\cite{mpmath}, and Mathematica~\cite{Mathematica}.
Figures were produced using matplotlib~\cite{Hunter:2007}, seaborn~\cite{Waskom2021}, and Mathematica~\cite{Mathematica}.

This manuscript has been authored by FermiForward Discovery Group, LLC under Contract No. 89243024CSC000002 with the U.S. Department of Energy, Office of Science, Office of High Energy Physics.

\end{acknowledgments}


\appendix

\section{Lanczos vector bi-orthogonality}
\label{app:ortho}

The orthogonality of Lanczos vectors within each block $j$ follows directly from the definition $\bm{\Delta}_j = \bm{\beta_j} \bm{\gamma_j}$ (\cref{eq:Delta-beta-gamma-decomp})
of $\bm{\beta}_j$ and $\bm{\gamma}_j$ in terms of the residual norm $\bm{\Delta}_j$,
\begin{equation}
  \begin{split}\label{eq:intrablock}
    \braket{v^L_{ja}}{v^R_{jb}} 
    &=  \sum_{cd} \beta^{-1}_{jac} \braket{ r^L_{jc}}{ r^R_{jd} } \gamma^{-1}_{jdb}\\
    &= [ \bm{\beta}^{-1}_j \bm{\Delta}_j \bm{\gamma^{-1}}_j ]_{ab} \\
    &= [ \bm{\beta}^{-1}_j \bm{\beta_j} \bm{\gamma_j} \bm{\gamma^{-1}}_j ]_{ab} \\
    &= \delta_{ab}
  \end{split}
\end{equation}
using \cref{eq:lr_norms} in the first equality and the definition $\braket{ r^L_{jc}}{ r^R_{jd} } = \Delta_{jcd}$ (\cref{eq:Delta-res-norm-def}) in the second equality.
Orthogonality between Lanzcos vectors from different blocks follows from the definitions of $\bm{\alpha}_j$, $\bm{\beta}_j$, and $\bm{\gamma}_j$ in Eq.~\eqref{eq:alpha_beta_gamma} as described next.

Starting with the orthogonality relation $\bigl< v^L_{ja} \big| v^R_{(j+1)b} \bigr> = 0$,
take the induction hypothesis $\bigl< v^L_{ia} \big| v^R_{i'b} \bigr> = \delta_{ii'} \delta_{ab}$ for all $i, i' \leq j$. 
The base case $j=1$ is established above.
By \cref{eq:lr_vecs,eq:lr_norms},
\begin{equation}
  \begin{split}
    \bigl< v^L_{ja} \big| v^R_{(j+1)b} \bigr>  
    &=  \sum_{cd} \left[ \delta_{cd} \bigl< v^L_{ja} \big| T \big| v^R_{jd} \bigr> \vphantom{\sum_c} 
    -  \bigl< v^L_{ja} \big| v^R_{jc} \bigr> \alpha_{jcd} 
    \right. \\ & \hspace{40pt} \left. 
    - \bigl< v^L_{ja} \big| v^R_{(j-1)c} \bigr> \beta_{jcd} \right] \gamma^{-1}_{(j+1)db}.
  \end{split}
\end{equation}
The third term vanishes and second term simplifies by the induction hypothesis, giving
\begin{equation}
  \begin{split}
    \bigl< v^L_{ja} \big| v^R_{(j+1)b} \bigr> &=  \sum_d \left[ \bigl< v^L_{ja} \big| T \big| v^R_{jd} \bigr>  - \alpha_{jad} \right] \gamma^{-1}_{(j+1)db} .
  \end{split}
\end{equation}
The orthogonality relation $\bigl< v^L_{ja} \big| v^R_{(j+1)b} \bigr> = 0$ is therefore achieved by defining $\alpha_{jab}$ as in Eq.~\eqref{eq:alpha_beta_gamma}.
The same definition leads to $\bigl< v^L_{(j+1)a} \big| v^R_{jb} \bigr> = 0$.

The corresponding orthogonality relation involving block $j-1$ is the vanishing of 
\begin{equation}
  \begin{split}
    \bigl< v^L_{(j-1)a} \big| v^R_{(j+1)b} \bigr> &=  \sum_{cd} \left[ \delta_{cd} \bigl< v^L_{(j-1)a} \big| T \big| v^R_{jd} \bigr> \vphantom{\sum_c} \right. \\
    & \hspace{10pt} \left. - \bigl< v^L_{(j-1)a} \big| v^R_{jc} \bigr> \alpha_{jcd} \right. \\
    & \hspace{10pt} \left. - \bigl< v^L_{(j-1)a} \big| v^R_{(j-1)c} \bigr> \beta_{jcd} \right] \gamma^{-1}_{(j+1)db}.
  \end{split}
\end{equation}
The second term vanishes by the first piece of the induction step, while the third term simplifies using  Eq.~\eqref{eq:intrablock}, which leads to
\begin{equation}
  \begin{split}
    \bigl< v^L_{(j-1)a} \big| v^R_{(j+1)b} \bigr> &= \sum_d \left[ \bigl< v^L_{(j-1)a} \big| T \big| v^R_{jd} \bigr> - \beta_{jad} \right] \gamma^{-1}_{(j+1)db} .
  \end{split}
\end{equation}
The orthogonality relation $\bigl< v^L_{(j-1)a} \big| v^R_{(j+1)b} \bigr> = 0$ is therefore achieved by defining $\beta_{jab}$ is in Eq.~\eqref{eq:alpha_beta_gamma}.
An analogous induction for $\bigl< v^L_{(j+1)a} \big| v^R_{(j-1)b} \bigr>$ shows this orthogonality relation is also achieved by Eq.~\eqref{eq:alpha_beta_gamma}.
Orthogonality with blocks $1,\ldots,j-2$ follows automatically from the fact that the recurrences for $T \bigl| v^R_{(j+1)a} \bigr>$ and $\bigl< v^L_{(j+1)a} \bigr| T$ only involve Lanczos vectors from blocks $j+1$, $j$, and $j-1$.
These definitions therefore ensure
\begin{equation}
  \begin{split}\label{eq:ortho}
    \braket{v^L_{ia}}{v^R_{jb}} &= \delta_{ij} \delta_{ab}.
  \end{split}
\end{equation}

Similarly applying the recursion relation to the normalization condition
\begin{equation}
  \begin{split}
    \bigl< v^L_{ja} \big| v^R_{jb} \bigr> &=  \sum_{cd} \left[ \delta_{cd} \bigl< v^L_{ja} \big| T \big| v^R_{(j-1)d} \bigr> \right. \\
    &\hspace{40pt} \left. -  \bigl< v^L_{ja} \big| v^R_{(j-1)c} \bigr> \alpha_{jcd} \right. \\
    & \hspace{40pt} \left. - \bigl< v^L_{ja} \big| v^R_{(j-2)c} \bigr> \beta_{jcd} \right] \gamma^{-1}_{jdb},
  \end{split}
\end{equation}
and noting that the second and third lines vanish by the orthogonality relations above gives
\begin{equation}
  \begin{split}
    \bigl< v^L_{ja} \big| v^R_{jb} \bigr> 
    &=  \sum_d \bigl< v^L_{ja} \big| T \big| v^R_{(j-1)d} \bigr>  \gamma^{-1}_{jdb} 
  \end{split}
\end{equation}
This is equal to $\delta_{ab}$ iff
\begin{equation}
    \gamma_{jab} = \bigl< v^L_{ja} \big| T \big| v^R_{(j-1)b} \bigr> .
\end{equation}
An analogous derivation shows that
\begin{equation}
\beta_{jab} = \bigl< v^L_{(j-1)a} \big| T \big| v^R_{jb} \bigr> .
\end{equation}
These relations, along with Eq.~\eqref{eq:lr_vecs} and Eq.~\eqref{eq:lr_norms}, can be used to prove the oblique block Lanczos three-term recurrences, Eq.~\eqref{eq:recurrence}.

\section{KPS bound}
\label{app:KPS}

In this section, we reproduce the block KPS bound of Ref.~\cite{Saad:1980}, suitably translated to the notation used in this work, and derive the form provided in the main text.
The derivation of Ref.~\cite{Saad:1980} applies for non-oblique block Lanczos only, so no $L/R$ distinctions are made and the result only applies in the infinite-statistics limit where the underlying Hermiticity of the transfer matrix is manifest in the data.
We thus necessarily assume an $r \times r$ Hermitian correlator matrix
\begin{equation}
    \mbraket{\psi_a}{T^t}{\psi_b} = C_{ab}(t) = C^*_{ba}(t)
\end{equation}
as in the rest of this work.
In the non-oblique block Lanczos algorithm there is no freedom of convention, so symmetric definitions 
apply, i.e.:
\begin{equation}
    C(0) = \bm{\beta}_{1} \bm{\gamma}_{1}
    = \bm{\beta}_{1}^2
\end{equation}
and
\begin{equation}
    \ket{v^R_j} = \ket{v^L_j} \equiv \ket{v_{1a}} = \sum_b \ket{\psi_b} \beta^{-1}_{1ba} ~.
\end{equation}

Consider the spectra of true eigenvalues $\lambda_k$ and Ritz values after $m$ steps $\lambda^{(m)}_k$, ordered such that\footnote{Note that our convention is that $\lambda_k$ are zero-indexed as $k = 0,1,\ldots$, which differs from the one-indexed convention $k = 1,2,\ldots$ of Ref.~\cite{Saad:1980}. All expressions here have been suitably translated to the zero-indexed convention.} 
\begin{equation}\begin{aligned}
    &\lambda_0 > \lambda_1 > \ldots > \lambda_\infty \, , \quad \text{and} \\
    &\lambda^{(m)}_0 > \lambda^{(m)}_1 > \ldots > \lambda^{(m)}_{r \hspace{0.5pt} m - 1} ~ .
\end{aligned}\end{equation}
Equation 3.10 of Ref.~\cite{Saad:1980} bounds the relative deviation of $\lambda^{(m)}_k$ from $\lambda_k$ as
\begin{equation}
    0 \leq \frac{\lambda_k - \lambda^{(m)}_k}{\lambda_k - \lambda_\infty} \leq \left[ \frac{K^{(m)}_k}{T_{m-k-1}(\Gamma_k)} \right]^2 || \ket{k} - \ket{\hat{x}_k} ||^2
\end{equation}
where $K^{(m)}_k$ and $\Gamma_k$ are as defined in \cref{eq:kps-gamma,eq:kps-K} above, $T_n$ is an $n$th order Chebyshev polynomial of the first kind, and $\ket{k}$ is a true eigenstate satisfying $T\ket{k} = \lambda_k \ket{k}$.
In what follows, we define and compute the final term $|| \ket{k} - \ket{\hat{x}_k} ||^2$.

As defined in Eqs.~3.1 and 3.2 of Ref.~\cite{Saad:1980}, the vector $\ket{\hat{x}_k}$ is the unique vector in the $r$-dimensional Krylov space $\mathcal{K}^{(1)}$ spanned by the initial states $\ket{v_{1a}}$ that satisfies
\begin{equation}
    \braket{\hat{x}_k}{l} = \delta_{kl}
    \label{eq:kps-xhat-implicit-def}
\end{equation}
for $l \in \{k, k+1, \ldots, k+r-1\}$, i.e.,
\begin{equation}\begin{aligned}
    \braket{\hat{x}_k}{k} &= 1 \\
    \braket{\hat{x}_k}{l} &= 0, \quad (k < l < k+r) ~ .
\end{aligned}\end{equation}
This vector may be written as\footnote{Note that there is a typographical error in the corresponding expression of Ref.~\cite{Saad:1980}.}
\begin{equation}
    \ket{\hat{x}_k} = \sum_{l=k}^{k+r-1} t^{(k)}_l \pi_1 \ket{l}
\end{equation}
in terms of $r$ coefficients $t^{(k)}_l$ and a projection operator $\pi_1$ onto $\mathcal{K}^{(1)}$; a convenient choice is
\begin{equation}
    \pi_1 = \sum_a \ket{v_{1a}} \bra{v_{1a}}
\end{equation}
noting the lack of $L/R$ distinction.
The coefficients $t^{(k)}_l$ may be obtained as the solution to the $r \times r$ linear system
\begin{equation}
    \sum_{l=k}^{k+r-1} \mbraket{j}{\pi_1}{l} t^{(k)}_l = \delta_{jk}
    \label{eq:kps-t-system}
\end{equation}
where the free index $j$ is considered only over the range $[k,k+r-1]$.
The RHS should not be thought of as a matrix but rather a length-$r$ vector with components $[1,0,\ldots,0]$.
The matrix defining the system may be rewritten as 
\begin{equation}\begin{aligned}
    \mbraket{j}{\pi_1}{l} 
    &= \sum_a \braket{j}{v_{1a}} \braket{v_{1a}}{l}  \\
    &= \sum_{abc} \braket{j}{\psi_{b}} \frac{\Omega_{1ba}}{\sqrt{\Lambda_{1a}}}
        \frac{\Omega^*_{1ca}}{\sqrt{\Lambda_{1a}}} \braket{\psi_c}{l} \\
    &= \sum_{bc} Z_{jb} [C^{-1}(0)]_{bc} Z^*_{lc}
    \\ &\equiv X_{jl}
\end{aligned}\end{equation}
recalling the definition $Z_{ja} = \braket{j}{\psi_a}$ of the overlap factors.
Although $X_{jl}$ 
can be considered over the full basis of states, \cref{eq:kps-t-system} only involves $j,l \in [k, k+r-1]$, so the relevant part is the $r \times r$ block on the diagonal beginning at $(j,l) = (k,k)$.
Denoting this block as $X_{(k,r)}$, the coefficients $t^{(k)}_l$ can be obtained by inverting
\begin{equation}
    t^{(k)}_l 
    = \sum_{j=k}^{k+r-1} \left[{X}_{(k,r)}^{-1}\right]_{lj} \, \delta_{jk} 
    = [{X}_{(k,r)}^{-1}]_{lk}
\end{equation}
i.e., $t^{(k)}$ are obtained as the first column of the inverted block ${X}^{-1}_{(k,r)}$, recalling that $j,l$ in the above expression are indexed starting at $k$.

Finally, we expand
\begin{equation}\begin{aligned}
    || \ket{k} - \ket{\hat{x}_k} ||^2 
    &= \braket{k}{k} - \braket{k}{\hat{x}_k} - \braket{\hat{x}_k}{k} + \braket{\hat{x}_k}{\hat{x}_k} \\
    &= \braket{\hat{x}_k}{\hat{x}_k} - 1
\end{aligned}\end{equation}
using $\braket{k}{\hat{x}_k} = 1$ by \cref{eq:kps-xhat-implicit-def}.
The remaining nontrivial term can be simplified as
\begin{equation}\begin{aligned}
    \braket{\hat{x}_k}{\hat{x}_k} 
    &= \sum_{j,l=k}^{k+r-1} t^{(k)*}_j \mbraket{j}{\pi_1^2}{l} t^{(k)}_l
    \\ &= \sum_{jl} t^{(k)*}_j \mbraket{j}{\pi_1}{l} t^{(k)}_l
    \\ &= \sum_j t^{(k)*}_j \delta_{jk} = t^{(k)*}_k
\end{aligned}\end{equation}
using projector idempotency $\pi_1^2 = \pi_1$ in the second equality and \cref{eq:kps-t-system} in the third.
We thus arrive at the result used in the main text, that
\begin{equation}
    t^{(k)*}_k = \left[{X}_{(k,r)}^{-1}\right]_{kk} ~,  
\end{equation}
i.e.~it is the upper-leftmost element of the inverted block ${X}_{(k,r)}^{-1}$ recalling the $k$-indexing, and that
\begin{equation}
    || \ket{k} - \ket{\hat{x}_k} ||^2 = \left[{X}_{(k,r)}^{-1}\right]_{kk} - 1 ~ .
\end{equation}

The block KPS bound reduces to the scalar version in the limit $r=1$.
As presented in Refs.~\cite{Wagman:2016bam,Hackett:2024xnx}, the scalar KPS bound reads
\begin{equation}
    0 \leq \frac{\lambda_k - \lambda^{(m)}_k}{\lambda_k - \lambda_\infty} \leq \left[ \frac{ K^{(m)}_k ~ \tan \arccos z_k}{T_{m-k-1}(\Gamma_k)} 
     \right]^2 
\end{equation}
where $z_k = \braket{k}{v_1} = Z_k / \sqrt{C(0)}$ is the normalized overlap.
The symbols $K^{(m)}_k$ and $\Gamma_k$ are identical with their scalar-case definitions when $r=1$, so it only remains to show that
\begin{equation}
    \left[{X}_{(k,r)}^{-1}\right]_{kk} - 1 ~ \xrightarrow{r=1} ~ \left[ \tan \arccos z_k \right]^2 ~.
    \label{eq:kps-scalar-reduction}
\end{equation}
Note that $[X^{-1}_{(k,1)}]_{kk}$ reduces to simply the reciprocal of the $k$th diagonal element of the matrix
\begin{equation}
    \sum_{bc} Z_{jb} [C^{-1}(0)]_{bc} Z^*_{lc}
    ~ \xrightarrow{r=1} ~
    \frac{Z_j Z_l^*}{C(0)} ~, 
\end{equation}
trivializing block indices, and so
\begin{equation}
    \left[{X}_{(k,r)}^{-1}\right]_{kk} \xrightarrow{r=1} ~ \frac{C(0)}{|Z_k|^2} = \frac{1}{z_k^2} ~ .
\end{equation}
The desired reduction \cref{eq:kps-scalar-reduction} finally follows from the identity $(\tan \arccos z)^2 = 1/z^2 - 1$.

\begin{figure*}[ht]
    \includegraphics[width=\linewidth]{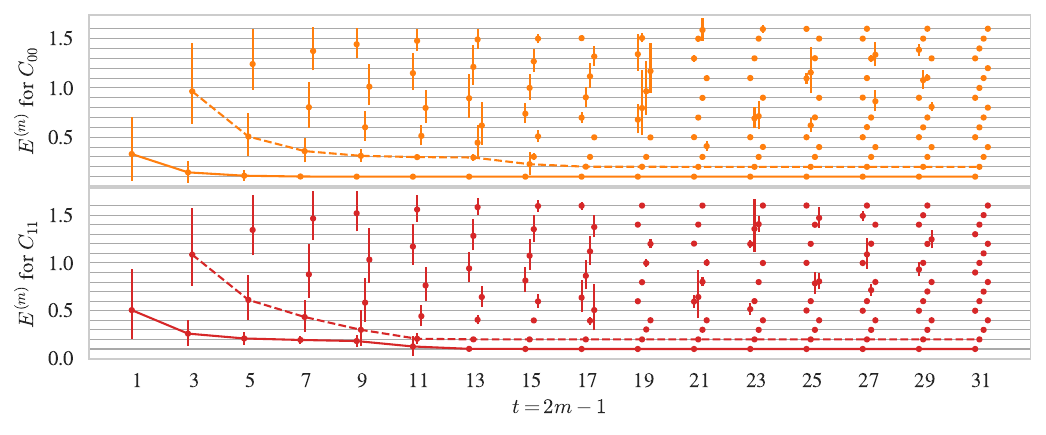}
    \caption{
        Demonstration of scalar Lanczos residual bounds for the noiseless example \cref{eq:noiseless:ex-def}, similar to \cref{fig:noiseless-spectrum-bounds}.
        Error bars are not statistical, but rather represent the extent of values allowed by the residual bound \cref{eq:block_bound}.
        The window of allowed values is computed for Ritz values and mapped through the logarithm, and thus asymmetric for energies.
        Lines connecting the lowest two energies at each $m$ are drawn to guide the eye.
    }
    \label{fig:noiseless-spectrum-scalar-bounds}
\end{figure*}


\section{Pseudo-plateaus and residual bounds for noiseless scalar Lanczos}
\label{app:scalar-res-bounds}

\Cref{sec:noiseless:spectrum} compares block and scalar Lanczos extractions of the spectrum on a noiseless example defined as \cref{eq:noiseless:ex-def}.
The results therein for scalar Lanczos---reproduced in the markers of \cref{fig:noiseless-spectrum-scalar-bounds}---raise the question of whether Lanczos can exhibit ``pseudo-plateau'' behavior.
This appendix addresses this concern.
To be concrete, consider two different definitions of a pseudo-plateau:
\begin{enumerate}[label=\alph*.,leftmargin=*]
    \item An apparent but false asymptote to a value unrelated to any energy in the true spectrum;
    \item After some finite number of iterations, an energy is converging to that of a different state than expected.
\end{enumerate}

In the example of \cref{sec:noiseless}, the two different interpolators have suppressed overlaps with different (even and odd) subsets of states.
This leads to scalar Lanczos initially yielding Ritz values which asymptote towards states with unsuppressed overlaps, then only later providing estimates for suppressed states.
For scalar Lanczos applied to $C_{11}$, this results in the largest Ritz value initially converging towards the first excited state, then later dropping to the ground state.
This satisfies definition b) above, if the largest Ritz value is expected to converge immediately to the ground state.
The general non-monotonicity of the scalar extractions results in various other examples satisfying b), given appropriate choices of expectations.
As always, cautious interpretation of results is required.

However, examination of the residual bounds shown in \cref{fig:noiseless-spectrum-scalar-bounds} provides evidence that the scalar results of \cref{sec:noiseless:spectrum} do not satisfy definition a).
For $C_{11}$, the residual bounds for the lowest energy tighten quickly; for all $m \geq 3$, the bounds indicate that the Ritz value near $E_1$ is consistent with \emph{only} the true $E_1$ and not with any other state.
Similar pictures apply for various other examples across both scalar Lanczos extractions of the spectrum.
Taken together, this suggests that the story told e.g.~by simply examining the sequence of lowest-lying energy estimates in the $C_{11}$ analysis---that the ground-state energy estimator initially converges to the wrong value---is not the best interpretation of these dynamics.
A better one is simply that scalar Lanczos provides no estimate at all for the ground-state energy until $m=6$, at which point a new, separate state which converges to the ground state appears.
Although state identification will always be an ambiguity, the value of the estimators in this example reliably correspond to physical values.
This is in contrast with the potentially more dangerous behavior indicated by definition a), wherein near cancellations result in false asymptotes and analyses which yield values with no physical meaning.

\section{Correlator matrix reality}
\label{app:real}

\subsection{$C$, $P$, and $R_{2\pi}$ symmetry}

Charge conjugation ($C$) transformations are defined for LQCD gauge fields as
\begin{equation}
    U(C) U_\mu(x) U(C)^\dagger = U_\mu(x)^*, 
\end{equation}
and for quark fields as
\begin{equation}
\begin{split}
    U(C) q(x) U(C)^\dagger &= C \overline{q}^T \\
    U(C) \overline{q}(x) U(C)^\dagger &= q^T C^\dagger,
    \end{split}
\end{equation}
where $C = \gamma_4 \gamma_2$ satisfies $C^T = C^\dagger = C^{-1} = - C$.
The Dirac-Pauli basis in which $\gamma_4 = \text{diag}(1,1,-1,-1)$ is used throughout this section; see Ref.~\cite{Detmold:2024ifm} for details.
Most LQCD actions possess a $C$ symmetry in the sense that the action is invariant under this transformation, and the existence of such as symmetry will be assumed throughout this work.
For more on $C$ transformation properties in LQCD see Refs.~\cite{Leinweber:1990dv,Melnitchouk:2002eg,Rinaldi:2019thf}.
The quark propagator therefore transforms as
\begin{equation}
\begin{split}
    U(C) S(x,y) U(C)^\dagger  &= C S(y,x)^T C^\dagger \\
    &= C \gamma_5 S(x,y)^* \gamma_5 C^\dagger,
\end{split}
\end{equation}
where $\gamma_5$-Hermiticity $S(y,x) = \gamma_5 S(x,y)^\dagger \gamma_5$ has been assumed in going to the second line.
This can be further re-expressed as
\begin{equation}
\begin{split}
    U(C) S(x,y) U(C)^\dagger  &= (iC \gamma_5) S(x,y)^* (i C \gamma_5)^\dagger \\
    &= (i \gamma_1 \gamma_3) S(x,y)^* (i \gamma_1 \gamma_3)^\dagger,
\end{split}
\end{equation}
where $\gamma_5 = \gamma_1 \gamma_2 \gamma_3 \gamma_4$ has been used in the last line.

In this form, the action of $C$ on quark propagators can be recognized as complex conjugation in conjugation with the action of a $2\pi$ rotation as follows.
The spinor representation $D(\vec{\omega})$ of a rotation---an element of the double cover of the cubic group $R(\vec{\omega}) \in O_h^D$---whose magnitude and direction are given by the vector $\vec{\omega}$ is
\begin{equation}
    D(\vec{\omega}) = \exp\left( -\frac{1}{8} \sum_k \omega_k \epsilon_{ijk} [\gamma_i,\gamma_j] \right).
\end{equation}
For more on cubic symmetry in LQCD see Refs.~\cite{Mandula:1982us,Mandula:1983ut,Basak:2005aq,Basak:2005ir,Luu:2011ep,Thomas:2011rh,Morningstar:2013bda,Prelovsek:2016iyo,Amarasinghe:2021lqa,Detmold:2024ifm}.
For the vector $\vec{\omega} = (0,\omega_2,0)$, this becomes
\begin{equation}
\begin{split}
    D(\omega_2 \hat{e}_2) &= \exp\left( -\frac{\omega_2}{4} \gamma_1 \gamma_3 \right) \\
    &= \cos\left( -\frac{\omega_2}{4}\right) - i  \gamma_1 \gamma_3 \sin\left( -\frac{\omega_2}{4}\right).
    \end{split}
\end{equation}
Choosing $\vec{\omega}_c = (0,-2\pi,0)$ then gives
\begin{equation}
    D(\vec{\omega}_c) = -i \gamma_1 \gamma_3 .
\end{equation}
Therefore, define $R_{2\pi} \in O_h^D$ as the corresponding $-2\pi \hat{e}_2$ rotation operator and  $U(R)$ as the Hilbert space operator whose action is given by $U(R_{2\pi}) S(x,y) U(R_{2\pi})^\dagger = D(\vec{\omega}_c) S(x,y) D(\vec{\omega}_c)^\dagger$.
We assume that $R_{2\pi}$ is a symmetry of the action.
The combined symmetry $CR_{2\pi}$ acts on quark propagators as
\begin{equation}
\begin{split}
    &U(CR_{2\pi}) S(x,y) U(CR_{2\pi})^\dagger \\
    &= (i \gamma_1 \gamma_3)  D(\vec{\omega}_c) S(x,y)^* D(\vec{\omega}_c)^\dagger (i \gamma_1 \gamma_3)^\dagger  \\
    &= S(x,y)^*.
\end{split}
\end{equation}
Since $R$ acts trivially on the gauge field, $U(C R_{2\pi}) U_\mu(x) U(C R_{2\pi})^\dagger = U_\mu(x)^*$.

The parity transformation $P \in O_h^D$ has the spinor representation $D(P) = \gamma_4$ and transforms propagators by
\begin{equation}
    U(P) S(x,y) U(P)^\dagger = \gamma_4  S((-\vec{x},x_4),(-\vec{y},y_4)) \gamma_4.
\end{equation}
The action of parity on the gauge field is
\begin{equation}
    U(P) U_{\mu}(x) U(P)^\dagger =  U_{-\mu}((-\vec{x},x_4)),
\end{equation}
where $U_{-\mu}(x) = U_\mu(x - \hat{e}_\mu)^\dagger$.
It will be convenient below to denote the set of gauge fields at time $t$ by 
\begin{equation}
    U_{(t)} = \{ U_\mu(x)\, | \, x_4 = t\},
\end{equation}
and their parity conjugates by
\begin{equation}
    U_{(t)}^P = \{ U(P) U_\mu(x) U(P)^\dagger \, | \, x_4 = t\}.
\end{equation}
Further denote the set of quark propagators starting and ending at time $t$ by
\begin{equation}
    S_{(t)} = \{ S(x,y)\, | \, x_4=y_4=t \},
\end{equation}
and their parity conjugates by
\begin{equation}
    S_{(t)}^P = \{ U(P) S(x,y) U(P)^\dagger \, | \, x_4=y_4=t \}.
\end{equation}

\subsection{Correlator matrix definitions}

Consider a generic correlator matrix
\begin{equation}
    \begin{split}
        C_{ab}(t) = \left< \chi_a(t) \overline{\psi}_b(0) \right>,
    \end{split}
\end{equation}
with source interpolating operator
\begin{equation}
    \begin{split}\label{eq:source}
        \overline{\psi}_b(0) &= \sum_{\vec{y}_1,\ldots,\vec{y}_{K}} \sum_{\vec{v}_1,\ldots,\vec{v}_{\overline{L}}}  e^{-i \left[ \sum_{i}  \vec{p}_i\cdot\vec{y}_i + \sum_{j}  \vec{k}_j\cdot\vec{v}_j \right]}  \\
        &\hspace{10pt} \times \overline{q}(\vec{y}_1,0) \cdots \overline{q}(\vec{y}_{K},0) \\
        &\hspace{10pt} \times q(\vec{v}_1,0)^T \cdots q(\vec{v}_{L},0)^T \\
        &\hspace{10pt} \times \psi_{b}(U_{(0)},\vec{y}_{1},\ldots,\vec{y}_{K},\vec{v}_1,\ldots,\vec{v}_{L}),
    \end{split}
\end{equation}
where $\psi_{b}(U_{(0)},\vec{y}_1,\ldots,\vec{y}_{K},\vec{v}_1,\ldots,\vec{v}_{L})$ is a rank $K+L$ spin-color-flavor tensor that can depend on the gauge field at the source timeslice, $U_{(0)}$.
The corresponding sink interpolating operator is
\begin{equation}
    \begin{split}\label{eq:sink}
        \chi_a(t) &= \sum_{\vec{x}_1,\ldots,\vec{x}_{K'}}  \sum_{\vec{u}_1,\ldots,\vec{u}_{L'}}  e^{i \left[ \sum_{i}  \vec{p}'_i\cdot\vec{x}_i + \sum_{j}  \vec{k}'_j\cdot\vec{u}_j \right]}  \\
        &\hspace{10pt} \times \chi_{a}(U_{(t)},\vec{x}_1,\ldots,\vec{x}_{K'},\vec{u}_1,\ldots,\vec{u}_{L'})^\dagger \\
        &\hspace{10pt} \times q(\vec{x}_{1},0)^T \cdots q(\vec{x}_{K'},0)^T \\
        &\hspace{10pt} \times \overline{q}(\vec{u}_{1},0) \cdots \overline{q}(\vec{u}_{L'},0),
    \end{split}
\end{equation}
where $\chi_{a}(U_{(t)},\vec{x}_1,\ldots,\vec{x}_{K'},\vec{u}_1,\ldots,\vec{u}_{L'})^\dagger $ is a rank $K' + L'$ spin-color-flavor tensor that can depend on the gauge field at the sink timeslice, $U_{(t)}$.
Note that time translation invariance can be used to trivially generalize this result to non-zero source times.
The total momentum carried by quark and antiquark fields in the initial and final state is therefore given by
\begin{equation}
\begin{split}
    \vec{P}_q &= \sum_{i=1}^K \vec{p}_i + \sum_{j=1}^{L} \vec{k}_j, \\
    \vec{P}'_q &= \sum_{i=1}^{K'} \vec{p}'_i + \sum_{j=1}^{L'} \vec{k}'_j .
    \end{split}
\end{equation}
The spatial translation properties of $\psi_b$ and $\chi_a$ encode the momenta $\vec{P}_g$ and $\vec{P}'_g$ carried by gluons in the initial and final states, respectively, 
and should be chosen to ensure total momentum conservation $\vec{P}_q + \vec{P}_g = \vec{P}'_q + \vec{P}'_g$.

The total number of quark fields, $F \equiv K + K'$, must equal the total number of antiquarks, $L+L'$, for all nonvanishing correlators in an $R_{2\pi}$-symmetric theory.
Wick's theorem allows the $F$ quark/antiquark fields in $C_{ab}(t)$ to be replaced by linear combinations of terms with $F$ quark propagators with coefficients given by products of permutation signatures.
The $F$ quark propagators can be partitioned into sets of $N$ quark propagators connecting source/sink sites, $M$ quark propagators connecting source/source sites, and $M'$ propagators connecting sink/sink sites, where $N + M + M' = F$.
Different such partitions
\begin{equation}
    \sigma \equiv \{(N, M, M') \in \mathbb{Z}_3 \ |\ N + M + M' = F \},
\end{equation}
of $F$ correspond to different linear combinations of Wick contractions.
This allows $C_{ab}(t)$ to be expressed as
\begin{equation}\label{eq:wick}
    C_{ab}(t) = \sum_{\sigma} \text{sig}(\sigma) \sum_{p \in S_N} \text{sig}(p) \, C^{\sigma p}_{ab}(t),
\end{equation}
where $\text{sig}(\sigma)$ is the signature of the permutation required to reorder the quark fields into the appropriate normal ordering for $N$ source/sink, $M$ source/source, and $M'$ sink/sink quark propagators,  $p \in S_N$ is an element of the permutation group of $N$ objects, $\text{sig}(p)$ is its signature, and
\begin{equation}
\begin{split}
    C^{\sigma p}_{ab}(t) &= \sum_{\vec{x}_1,\ldots,\vec{x}_{K'}}  \sum_{\vec{u}_1,\ldots,\vec{u}_{L'}} \sum_{\vec{y}_1,\ldots,\vec{y}_{K}} \sum_{\vec{v}_1,\ldots,\vec{v}_{L}} \\
    &\hspace{10pt} \times e^{i \left[ \sum_{i}  \vec{p}'_i\cdot\vec{x}_i + \sum_{j}  \vec{k}'_j\cdot\vec{u}_j \right]}  e^{-i \left[ \sum_{i}  \vec{p}_i\cdot\vec{y}_{i} + \sum_{j}  \vec{k}_j\cdot\vec{v}_{j} \right]}  \\
    &\hspace{10pt} \times \bigl< \Xi_{a}(U_{(t)},S_{(t)},\vec{x}_1,\ldots,\vec{x}_{K'},\vec{u}_1,\ldots,\vec{u}_{L'})^\dagger \\
    &\hspace{25pt} \times \bigotimes_{i,j=1}^N S((\vec{x}_{i},t),(\vec{y}_{p(j)},0))  \\
    &\hspace{25pt} \times \Psi_{b}(U_{(0)},S_{(0)},\vec{y}_1,\ldots,\vec{y}_{K},\vec{v}_1,\ldots,\vec{v}_{L}) \bigr>,
    \end{split}
\end{equation}
where the spin-color-flavor tensor indices of the quark propagator product corresponding to the $\overline{q}(\vec{y}_i)$ fields are permuted into the same order as the source spatial index labels.
Here, the source wavefunction including source/source propagator factors is defined by
\begin{equation}
    \begin{split}
        &\Psi_{b}(U_{(0)},S_{(0)},\vec{y}_1,\ldots,\vec{y}_{K},\vec{v}_1,\ldots,\vec{v}_{L}) \\
        &\equiv \sum_{p \in S_M} \text{sig}(p) \bigotimes_{i=1}^M S((\vec{v}_{i},0),(\vec{y}_{N+p(i)},0)) \\
        &\hspace{10pt} \times \psi_{b}(U_{(0)},\vec{y}_{1},\ldots,\vec{y}_{K},\vec{v}_{1},\ldots,\vec{v}_{L}),
    \end{split}
\end{equation}
and the corresponding sink wavefunction is defined by
\begin{equation}
    \begin{split}
        &\Xi_{a}(U_{(t)},S_{(t)},\vec{x}_1,\ldots,\vec{x}_{K'},\vec{u}_1,\ldots,\vec{u}_{L'})^\dagger \\
        &\equiv \chi_{a}(U_{(t)},\vec{x}_1,\ldots,\vec{x}_{K'},\vec{u}_1,\ldots,\vec{u}_{L'})^\dagger \\
        &\hspace{10pt} \times \sum_{p \in S_{M'}} \text{sig}(p) \bigotimes_{i=1}^{M'} S((\vec{x}_{N+i},0),(\vec{u}_{p(i)},0)).
    \end{split}
\end{equation}
Since the coefficients appearing Eq.~\eqref{eq:wick} are $\pm 1$ permutation signatures, it is sufficient to prove that $C^{\sigma p}_{ab}(t) \in \mathbb{R}$ for arbitrary $\sigma, p$ in order to show that $C_{ab}(t) \in \mathbb{R}$.

\subsection{Correlator matrix transformations}

Applying a $CR_{2\pi}$ transformation to $C^{\sigma p}_{ab}(t)$ gives
\begin{equation}
\begin{split}
     &U(CR_{2\pi}) C_{ab}^{\sigma p}(t) U(CR_{2\pi})^\dagger \\
    &= \sum_{\vec{x}_1,\ldots,\vec{x}_{K'}}  \sum_{\vec{u}_1,\ldots,\vec{u}_{L'}} \sum_{\vec{y}_1,\ldots,\vec{y}_{K}} \sum_{\vec{v}_1,\ldots,\vec{v}_{L}}\\
    &\hspace{10pt} \times e^{i \left[ \sum_{i}  \vec{p}'_i\cdot\vec{x}_i + \sum_{j}  \vec{k}'_j\cdot\vec{u}_j \right]}  e^{-i \left[ \sum_{i}  \vec{p}_i\cdot\vec{y}_i + \sum_{j}  \vec{k}_j\cdot\vec{v}_j \right]} \\
    &\hspace{10pt} \times \bigl< \Xi_{a}(U_{(t)}^*,S_{(t)}^*,\vec{x}_1,\ldots,\vec{x}_{N},\vec{u}_1,\ldots,\vec{u}_{M'})^\dagger \\
    &\hspace{25pt} \times \bigotimes_{i,j=1}^N  S((\vec{x}_{i},t),(\vec{y}_{p(j)},0))^*  \\
    &\hspace{25pt} \times \Psi_{b}(U_{(0)}^*,S_{(0)}^*,\vec{y}_1,\ldots,\vec{y}_{N},\vec{v}_1,\ldots,\vec{v}_{M}) \bigr>.
    \end{split}
\end{equation}
Further including a $P$ transformation gives
\begin{equation}
\begin{split}
     &U(CPR_{2\pi}) C_{ab}^{\sigma p}(t) U(CPR_{2\pi})^\dagger \\
    &= \sum_{\vec{x}_1,\ldots,\vec{x}_{K'}}  \sum_{\vec{u}_1,\ldots,\vec{u}_{L'}} \sum_{\vec{y}_1,\ldots,\vec{y}_{K}} \sum_{\vec{v}_1,\ldots,\vec{v}_{L}} \\
    &\hspace{10pt} \times e^{i \left[ \sum_{i}  \vec{p}'_i\cdot\vec{x}_i + \sum_{j}  \vec{k}'_j\cdot\vec{u}_j \right]}  e^{-i \left[ \sum_{i}  \vec{p}_i\cdot\vec{y}_i + \sum_{j}  \vec{k}_j\cdot\vec{v}_j \right]} \\
    &\hspace{10pt} \times \bigl< \Xi_{a}(U_{(t)}^{P*},S_{(t)}^{P*},\vec{x}_1,\ldots,\vec{x}_{N},\vec{u}_1,\ldots,\vec{u}_{M'})^\dagger \gamma_4 \\
    &\hspace{25pt} \times \bigotimes_{i,j=1}^N S((-\vec{x}_{i},t),(-\vec{y}_{p(j)},0))^*  \\
    &\hspace{25pt} \times \gamma_4 \Psi_{b}(U_{(0)}^{P*},S_{(0)}^{P*},\vec{y}_1,\ldots,\vec{y}_{N},\vec{v}_1,\ldots,\vec{v}_{M}) \bigr>.
    \end{split}
\end{equation}
Changing summation variables to reverse the orientations of all momenta, e.g.~$\vec{x}_i \rightarrow -\vec{x}_i$, allows this to be expressed as
\begin{equation}
\begin{split}
    &U(CPR_{2\pi}) C_{ab}^{\sigma p}(t) U(CPR_{2\pi})^\dagger \\
    &= \sum_{\vec{x}_1,\ldots,\vec{x}_{K'}}  \sum_{\vec{u}_1,\ldots,\vec{u}_{L'}} \sum_{\vec{y}_1,\ldots,\vec{y}_{K}} \sum_{\vec{v}_1,\ldots,\vec{v}_{L}} \\
    &\hspace{10pt} \times \left( e^{i \left[ \sum_{i}  \vec{p}'_i\cdot\vec{x}_i + \sum_{j}  \vec{k}'_j\cdot\vec{u}_j \right]}  e^{-i \left[ \sum_{i}  \vec{p}_i\cdot\vec{y}_i + \sum_{j}  \vec{k}_j\cdot\vec{v}_j \right]}\right)^* \\
    &\hspace{10pt} \times \bigl< \Xi_{a}(U_{(t)}^{*}, \gamma_4 S_{(t)}^{*} \gamma_4,-\vec{x}_1,\ldots,-\vec{x}_{N},-\vec{u}_1,\ldots,-\vec{u}_{M'})^\dagger \\
    &\hspace{25pt} \times \bigotimes_{i,j=1}^N \gamma_4  S((\vec{x}_{i},t),(\vec{y}_{p(j)},0))^*  \gamma_4  \\
    &\hspace{25pt} \times\Psi_{b}(U_{(0)}^{*}, \gamma_4 S_{(0)}^{*} \gamma_4,-\vec{y}_1,\ldots,-\vec{y}_{N},-\vec{v}_1,\ldots,-\vec{v}_{M}) \bigr>,
    \end{split}
\end{equation}
where $\gamma_4 S_{(t)}^* \gamma_4 \equiv  \{ \gamma_4 S(x,y)^* \gamma_4 \, | \, x_4=y_4=t \}$.

To complete the proof of correlator matrix reality, it is convenient to decompose quark propagators into positive- and negative-parity components using the parity projectors $(1 \pm \gamma_4)/2$.
Propagators can be decomposed as
\begin{equation}
    S(x,y) = \sum_{\rho,\rho' = \pm 1} S_{\rho' \rho}(x,y),
\end{equation}
where
\begin{equation}
    S_{\rho' \rho}(x,y) \equiv \left( \frac{1 + \rho' \gamma_4}{2} \right) S(x,y) \left( \frac{1 + \rho \gamma_4}{2} \right).
\end{equation}
Wavefunctions can similarly be decomposed as
\begin{equation}
    \begin{split}
         \Xi_{a}^T = \sum_{\vec{\rho} \in Z_2^N} \Xi_{a \vec{\rho}}^T, \hspace{20pt} \Psi_{b} &= \sum_{\vec{\rho} \in Z_2^N} \Psi_{b \vec{\rho}},
    \end{split}
\end{equation}
where
\begin{equation}
    \begin{split}
       &\Xi_{a \vec{\rho}}^T  \equiv \bigotimes_{i=1}^N \left( \frac{1 + \rho_i \gamma_4}{2} \right) \Xi_{a}^T, \\
       &\Psi_{b \vec{\rho}} \equiv \bigotimes_{i=1}^N \left( \frac{1 + \rho_i \gamma_4}{2} \right) \Psi_{b } .
    \end{split}
\end{equation}
Since $\left( \frac{1 + \gamma_4}{2} \right) \left( \frac{1 - \gamma_4}{2} \right) = 0$,
this decomposition allows the original correlation function to be expressed as
\begin{equation}
\begin{split}
    C^{\sigma p}_{ab}(t) &= \sum_{\vec{\rho},\vec{\rho'} \in Z_2^N} \sum_{\vec{x}_1,\ldots,\vec{x}_{K'}}  \sum_{\vec{u}_1,\ldots,\vec{u}_{L'}} \sum_{\vec{y}_1,\ldots,\vec{y}_{K}} \sum_{\vec{v}_1,\ldots,\vec{v}_{L}} \\
    &\hspace{10pt} \times e^{i \left[ \sum_{i}  \vec{p}'_i\cdot\vec{x}_i + \sum_{j}  \vec{k}'_j\cdot\vec{u}_j \right]}  e^{-i \left[ \sum_{i}  \vec{p}_i\cdot\vec{y}_i + \sum_{j}  \vec{k}_j\cdot\vec{v}_j \right]} \\
    &\hspace{10pt} \times \bigl< \Xi_{a\vec{\rho}'}(U_{(t)},S_{(t)},\vec{x}_1,\ldots,\vec{x}_{N},\vec{u}_1,\ldots,\vec{u}_{2M'})^\dagger \\
    &\hspace{25pt} \times \bigotimes_{i,j=1}^N  S_{\rho'_i \rho_j}((\vec{x}_{i},t),(\vec{y}_{p(j)},0))  \\
    &\hspace{25pt} \times \Psi_{b\vec{\rho}}(U_{(0)},S_{(0)},\vec{y}_1,\ldots,\vec{y}_{N},\vec{v}_1,\ldots,\vec{v}_{2M}) \bigr>.
    \end{split}
\end{equation}
Applying the same decomposition to the transformed correlation function gives
\begin{equation}
\begin{split}
    &U(CPR_{2\pi}) C_{ab}^{\sigma p}(t) U(CPR_{2\pi})^\dagger \\
    &= \sum_{\vec{\rho},\vec{\rho'} \in Z_2^N} \sum_{\vec{x}_1,\ldots,\vec{x}_{K'}}  \sum_{\vec{u}_1,\ldots,\vec{u}_{L'}} \sum_{\vec{y}_1,\ldots,\vec{y}_{K}} \sum_{\vec{v}_1,\ldots,\vec{v}_{L}} \\
    &\hspace{10pt} \times \left( e^{i \left[ \sum_{i}  \vec{p}'_i\cdot\vec{x}_i + \sum_{j}  \vec{k}'_j\cdot\vec{u}_j \right]}  e^{-i \left[ \sum_{i}  \vec{p}_i\cdot\vec{y}_i + \sum_{j}  \vec{k}_j\cdot\vec{v}_j \right]} \right)^* \\
    &\hspace{10pt} \times \bigl< \Xi_{a\vec{\rho}'}(U_{(t)}^{*}, \gamma_4 S_{(t)}^{*} \gamma_4,-\vec{x}_1,\ldots,-\vec{x}_{N},-\vec{u}_1,\ldots,-\vec{u}_{M'})^\dagger \gamma_4 \\
    &\hspace{25pt} \times \bigotimes_{i,j=1}^N  S_{\rho'_i \rho_j}((\vec{x}_{i},t),(\vec{y}_{p(j)},0))^*  \\
    &\hspace{25pt} \times \gamma_4 \Psi_{b\vec{\rho}}(U_{(0)}^{*}, \gamma_4 S_{(0)}^{*} \gamma_4,-\vec{y}_1,\ldots,-\vec{y}_{N},-\vec{v}_1,\ldots,-\vec{v}_{M}) \bigr>.
    \end{split}
\end{equation}
The action of $\gamma_4$ on $\chi_{a\vec{\rho}'}$ and $\psi_{b\vec{\rho}}$ then simply results in sign factors, 
\begin{equation}
\begin{split}
    &U(CPR_{2\pi}) C_{ab}^{\sigma p}(t) U(CPR_{2\pi})^\dagger \\
    &= \sum_{\vec{\rho},\vec{\rho'} \in Z_2^N} \det(\vec{\rho}'\vec{\rho}^T) \sum_{\vec{x}_1,\ldots,\vec{x}_{K'}}  \sum_{\vec{u}_1,\ldots,\vec{u}_{L'}} \sum_{\vec{y}_1,\ldots,\vec{y}_{K}} \sum_{\vec{v}_1,\ldots,\vec{v}_{L}} \\
    &\hspace{10pt} \times \left( e^{i \left[ \sum_{i}  \vec{p}'_i\cdot\vec{x}_i + \sum_{j}  \vec{k}'_j\cdot\vec{u}_j \right]}  e^{-i \left[ \sum_{i}  \vec{p}_i\cdot\vec{y}_i + \sum_{j}  \vec{k}_j\cdot\vec{v}_j \right]} \right)^* \\
    &\hspace{10pt} \times \bigl< \Xi_{a\vec{\rho}'}(U_{(t)}^{*}, \gamma_4 S_{(t)}^{*} \gamma_4,-\vec{x}_1,\ldots,-\vec{x}_{N},-\vec{u}_1,\ldots,-\vec{u}_{M'})^\dagger \\
    &\hspace{25pt} \times \bigotimes_{i,j=1}^N  S_{\rho'_i \rho_j}((\vec{x}_{i},t),(\vec{y}_{p(j)},0))^*  \\
    &\hspace{25pt} \times \Psi_{b\vec{\rho}}(U_{(0)}^{*}, \gamma_4 S_{(0)}^{*} \gamma_4,-\vec{y}_1,\ldots,-\vec{y}_{N},-\vec{v}_1,\ldots,-\vec{v}_{M}) \bigr>.
    \end{split}
\end{equation}
The same decomposition can be applied to the propagators appearing in disconnected quark loops.
Defining
\begin{equation}
\begin{split}
  S_{(t)}^{\vec{\xi}' \widetilde{\vec{\xi}}'} &\equiv \{ S_{{\xi'_i \widetilde{\xi}'_i}}(x,y) \, | \, x_4=y_4=t,\ i=1,M' \}, \\
  S_{(0)}^{\vec{\xi} \widetilde{\vec{\xi}}} &\equiv \{ S_{{\xi_i \widetilde{\xi}_i}}(x,y) \, | \, x_4=y_4=t,\ i=1,M \}, 
  \end{split}
\end{equation}
the original correlation function can be expressed as
\begin{equation}
\begin{split}
  C^{\sigma p}_{ab}(t) &= \sum_{\vec{\rho},\vec{\rho'} \in Z_2^N} \sum_{\vec{x}_1,\ldots,\vec{x}_{K'}}  \sum_{\vec{u}_1,\ldots,\vec{u}_{L'}} \sum_{\vec{y}_1,\ldots,\vec{y}_{K}} \sum_{\vec{v}_1,\ldots,\vec{v}_{L}} \\
  &\hspace{10pt}\sum_{\vec{x}_1,\ldots,\vec{x}_{N}}  \sum_{\vec{u}_1,\ldots,\vec{u}_{M'}} \sum_{\vec{y}_1,\ldots,\vec{y}_{N}} \sum_{\vec{v}_1,\ldots,\vec{v}_{M}} \\
    &\hspace{10pt} \times e^{i \left[ \sum_{i}  \vec{p}'_i\cdot\vec{x}_i + \sum_{j}  \vec{k}'_j\cdot\vec{u}_j \right]}  e^{-i \left[ \sum_{i}  \vec{p}_i\cdot\vec{y}_i + \sum_{j}  \vec{k}_j\cdot\vec{v}_j \right]} \\
    &\hspace{10pt} \times \bigl< \Xi_{a\vec{\rho}'}(U_{(t)}, S^{\vec{\xi}'\widetilde{\vec{\xi}}'}_{(t)},\vec{x}_1,\ldots,\vec{x}_{N},\vec{u}_1,\ldots,\vec{u}_{2M'})^\dagger \\
    &\hspace{25pt} \times \bigotimes_{i,j=1}^N  S_{\rho'_i \rho_j}((\vec{x}_{i},t),(\vec{y}_{p(j)},0))  \\
    &\hspace{25pt} \times \Psi_{b\vec{\rho}}(U_{(0)}, S^{\vec{\xi}\widetilde{\vec{\xi}}}_{(0)},\vec{y}_1,\ldots,\vec{y}_{N},\vec{v}_1,\ldots,\vec{v}_{2M}) \bigr>.
    \end{split}
\end{equation}
Applying the same decomposition to the transformed correlation function and simplifying the action of $\gamma_4$,
\begin{equation}
\begin{split}
    &U(CPR_{2\pi}) C_{ab}^{\sigma p}(t) U(CPR_{2\pi})^\dagger \\
    &= \sum_{\vec{\rho},\vec{\rho'} \in Z_2^N}  \sum_{\xi_1,\ldots,\xi_M} \sum_{\widetilde{\xi}_1,\ldots,\widetilde{\xi}_M} \sum_{\xi'_1,\ldots,\xi'_M} \sum_{\widetilde{\xi'}_1,\ldots,\widetilde{\xi'}_M}  \Pi_S \Pi'_S \\
    &\hspace{10pt} \sum_{\vec{x}_1,\ldots,\vec{x}_{K'}}  \sum_{\vec{u}_1,\ldots,\vec{u}_{L'}} \sum_{\vec{y}_1,\ldots,\vec{y}_{K}} \sum_{\vec{v}_1,\ldots,\vec{v}_{L}} \\
    &\hspace{10pt} \times \left( e^{i \left[ \sum_{i}  \vec{p}'_i\cdot\vec{x}_i + \sum_{j}  \vec{k}'_j\cdot\vec{u}_j \right]}  e^{-i \left[ \sum_{i}  \vec{p}_i\cdot\vec{y}_i + \sum_{j}  \vec{k}_j\cdot\vec{v}_j \right]} \right)^* \\
    &\hspace{10pt} \times \bigl< \Xi_{a\vec{\rho}'}(U_{(t)}^{*}, [S^{\vec{\xi}'\widetilde{\vec{\xi}}'}_{(t)}]^* ,-\vec{x}_1,\ldots,-\vec{x}_{N},-\vec{u}_1,\ldots,-\vec{u}_{M'})^\dagger \\
    &\hspace{25pt} \times \bigotimes_{i,j=1}^N  S_{\rho'_i \rho_j}((\vec{x}_{i},t),(\vec{y}_{p(j)},0))^*  \\
    &\hspace{25pt} \times \Psi_{b\vec{\rho}}(U_{(0)}^{*}, [S^{\vec{\xi}\widetilde{\vec{\xi}}}_{(0)}]^* ,-\vec{y}_1,\ldots,-\vec{y}_{N},-\vec{v}_1,\ldots,-\vec{v}_{M}) \bigr>,
    \end{split}
\end{equation}
where the total parities of all source/sink quark propagator factors are given by
\begin{equation}
  \begin{split}
    \Pi_S = \prod_{i=i}^N \rho_i \prod_{j=1}^M \xi_j \widetilde{\xi}_j  , \hspace{20pt}
    \Pi'_S = \prod_{i=i}^N \rho'_i\prod_{j=1}^{M'}  \xi'_j \widetilde{\xi}'_j. 
    \end{split}
\end{equation}
It is therefore sufficient for the source and sink wavefunction to satisfy
\begin{equation}\label{eq:real_source}
\begin{split}
    &\Psi_{b \vec{\rho}}(U_{(0)}^{*}, [S^{\vec{\xi}\widetilde{\vec{\xi}}}_{(0)}]^* ,-\vec{y}_1,\ldots,-\vec{y}_{K},-\vec{v}_1,\ldots,-\vec{v}_{L}) \\
    &= \Pi_S \Psi_{b \vec{\rho} }(U_{(0)},S^{\vec{\xi}\widetilde{\vec{\xi}}}_{(0)},\vec{y}_1,\ldots,\vec{y}_{K},\vec{v}_1,\ldots,\vec{v}_{L})^*,
    \end{split}
\end{equation}
and
\begin{equation}\label{eq:real_sink}
    \begin{split}
        &\Xi_{a\vec{\rho}'}(U_{(t)}^{*}, [S^{\vec{\xi}'\widetilde{\vec{\xi}}'}_{(t)}]^* ,-\vec{x}_1,\ldots,-\vec{x}_{K'},-\vec{u}_1,\ldots,-\vec{u}_{L'})^\dagger \\
        &= \Pi'_S \Xi_{a \vec{\rho}'}(U_{(t)}, S^{\vec{\xi}'\widetilde{\vec{\xi}}'}_{(t)}, \vec{x}_1,\ldots,\vec{x}_{K'},\vec{u}_1,\ldots,\vec{u}_{L'})^T,
    \end{split}
\end{equation}
in order to guarantee that
\begin{equation}
    U(CPR_{2\pi}) C_{ab}^{\sigma p}(t) U(CPR_{2\pi})^\dagger = C_{ab}^{\sigma p}(t)^*.
\end{equation}
Because there are path integral changes of variables that induce $CPR_{2\pi}$ transformations and they have equal weight by $CPR_{2\pi}$ symmetry of the action, averaging over all quark-and-gluon field configurations leads to the averaging of $C_{ab}^{\sigma p}(t)$ with $U(CPR_{2\pi}) C_{ab}^{\sigma p}(t) U(CPR_{2\pi})^\dagger = C_{ab}^{\sigma p}(t)^*$.
The imaginary part of $C_{ab}^{\sigma p}(t)$ therefore vanishes, or in other words Eqs.~\eqref{eq:real_source}-\eqref{eq:real_sink} are sufficient to ensure that $C_{ab}^{\sigma p}(t) \in \mathbb{R}$ and therefore that $C_{ab}(t) \in \mathbb{R}$.

\subsection{Correlator matrix reality conditions}

It is straightforward to obtain corresponding reality conditions for the quark-field coefficient functions appearing in the interpolating operator definitions Eqs.~\eqref{eq:source}-\eqref{eq:sink}.
Define a quark-field-parity decomposition of these wavefunctions by
\begin{equation}
    \begin{split}
         \chi_{a}^T = \sum_{\vec{\tau} \in Z_2^{K'+L'}} \chi_{a \vec{\tau}}^T, \hspace{20pt} \psi_{b} &= \sum_{\vec{\tau} \in Z_2^{K+L}} \psi_{b \vec{\tau}},
    \end{split}
\end{equation}
where
\begin{equation}
    \begin{split}
       &\chi_{a \vec{\tau}}^T  \equiv \bigotimes_{i=1}^{K'+L'} \left( \frac{1 + \tau_i \gamma_4}{2} \right) \chi_{a}^T, \\
       &\psi_{b \vec{\tau}}\equiv \bigotimes_{i=1}^{K+L} \left( \frac{1 + \tau_i \gamma_4}{2} \right) \psi_{b } .
    \end{split}
\end{equation}
The fact that the total fermion-field parity
\begin{equation}
\begin{split}
    \Pi &\equiv \Pi_S \Pi_{S'} = \prod_{i=1}^N \rho_i \rho'_i \prod_{j=1}^M \xi_j \widetilde{\xi}_j \prod_{k=1}^M \xi'_j \widetilde{\xi}'_j \\
    &= \prod_{i=1}^{K+L} \tau_i \prod_{j=1}^{K'+L'} \tau'_j,
    \end{split}
\end{equation}
is independent of how the quark fields are partitioned into source/sink, source/source, and sink/sink propagators implies that the pair of conditions Eq.~\eqref{eq:real_source}-\eqref{eq:real_sink} is equivalent to
\begin{equation}\label{eq:real_first}
\begin{split}
    &\psi_{b \vec{\tau}}(U_{(0)}^{*} ,-\vec{y}_1,\ldots,-\vec{y}_{K},-\vec{v}_1,\ldots,-\vec{v}_{L}) \\
    &= \eta\, \psi_{b \vec{\tau} }(U_{(0)}, \vec{y}_1,\ldots,\vec{y}_{K},\vec{v}_1,\ldots,\vec{v}_{L})^*,
    \end{split}
\end{equation}
and
\begin{equation}\label{eq:real_middle}
    \begin{split}
        &\chi_{a\vec{\tau}'}(U_{(t)}^{*}, -\vec{x}_1,\ldots,-\vec{x}_{K'},-\vec{u}_1,\ldots,-\vec{u}_{L'}) \\
        &= \eta'\, \chi_{a \vec{\tau}'}(U_{(t)}, \vec{x}_1,\ldots,\vec{x}_{K'},\vec{u}_1,\ldots,\vec{u}_{L'})^*,
    \end{split}
\end{equation}
where $\eta,\eta' \in \mathbb{C}$ are any normalization factors satisfying
\begin{equation}\label{eq:real_last}
    \eta \eta' = \Pi.
\end{equation}
The conditions Eqs.~\eqref{eq:real_first}-\eqref{eq:real_last} are sufficient to ensure that $C_{ab}(t) = C_{ab}(t)^*$ is real, and will be assumed for all correlator matrices studied in this work.
This means that Hermitian correlator matrices for which $\chi = \psi$ are real, symmetric matrices.

The use of standard source and sink interpolating operators leads to wavefunctions satisfying Eqs.~\eqref{eq:real_first}-\eqref{eq:real_last}.
In particular, Gaussian (momentum-)smeared~\cite{Gusken:1989ad,Gusken:1989qx,Bali:2016lva} sources and sinks are complex conjugated by replacing $U_\mu(x)$ with $U_\mu(x)^*$ and simultaneously replacing $\vec{x}$ with $-\vec{x}$ in momentum-smearing phases.
To determine whether the spin-structure of a set of interpolating operators is compatible with Eqs.~\eqref{eq:real_first}-\eqref{eq:real_last}, it is helpful to note that a sufficient (although not strictly necessary) set of conditions for satisfying Eqs.~\eqref{eq:real_first}-\eqref{eq:real_last} is
\begin{itemize}
    \item The spin-color-flavor weights multiplying all quark and antiquark fields in the source/sink interpolating operators are real in the Dirac-Pauli basis where $\gamma_4 = \text{diag}(1,1,-1,-1)$.
    \item All spatial wavefunctions (including iterative smearing kernels) are complex conjugated by replacing $U_\mu(x)$ with $U_\mu(x)^*$ and simultaneously replacing all spatial coordinates $\vec{x}$ with $-\vec{x}$.
    \item The product of source and sink interpolating operators includes an even number of odd-parity quark-field components.
\end{itemize}
The last condition will be true, for instance, for any correlation function formed from products of nucleon fields with $(q^T C \gamma_5 q) q$ spin structures times an even number of total pion fields (note those involving an odd number of pion fields vanish in an isospin- and $G$-parity symmetric theory).

It is also possible to obtain real correlation functions when there are an odd number of odd-parity quark-field components if the spatial wavefunction is replaced by minus its complex conjugate when the spatial coordinates are inverted, $\vec{x}$ with $-\vec{x}$.
If either but not both of the net quark-field parity or the spatial inversion parity are odd, then a correlation function will be purely imaginary.

\section{Left GEVP}
\label{app:leftGEVP}

As presented in \cref{sec:formalism}, block Lanczos applied to noisy correlator data involves distinct $L/R$ quantities with potentially different values.
The identification of GEVP as one step of block Lanczos in \cref{sec:gevp:vs-block} makes clear that the same $L/R$ distinction must apply for GEVP quantities as well, even when the correlator matrices are Hermitian or symmetric and real.
The moving-pivot estimators \cref{eq:gevp-mp-Z,eq:gevp-mp-Jeff} correspond to one-step block Lanczos $R$ estimators, assuming $G^R$ is real. Inequivalent $L$ quantities may be defined in terms of the left eigenvectors $G^L$ that arise in the left GEVP
\begin{equation}
    \sum_{b} G^L_{kb} C_{ba}(t_d) = \lambda_k^L(t_d, t_0) \sum_b  G^L_{kb} C^{-1}_{ba}(t_0) ~ .
\end{equation}
For Hermitian $\bm{C}(t)$, in the absence of statistical noise, all $\lambda_k^L = \lambda_k$ are real, and $G^L_{ka} = (G^R_{ak})^*$ such that all $R$ and $L$ GEVP estimators are identical.
While this is known to hold asymptotically at large $t_0, t_d$, this is nontrivial to prove algebraically from the GEVP perspective~\cite{Luscher:1990ck}; it is obvious from the Lanczos perspective.

Statistical noise at large $t$ may generically result in complex eigenvalues and break the $L/R$ coincidence, per the discussion of Hermitian subspaces in \cref{sec:Hermitian}.
It is clear from the Lanczos perspective that in general there is only a single set of (complex) eigenvalues $\lambda_k^L = \lambda_k = \bar{\lambda}_k^{(1)}$, the right GEVP eigenvectors are given by Eq.~\eqref{eq:GEVP=block1}, and the left GEVP eigenvectors are given analogously by
\begin{equation}
    G^L_{ak}(t_d,t_0) =  \sum_b  (\bar{\omega}^{-1})^{(1)}_{k1b} \bar{\beta}_{1ba}^{-1}.
\end{equation}
For complex eigenvalues, $G^L \neq G^R$ in general, but they may also differ for real eigenvalues; in either case, this leads to inequivalent $L/R$ estimators for overlaps and matrix elements.
From the Lanczos perspective, this is a clear indication of a spurious state outside the Hermitian subspace that should be discarded.

\section{Confidence interval estimation}
\label{app:ci}

\begin{figure}
    \includegraphics[width=\linewidth]{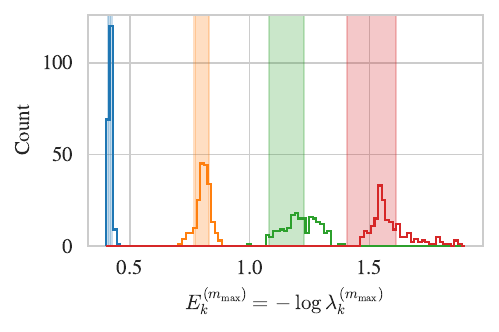}
    \caption{
       Histogram of the union of the bootstrap distributions of $E_n^{(m_{\rm max})}$ for all non-spurious states computed using bootstrap-median estimators. Shaded regions show Gaussian estimates for the 1$\sigma$ confidence interval, $E_n^{(m_{\rm max})} \pm \delta E_n^{(m_{\rm max})}$.
    }
    \label{fig:boot-last-hist}
\end{figure}

Histograms of Lanczos energy estimators in \cref{fig:boot-last-hist} clearly show qualitative signs of non-Gaussianity that are much larger for sample-mean estimators but still present for bootstrap-median estimators.
These deviations from Gaussianity can be quantified using the Kolmogorov-Smirnov and Shapiro-Wilk tests.
As shown in \cref{tab:gauss}, the sample-mean estimators are highly non-Gaussian, while the bootstrap-median estimators are much closer to Gaussian but still display clear signs of non-Gaussianity.

\begin{figure}
    \includegraphics[width=\linewidth]{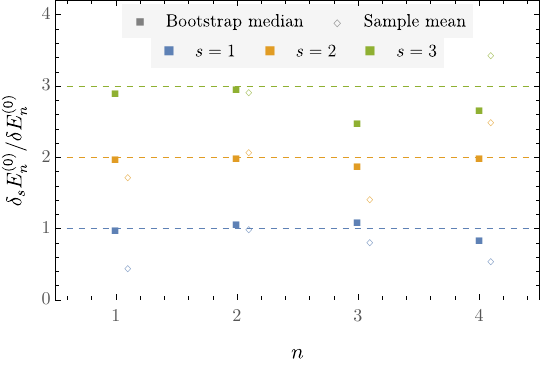}
    \caption{
        Comparisons of $s \sigma$ empirical bootstrap confidence intervals $\delta_s E_n^{(m)}$ with $s \in \{1,2,3\}$ with the Gaussian approximations $s \delta E_n^{(m)}$ involving the bootstrap variance. Open (closed) markers show results obtained using idiomatic bootstrap uncertainties for sample-mean (bootstrap-median) estimators.
    }
    \label{fig:boot-ci}
\end{figure}

Empirical bootstrap confidence intervals provide a standard estimator for computing bootstrap uncertainties that is more outlier robust than the bootstrap variance and applicable to non-Gaussian distributions~\cite{Efron:1982,Davison:1997}.
The empirical bootstrap formula providing $s\sigma$ confidence intervals $\delta_s X$ for any quantity $X$ with bootstrap samples $X^{(b)}$ is\footnote{Empirical and percentile bootstrap constructions give identical expressions for the width of the symmetric confidence interval~\cite{Davison:1997}.}
\begin{equation}\label{eq:CI}
\begin{split}
    \delta_s X &= \frac{1}{2}\left[ \text{Quantile}\left(X^{(b)}, \frac{1 + \text{erf}(s/\sqrt{2})}{2}\right) \right. \\
    &\hspace{30pt} \left. - \text{Quantile}\left(X^{(b)}, \frac{1 - \text{erf}(s/\sqrt{2})}{2}\right) \right].
    \end{split}
\end{equation}
Results for 1-3$\sigma$ confidence intervals can be computed using this definition; with $N_{\rm boot} = 200$ the 3$\sigma$ interval may be significantly impacted by finite $N_{\rm boot}$ effects and larger confidence intervals are likely to be unreliable.

These empirical bootstrap confidence interval results for $\delta_s X$ can be compared with the expectation of Gaussian uncertainties: $\delta_s X \approx s \delta X$ where $\delta X$ is defined from the bootstrap variance in \cref{eq:var}.
As shown in \cref{fig:boot-ci} for $E_n^{(m_{\rm max})}$, the approximate equality of these confidence interval estimates holds within 6\% for $n\in \{0,1\}$ states for $s\sigma$ confidence intervals with $s \in \{1,2,3\}$.
These deviations may be due in part to statistical fluctuations of the bootstrap estimators, which we do not quantify here but are proportional to $1/\sqrt{N_{\rm boot}}$.
For $n \in \{2,3\}$ the agreement is less good and deviations of up to 20\% are seen.
Identification of $s\sigma$ confidence intervals with $s \delta E_n^{(m_{\rm max})}$ therefore includes systematic uncertainties of these sizes.

\begin{table}[t]
    \centering
    \begin{ruledtabular}
    Sample mean
    \begin{tabular}{cllll}
    Gaussianity test & $E_0$ & $E_1$ & $E_2$ & $E_3$ \\ \hline \vspace{0.2em}
    Kolmogorov-Smirnov  & 
        $0$ & $0.001$ & $10^{-5}$ & $0$ \\
    Shapiro-Wilk  & 
        $10^{-26}$ & $10^{-5}$ & $10^{-17}$ & $10^{-15}$\\
    \end{tabular}
    \end{ruledtabular}\vspace{15pt}
    \begin{ruledtabular}
    Bootstrap median
    \begin{tabular}{cllll}
    Gaussianity test & $E_0$ & $E_1$ & $E_2$ & $E_3$ \\ \hline \vspace{0.2em}
    Kolmogorov-Smirnov  & 
        $0.350$ & $0.191$ & $0.501$ & $10^{-4}$ \\
    Shapiro-Wilk  & 
        $0.005$ & $0.191$ & $0.487$ & $10^{-7}$\\
    \end{tabular}
    \end{ruledtabular}
    \caption{
        Results of the Kolmogorov-Smirnov and Shapiro-Wilk tests for Gaussianity for the bootstrap distributions of bootstrap median block Lanczos energy estimators in \cref{fig:boot-last-hist} (bottom) and their sample mean counterparts (top).
    }
    \label{tab:gauss}
\end{table}

\section{Inequivalence of block Lanczos and GEVM/PGEVM}\label{app:PGEVM}

Ref.~\cite{Ostmeyer:2024qgu} asserts that block Lanczos is equivalent to ``the block PGEVM approach introduced in Ref.~\cite{Fischer:2020bgv}.'' 
That work discusses correlator matrix analysis via GEVP solutions, which they refer to as applications of the generalized eigenvalue method (GEVM).
The only discussion of generalizing PGEVM to correlator matrices present in Ref.~\cite{Fischer:2020bgv} (in which the word ``block'' does not appear) is ``There is one straightforward way to combine GEVM and PGEVM: we noted already above that the principal correlators of the GEVM are again a sum of exponentials, and, hence, the PGEVM can be applied to them.
This means a sequential application of first the GEVM with a correlator matrix of size $n_0$ to determine principal correlators $\lambda_k$ and then of the PGEVM with size $n_1$ and the $\lambda_k$’s as input, which we denote as GEVM/PGEVM.'' 

\begin{figure}
    \includegraphics[width=\linewidth]{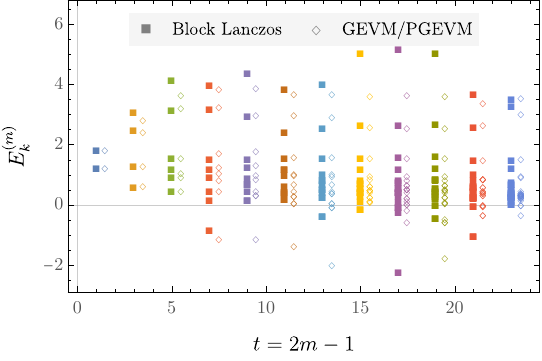}
    \caption{Comparison of block Lanczos with the GEVM/PGEVM method~\cite{Fischer:2020bgv} applied to the $2\times 2$ proton correlator matrix discussed in the main text. Both methods coincide with GEVP for $m=1$. Sample-mean estimators are shown; no spurious state filtering is performed.
    }
    \label{fig:block-gevmpgevm}
\end{figure}

\begin{figure}
    \includegraphics[width=\linewidth]{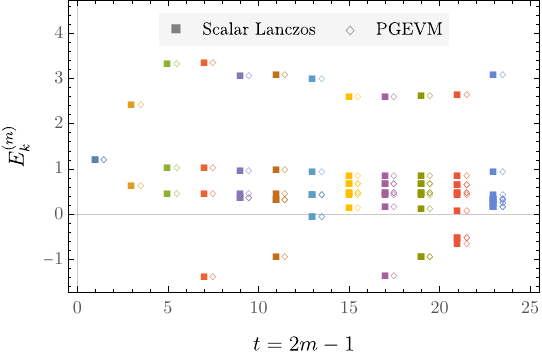}
    \caption{
        Comparison of scalar Lanczos with the PGEVM method~\cite{Fischer:2020bgv} applied to the $WW$ diagonal element of the proton correlator matrix discussed in the main text.  Sample-mean estimators are shown; no spurious state filtering is performed.
    }
    \label{fig:scalar-pgevm}
\end{figure}

It was first pointed out in Ref.~\cite{Wagman:2024rid} that ``the Ritz values are numerically identical to the solutions of the Prony generalized eigenvalue method (PGEVM) with $n=m$, $\tau_0 = 0$, and $\Delta = \delta t = 1$, which by construction coincide with the roots from Prony's method.''
Therefore, the GEVM/PGEVM method is equivalent to applying scalar Lanczos (without spurious state filtering) to the principal correlators $\lambda_k$ in Ref.~\cite{Fischer:2020bgv}, which according to the GEVP definition in Eq. (6) of that work are simply the GEVP eigenvalues that are also denoted $\lambda_k(t,t_0)$ here.

Applying the GEVM/PGEVM method for $m$ steps to a rank $r$ correlator matrix is therefore equivalent to separate Lanczos analysis of $r$ Krylov spaces, each of dimension $m$, generated by individually analyzing each $\lambda_k$. 
For $m=1$, both methods involve analyzing the same $r$-dimensional Krylov space as GEVP, and both are therefore equivalent to GEVP and hence one another for $m=1$.
For $m>1$, the $rm$ dimensional Krylov space analysis provided by block Lanczos is not equivalent to (and is strictly more optimal than) the Lanczos analyses of $r$ individual $m$-dimensional Krylov spaces provided by GEVM/PGEVM. This proves that GEVM/PGEVM cannot be equivalent to block Lanczos. 

To verify this result numerically, we show results for block Lanczos and GEVM/PGEVM applied to the central values of the $2\times 2$ proton correlator matrix discussed in \cref{sec:lattice,sec:gevp} in \cref{fig:block-gevmpgevm}.
The results are not identical for $m>1$.
This can be contrasted with the identical results for scalar Lanczos and PGEVM applied to the central values of a single diagonal correlator matrix entry shown in \cref{fig:scalar-pgevm}.


\bibliography{Lanczos_bib}

\end{document}